\documentclass[12pt, draftclsnofoot, onecolumn]{IEEEtran}
\usepackage{epsfig,graphicx,subfigure,psfrag,amsmath,cases}
\usepackage{latexsym,amssymb,amsmath,epsfig,subfigure,algorithm,mathtools}
\usepackage{algorithmic}
\usepackage{color}
\usepackage{url}
\usepackage{scrtime}
\usepackage{cite}
\usepackage{hyperref}
\usepackage{subfigure}
\usepackage{bbding}
\usepackage{bm}
\usepackage{multicol}
\author{Zhiqiang Wei, Yuanxin Cai, Zhuo Sun, Derrick Wing Kwan Ng, \\ Jinhong Yuan, Mingyu Zhou, and Lixin Sun
\thanks{Z. Wei, Y. Cai, D. W. K. Ng, and J. Yuan are with the School of Electrical Engineering and Telecommunications, the University of New South Wales, Australia; Z. Sun is with the Research School of Engineering, Australian National
University, Australia; M. Zhou and L. Sun are with Baicells Technologies, China (e-mail: \{zhiqiang.wei, yuanxin.cai\}@unsw.edu.au, zhuo.sun@anu.edu.au, \{w.k.ng, j.yuan\}@unsw.edu.au, \{Zhoumingyu, sunlixin\}@baicells.com).}\vspace{-17mm}}

\title{Sum-Rate Maximization for IRS-Assisted UAV OFDMA Communication Systems}

\newtheorem{T-Prob}{Transformed Problem}

\DeclareMathOperator{\maxo}{maximize}
\DeclareMathOperator{\mino}{minimize}
\DeclareMathOperator{\diag}{\mathrm{diag}}

\newtheorem{Remark}{Remark}

\newcommand{\abs}[1]{\lvert#1\rvert}

\begin{document}
\maketitle
\begin{abstract}
In this paper, we consider the application of intelligent reflecting surface (IRS) in unmanned aerial vehicle (UAV)-based orthogonal frequency division multiple access (OFDMA) communication systems, which exploits both the significant beamforming gain brought by the IRS and the high mobility of UAV for improving the system sum-rate.
%\cite{huang2019holographic,BasarReview,di2019smart,Ntontin2019RIS,qingqing2019towards,de2019non,WeiPerformanceGain,WuIRS,nadeem2019large}\cite{DMTadeoff}\cite{Sun2019,HeIRSChannelEstimation,lindstrom2019optimal}\cite{guo2019weighted,QiangLi2019,Zhao2019,han2019large,TangWanKaiIRS}\cite{WeiTCOM2017}\cite{TanFirstIRS,WeiTSSA}\cite{WeiPerformanceGain}\cite{tang2019wireless}\cite{SunMCNOMA}\cite{CuiWirelessScheduling,hu2018capacity}
%{Weijie IVT:
	%
	%Wireless communication is the key enabler for realizing the potentials of IVN in the future B5G wireless networks. In particular, millimeter wave mmWave communication \cite{ZhaoLouMMwAVE,wei2018multibeam,WeiBeamWidthControl} has been widely recognized as an enabling technology to meet the needs of data-intensive applications in future in-vehicle infotainment systems. Besides, the abundant spectrum in the mmWave frequency band can be exploited for accurate localization and tracking of vehicle. New distributed mmWave systems \cite{WeiTSSA,ZhaoDAS} should be developed for catering the inherent distributed property in signal processing and communications of IVN. In addition, to accommodate the massive number of vehicles in IVN with limited resources, novel multiple access schemes, such as non-orthogonal multiple access \cite{WeiTCOM2017,WeiPerformanceGain,WeiLetter2018} and random access \cite{SunALOHA,SunPNC,SunJUICETCOM}, need to be proposed for facilitating spectrally and energy-efficient communications.}
%
The joint design of UAV's trajectory, IRS scheduling, and communication resource allocation for the proposed system is formulated as a non-convex optimization problem to maximize the system sum-rate while taking into account the heterogeneous quality-of-service (QoS) requirement of each user.
The existence of an IRS introduces both frequency-selectivity and spatial-selectivity in the fading of the composite channel from the UAV to ground users.
To facilitate the design, we first derive the expression of the composite channels and propose a parametric approximation approach to establish an upper and a lower bound for the formulated problem.
An alternating optimization algorithm is devised to handle the lower bound optimization problem and its performance is compared with the benchmark performance achieved by solving the upper bound problem.
Simulation results unveil the small gap between the developed bounds and the promising sum-rate gain achieved by the deployment of an IRS in UAV-based communication systems.
\end{abstract}

\vspace{-6mm}
\section{Introduction}
The recent advancement of unmanned aerial vehicles (UAV) manufacturing technologies and substantial cost reduction have motivated extensive studies on the amalgamation between UAV and wireless communication systems\cite{zeng2019accessing}.
In particular, UAV-enabled wireless communication is expected to serve as a building block for the upcoming fifth-generation (5G) and beyond 5G (B5G) networks, which is potential to provide high data rate communications and to support massive access over a large area \cite{zeng2019accessing}.
For instance, serving as an aerial base station (BS), UAV-enabled communications provide an effective approach to combat channel fading owing to its high probability of establishing line-of-sight (LoS) links to ground users\cite{zeng2019accessing}.
Thanks to the high flexibility and the low cost deployment of UAVs, efficient traffic offloading for terrestrial cellular networks can be performed which relieves system performance bottlenecks due to overloaded traffic or blocked links.
Therefore, UAV-enabled wireless communications have drawn significant attention from both academia and industry lately\cite{ZengThroughMaxi,CaiUAVEESecure,ZengyongEnergyMinimization,QingqingWuUAV,SunUAV3D,DongfangUAV}.

Benefiting from its high maneuverability, UAV's trajectory can be designed to adapt to the actual propagation environment and the traffic demanding of the networks, which provides additional design degrees of freedom to improve the system performance.
As a result, the joint trajectory and resource allocation design for UAV communication systems has been extensively studied in the literature.
In \cite{ZengThroughMaxi}, the authors proposed to deploy a UAV to serve as a mobile relay and optimized its trajectory as well as the communication resource allocation so as to maximize the end-to-end system throughput.
Compared to static relaying systems, a substantial throughput gain can be achieved, which demonstrates the potentials of applying UAV in wireless communications.
Furthermore, due to the limited onboard battery capacity of UAVs, the energy efficiency maximization and the energy consumption minimization problems were studied in \cite{CaiUAVEESecure} and \cite{ZengyongEnergyMinimization}, respectively.
Extending to a multi-UAV network, the authors in \cite{QingqingWuUAV} jointly designed the user scheduling, the UAV's trajectory, and the power allocation to maximize the minimum average data rate among all the users.
Different from the existing works on trajectory design with a fixed altitude, the authors in \cite{SunUAV3D} proposed an optimal three-dimensional (3D) trajectory design for a solar-powered UAV communication system.
Additionally, the authors in \cite{DongfangUAV} mounted a multi-antenna array on the UAV and jointly designed the trajectory and precoder to minimize the total transmit power by taking into account practical UAV's jittering and user location's uncertainty.
Despite the fruitful results in the literature, the performance of UAV-based communication systems is still restrained by the limited service duration and the users with weak communication links.
As a result, there is an emerging need for the deployment of new technologies to fully unleash the potentials of UAV-based communications.

%However, employing intelligent reflecting surface (IRS) into UAV-enabled wireless communications has rarely been discussed in the literature, which motivates us to further investigate in this work.

Recently, intelligent reflecting surface (IRS) has attracted extensive attention in the wireless communication research community, due to its capability of shaping wireless propagation and establishing a programmable radio environment\cite{di2019smart,zhang2019multiple}.
In particular, an IRS is a meta-surface constituted by many meta-atoms, which are engineered to implement different interactive functions, such as absorption, reflection, refraction, and polarization, for the incoming electromagnetic waves shined on them\cite{di2019smart,Ntontin2019RIS}.
To be more specific, programmable integrated circuits (ICs) are introduced to manipulate the meta-atoms such that their impedance characteristics can be altered by an external IRS controller to adjust the amplitude and phase of the reflected signals\cite{qingqing2019towards}.
%
%One possible way to interact with the impinging waves is anomalous reflection, in which the impinging wave is defected away from the specular direction according to the generalized Snell's law\cite{Ntontin2019RIS}.
%
For instance, the authors in \cite{WuIRS} formulated the joint active and passive beamforming design problem to minimize the total transmit power and further extended to a practical case with a discrete phase control at an IRS \cite{WuIRSDiscrete}.
Besides, the authors in \cite{ChongWenIRS} investigated the possibility of deploying an IRS to improve the system energy efficiency and they demonstrated that a significant energy efficiency gain can be realized even though a low-resolution phase shifter is equipped at the IRS.
{Furthermore, the authors in \cite{Di2019reflection} proposed an analytical framework to quantify the performance limits of IRS in large-scale wireless networks and the LoS probability improvement with the large-scale deployment of IRSs were analyzed in \cite{kishk2020exploitingV2}.}
Also, promising performance gains can be brought by IRSs in terms of communication security\cite{yu2019robust}.
%The potentials of applying IRS to secure wireless communications have been investigated in \cite{chen2019intelligent} and \cite{yu2019robust}, which demonstrated the huge potentials of IRS to improve the communication security.
%
Most recently, different from existing works considering only narrow-band IRS communications, the authors in \cite{zheng2019intelligent} investigated the channel estimation and reflection coefficient optimization for IRS-enhanced multi-carrier orthogonal frequency-division multiplexing (OFDM) communication systems.
However, most of the existing works focused on applying the IRS technology in terrestrial communications and their results cannot directly apply to emerging applications with aerial communication nodes.
%
%In fact, the integration of aerial and terrestrial communications offers promising opportunities for B5G networks\cite{SekanderB5G,LiIoT}, which motivates us to explore the possibility of deploying an IRS in UAV communication systems.

The integration between the terrestrial IRS and UAV paves the way for the development of the B5G network to offer ubiquitous communication services \cite{SekanderB5G,LiIoT}.
It is well-known that mounting multiple antennas at wireless transceivers can further improve the communication system performance significantly\cite{DMTadeoff}, due to the potentials in exploiting multiplexing gains offered by the spatial degrees of freedom.
%
%However, most works in the literature focused on designing communication systems with the aid of a single-antenna UAV\cite{ZengThroughMaxi,CaiUAVEESecure,ZengyongEnergyMinimization,QingqingWuUAV,SunUAV3D}.
%
{However, the size, weight, and power (SWAP) constraints of UAVs hinder the deployment of advanced multiple-input multiple-output (MIMO) techniques for mitigating the detrimental fading effects.
On the other hand, the precoding design in the multi-antenna setting is coupled with the UAV's trajectory design which is challenging, since the effective channel gains between the UAV and ground users depend on both its trajectory and precoding strategy, resulting in highly non-convex functions \cite{DongfangUAV}.
In contrast, the IRS technology provides a promising but inexpensive solution to handle this dilemma, which can mimic the massive MIMO gain with a small number of active antennas \cite{ShaHuLIS,NadeemIRS}.
As a result, single-antenna UAV-assisted communications have been heavily studied in the literature \cite{RuideLi,You3DUAV,SunUAV3D}.}
%
%Hence, it is not cost-effective to allow a UAV equipping multiple antennas, not to mention equipping a power-hungry massive antenna array.
%
%On the other hand, the trajectory design for a multi-antenna UAV becomes very challenging since the channel gains between the UAV and ground users depend on both its trajectory and its precoding strategy which is highly non-convex\cite{DongfangUAV}.
%
%In contrast, the IRS technology provides a promising but inexpensive solution to handle this dilemma.
%
Nevertheless, an IRS offers a high passive beamforming gain via adjusting its reflection coefficients intelligently, without the need in deploying multiple antennas on UAV.
Therefore, the IRS can help ``recycling'' part of the dissipated signals by reflecting them back to the desired users, which is one of the main motivations of this work.
Secondly, deploying an IRS in UAV-enabled communication systems can improve the flexibility in designing UAV's trajectory.
For example, if a user is far away from the UAV but is close to an IRS, the UAV does not have to deliberately alter its route and fly close to this user to establish strong communication links, which is usually time- and energy-consuming.
Instead, an IRS can perform beamforming on the reflected signals jointly with the UAV to improve the received signal strength at the far ground user such that it can enjoy an acceptable data rate.

In practice, introducing an IRS into UAV-enabled communication systems brings both opportunities and challenges for its joint trajectory and resource allocation design.
Specifically, due to the existence of the IRS, the composite channel power gain compositing the direct link from the UAV to ground users and the reflected link via IRS is a complicated function of the UAV's trajectory.
Furthermore, how to efficiently schedule users to be assisted by the IRS is still unknown and deserves our efforts to explore.
Thirdly, as broadband communications have been widely adopted in current cellular networks, the additional reflected path of IRS indeed causes a frequency- and spatial-selective fading channel imposing a significant challenge for the trajectory design of UAV, which was overlooked by existing works based on frequency-flat channel models \cite{ZengThroughMaxi,CaiUAVEESecure,ZengyongEnergyMinimization,QingqingWuUAV,SunUAV3D,DongfangUAV}.
{Although a multi-carrier channel model was built for IRS-assisted communications in \cite{zheng2019intelligent}, it is not applicable to the UAV communication systems as it does not take into account the UAV's mobility.}
{At the time of writing, to the best of the authors' knowledge, there are three related works of applying IRS into UAV communication systems \cite{zhang2019reflections,li2019reconfigurable,GeIRSUAV}.
	Specifically, the authors in \cite{zhang2019reflections} equipped an IRS on UAV to
	improve the reliability of terrestrial millimeter-wave communication systems.
	To maximize the system sum-rate, a reinforcement-based learning method was applied to optimize the position of a UAV and the reflection coefficients of the IRS.
	Also, in \cite{li2019reconfigurable}, the IRS was mounted on a building surface and was
	treated as a passive relay to assist UAV communication systems.
	In particular, the reflection coefficients and the trajectory were designed jointly to maximize the system sum-rate.
	Furthermore, the authors in \cite{GeIRSUAV} proposed a joint design to maximize the received power for a multi-IRS-assisted UAV communication system.
	However, all these works \cite{zhang2019reflections,li2019reconfigurable,GeIRSUAV} considered a narrow-band channel model and their results do not valid for wideband systems.
	In addition, the works \cite{li2019reconfigurable,GeIRSUAV} considered only a simple single-user case, while has limited application scenarios in nowadays wireless systems.
	More importantly, applying the existing results of \cite{li2019reconfigurable,GeIRSUAV}
	to multi-user wideband systems may result in unsatisfactory performance.}

In this paper, we investigate the application of an IRS to UAV-based orthogonal frequency division multiple access (OFDMA) communication systems by studying the joint trajectory and resource allocation design to maximize the system sum-rate.
The main contributions of this work are summarized as follows:
\begin{enumerate}
	\item We propose a novel IRS-assisted UAV OFDMA communication system, which enjoys both the high beamforming gain of the IRS and the high mobility of the UAV.
	To support simultaneous multi-user communications, OFDMA is adopted for the proposed IRS-assisted UAV system, while the considered system model is fundamentally different from narrow-band IRS systems considered in the literature\cite{WuIRS,WuIRSDiscrete,ChongWenIRS,yu2019robust,zhang2019reflections,li2019reconfigurable}.
	\item Due to the additionally reflected propagation path introduced by the IRS, the composite channel gains from the UAV to ground users becomes both frequency- and spatial-selective which complicates the trajectory design of the UAV.
	To start with,	we first characterize the composite fading channels.
	Subsequently, based on the LoS component in Rician fading channels, we optimize the phase control strategy at the IRS to maximize the composite channel gain.
	Then, the joint trajectory, IRS scheduling, and resource allocation design for the proposed system is formulated as a non-convex optimization problem to maximize the system sum-rate while taking into account the heterogeneous quality-of-service (QoS) requirement of each user.
	\item Via exploiting the cosine fading pattern in the composite channel power gains, we propose a parametric approximation method to obtain an upper and a lower bound for the formulated problem.
	We focus on the practical solution design for the lower bound problem, while the performance achieved by solving the upper bound problem serves as a benchmark.
	An alternating optimization approach is adopted to facilitate the development of an iterative algorithm to achieve a suboptimal solution of the lower bound problem.
	\item {Extensive simulations are conducted to demonstrate the performance gain of the proposed scheme.
	The performance gap between the proposed parametric upper bound and lower bound problems is revealed, which can be reduced by an optimal selection of the approximation parameter at the expense of a higher complexity.
	We demonstrate that employing an IRS in UAV OFDMA communication systems can substantially improve the system sum-rate.}
\end{enumerate}

\begin{table*}[t]
	\scriptsize
	\caption{Notations for Main System Parameters.} \label{notations}
	\vspace{-8mm}
	\begin{center}
		\begin{tabular}{ c | c |c |c }
			\hline			
			Notations   & Physical meaning          & Notations   & Physical meaning \\ \hline
			$K$         & Total number of ground users    & $\phi_{m_{\rm r}, m_{\rm c}}$ & Phase control at PRU $\left(m_{\rm r}, m_{\rm c}\right)$  \\
			$M_{\rm c}$ & Total number of PRUs in each column of the IRS & $M_{\rm r}$  & Number of PRUs in each row of the IRS\\
			$N$         & Total number of time slots     & ${\bf{q}}\left[ n \right]$           & UAV's trajectory \\
			$\mathbf{w}_k$  & Location of user $k$      &  $\mathbf{w}_{\mathrm{R}}$  & Location of the IRS    \\
			$d_k^{{\rm{UG}}}\left[ n \right]$           & Distance between the UAV and ground user $k$ in time slot $n$ &   $d^{{\rm{UR}}}\left[ n \right]$          & Distance between the UAV and the IRS in time slot $n$ \\
			$d_k^{{\rm{RG}}}$    & Distance between the IRS and ground user $k$ & $N_{0}$         & Power spectrum density of thermal noise   \\
			$N_{\mathrm{F}}$    & Total number of subcarriers & $\Delta f$         & Subcarrier bandwidth	\\
			${\beta_0}$ & Channel power gain at the reference distance $d_0 = 1$ m & ${f_\mathrm{c}}$ & Carrier frequency \\
			$\theta^{\mathrm{UR}}[n]$  & 	Vertical AoA from the UAV to the IRS & $\xi^{\mathrm{UR}}[n]$  & Horizontal AoA from the UAV to the IRS \\
			$\theta_{k}^{\mathrm{RG}}$ & Vertical AoD from the IRS to ground user $k$ & $\xi_{k}^{\mathrm{RG}}$ & Horizontal AoA from the IRS to ground user $k$ \\
			$\alpha^{\mathrm{RG}}_k$& Path loss exponent of the IRS-to-user link for user $k$ & $\kappa^{\mathrm{RG}}_k$&  Rician factor of the IRS-to-user link for user $k$ \\
			$\alpha^{\mathrm{UG}}_k$& Path loss exponent of the UAV-to-user link for user $k$ & $\kappa^{\mathrm{UG}}_k$& Rician factor of the UAV-to-user link for user $k$\\
			$p_{\mathrm{max}}$ & Maximum transmission power in each time slot & ${p_{k,i}}\left[ n \right]$ & Power allocation variable \\
			${u_{k,i}}\left[ n \right]$ & User scheduling variable & $s_k \left[n\right]$ & IRS scheduling variable\\	\hline
		\end{tabular}
	\end{center}
	\vspace{-12mm}
\end{table*}

Notations used in this paper are listed as follows. Boldface capital and lower case letters are reserved for matrices and vectors, respectively. $\mathbb{C}^{M\times N}$ denotes the set of all $M\times N$ matrices with complex entries; ${\left( \cdot \right)^{\mathrm{T}}}$ denotes the transpose of a vector or a matrix and ${\left( \cdot \right)^{\mathrm{H}}}$ denotes the Hermitian transpose of a vector or a matrix;
$\abs{\cdot}$ denotes the absolute value of a complex scalar or the cardinality of a set;
and $\left\|\cdot\right\|$ denotes the Euclidean norm of a vector.
$\mathbf{X} \otimes \mathbf{Y}$ represents the Kronecker product of two matrices $\mathbf{X}$ and $\mathbf{Y}$; $\diag(\mathbf{x})$ denotes a diagonal matrix whose diagonal elements are given by its input vector $\mathbf{x}$.
The circularly symmetric complex Gaussian distribution with mean $\mu$ and variance $\sigma^2$ is denoted by ${\cal CN}(\mu,\sigma^2)$.
{For clarity, we first summarize the adopted notations of this paper in Table I.}

\vspace{-4mm}
\section{System Model}
{In this section, we first present the system and channel models of the considered IRS-assisted UAV communication system and then introduce the resource allocation and IRS scheduling variables.}

\begin{figure}[t]
	\centering\vspace{-5mm}
	\includegraphics[width=3.8in]{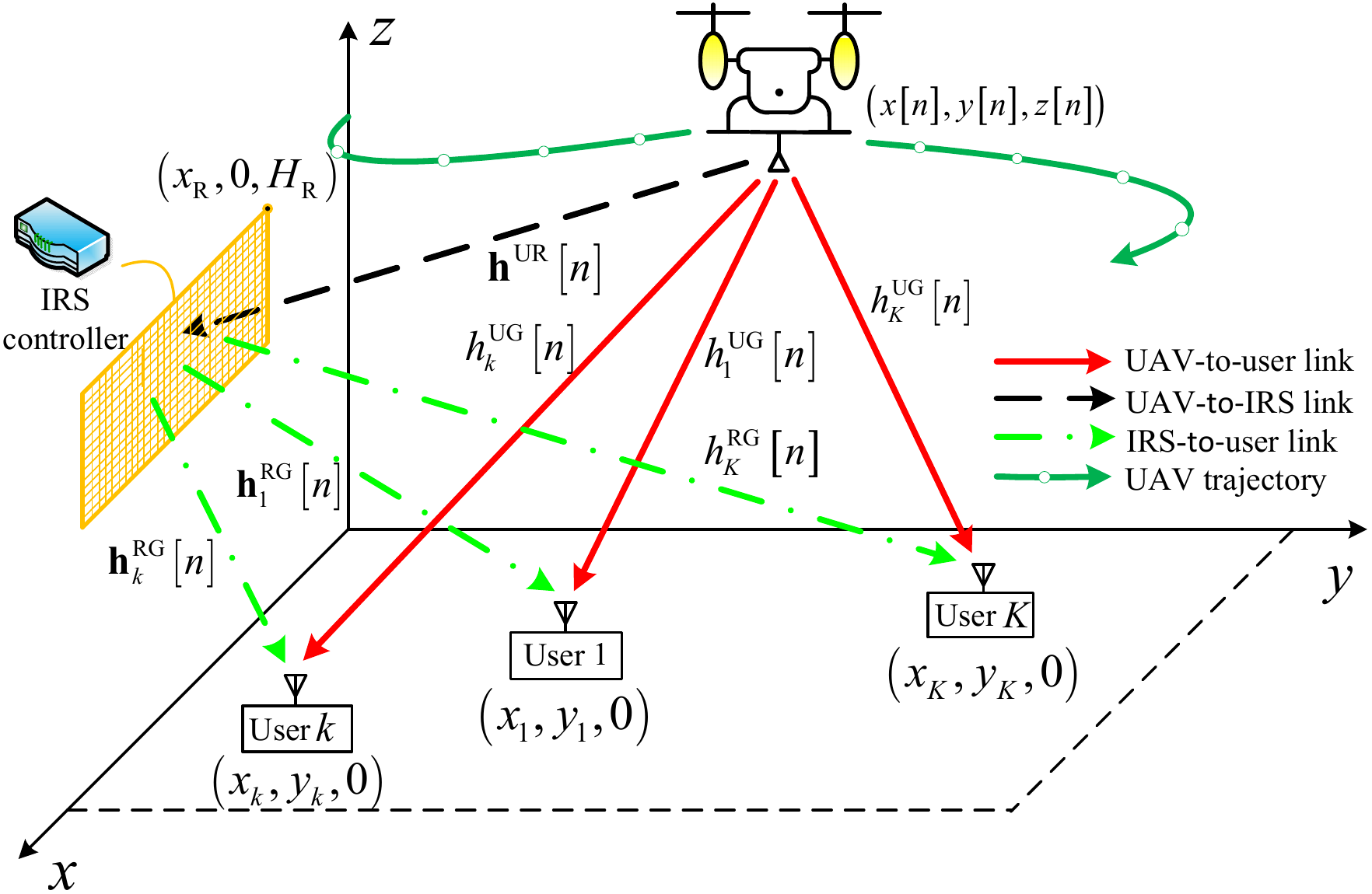}\vspace{-7mm}
	\caption{The system model of IRS-Assisted UAV communication systems.}\vspace{-10mm}
	\label{SystemModel}%
\end{figure}

\vspace{-4mm}
\subsection{System Setup}

We consider a single UAV\footnote{Depending on the applications and the types of UAV, the data intended for ground users can be downloaded or transmitted to the UAV in an offline manner or an online manner via out-of-band communication links \cite{zeng2019accessing}, respectively.} serving as an aerial BS  providing downlink communications to $K$ ground users within a considered area, as shown in Fig. \ref{SystemModel}.
Both the UAV and ground users are equipped with a single-antenna.
However, the single-antenna UAV is assisted by an intelligent reflection surface (IRS).
{To guarantee the IRS in the sight of both the UAV and ground users, we deploy the IRS at the boundary of the service area facing all the ground users\cite{PanMulticellIRS}.}
The IRS consists of $M_{\rm c} \times M_{\rm r}$ passive reflection units (PRUs), which are spanned as a uniform planar array (UPA).
In particular, each column of the UPA has $M_{\rm c}$ PRUs with an equal spacing of $d_{\mathrm{c}}$ meters and each row of the UPA consists of $M_{\rm r}$ PRUs with an equal spacing of $d_{\mathrm{r}}$ meters.
In particular, each PRU can re-scatter its incident signal with an independent reflection coefficient, which consists of an amplitude $a \in \left[0,1\right]$ and a phase shift $\phi_{m_{\rm r}, m_{\rm c}} \in \left[-\pi,\pi\right)$, i.e., $r_{m_{\rm r}, m_{\rm c}} = a e^{j\phi_{m_{\rm r}, m_{\rm c}}}$, $\forall m_{\rm r} \in \{1,2,\ldots,M_{\rm r }\}$, and $\forall m_{\rm c} \in \{1,2,\ldots,M_{\rm c }\}$.
Note that variable $a$ models the fixed reflection loss of IRS and $\phi_{m_{\rm r}, m_{\rm c}}$ denotes the phase shift inserted at PRU $\left({m_{\rm r}, m_{\rm c}}\right)$, which can be adjusted by the IRS controller\footnote{In this work, we consider an infinite resolution for the phase shifters at the IRS and the obtained performance serves as a performance upper bound of the one adopting finite resolution phase shifters\cite{ChongWenIRS,DiFinitePS,Sohrabi2016}.
Note that unlike existing works \cite{WuIRS,WuIRSDiscrete} considering the application of IRS without the UAV, optimizing the phase control strategy in this paper is very challenging due to the discrete constraint as it is coupled with the trajectory of the UAV, which will be investigated in our future work.}.
	%
	%Note that 2- or 3-bit phase shifters are able to achieve the average concatenation channel gain with infinite resolution phase shifters, thus only leading to a marginal performance degradation as commonly shown in the literature \cite{ChongWenIRS,DiFinitePS,Sohrabi2016}.

To facilitate the trajectory design, the total flying time, $T$, is discretized into $N$ time slots with an equal time interval, i.e., $\delta_{\mathrm{t}} = \frac{T}{N}$.
%
%We assume that the distance between the UAV and each ground user is invariant within each time slot since it is much longer compared to the displacement of the UAV during $\delta_{\mathrm{t}}$ \cite{ZengThroughMaxi,CaiUAVEESecure,ZengyongEnergyMinimization,QingqingWuUAV,SunUAV3D}.
%
The three dimensional (3D) trajectory of UAV can be denoted as a sequence $\left\{ {{\bf{q}}\left[ n \right] = {{\left[ {x\left[ n \right],y\left[ n \right],z\left[n\right]} \right]}^{\rm{T}}}} \right\}_{n = 1}^N$, where ${{\bf{q}}\left[ n \right] = {{\left[ {x\left[ n \right],y\left[ n \right],z\left[n\right]} \right]}^{\rm{T}}}}$ denotes the 3D coordinate of the UAV in time slot $n$.
In practice, we need to satisfy the minimum and maximum flight altitudes for the UAV due to some safety regulations, i.e., $H^{\mathrm{min}}_{\mathrm{U}} \le z\left[n\right] \le H^{\mathrm{max}}_{\mathrm{U}}$.
The location of ground user $k$ is assumed to be fixed and is denoted as $\mathbf{w}_k = {\left[ {x_k,y_k,0} \right]}^{\rm{T}}$.
The IRS is coated/installed on the surface of a building wall with a certain altitude $H_{\mathrm{R}}$, i.e., $\mathbf{w}_{\mathrm{R}} = \left[x_{\mathrm{R}},0,H_{\mathrm{R}}\right]^{\rm{T}}$.
In time slot $n$, the distance between the UAV and ground user $k$ is given by
$d_k^{{\rm{UG}}}\left[ n \right] =  {{{\left\| {{\bf{q}}\left[ n \right] - {{\bf{w}}_k}} \right\|}}}$
and the distance between the UAV and the IRS is given by
$d^{{\rm{UR}}}\left[ n \right] =  {{{\left\| {{\bf{q}}\left[ n \right] - {{\bf{w}}_{\rm{R}}}} \right\|}}}$.
{We assume that the distances $d^{{\rm{UR}}}\left[ n \right]$ and $d_k^{{\rm{UG}}}\left[ n \right]$ are invariant within each time slot $\delta_{\mathrm{t}}$ since UAV's displacement during $\delta_{\mathrm{t}}$ is much smaller than $d^{{\rm{UR}}}\left[ n \right]$ and $d_k^{{\rm{UG}}}\left[ n \right]$.}
In addition, the distance between the IRS and ground user $k$ is given by
$d_k^{{\rm{RG}}} =  { {{\left\| {{{\bf{w}}_{\rm{R}}} - {{\bf{w}}_{{k}}}} \right\|}}}$,
which is assumed to be fixed in the considered system.
Due to significant path loss and reflection loss, we assume that the power of the signals that are reflected by the IRS two or more times is negligible and thus ignored\cite{WuIRS,WuIRSDiscrete}.

\vspace{-4mm}
\subsection{Channel Model for IRS-assisted UAV OFDMA Communication Systems}
OFDMA has been widely adopted in practice to support multi-user communications owing to its flexibility in resource allocation design and the possibility of exploiting multi-user diversity\cite{DerrickEESWIPT}.
In the considered system, the total system bandwidth $B$ is divided into $N_{\mathrm{F}}$ subcarriers with subcarrier spacing $\Delta f = \frac{B}{N_{\mathrm{F}}}$.
In the following, we present the broadband channel model for the proposed system.
To facilitate the joint trajectory and resource allocation design, we assume LoS-dominated propagation among the UAV, the IRS, and ground users\cite{ZengThroughMaxi,CaiUAVEESecure,ZengyongEnergyMinimization,QingqingWuUAV,SunUAV3D,DongfangUAV}.
{Since the IRS is deployed in the higher altitude of $H_{\mathrm{R}} = 30$ m, the signal propagation of the UAV-to-IRS link typically occurs in clear airspace where the obstruction or reflection effects diminish.
	Therefore, as commonly adopted in the literature \cite{WuMultiUAV,SunUAV3D,CaiUAVEESecure}, we adopt a simple yet reasonably accurate LoS channel model between the UAV and the IRS to offer better insights into the performance and the design of this system.
	Note that considering a height-dependent path loss exponent model \cite{NguyenUAV,SimunekUAV} is an interesting but challenging problem, which will be investigated in our future work.}
{In time slot $n$, the channel vector between the UAV and the IRS on subcarrier $i$ is given by \cite{AlkhateebWideBandChannel}
\vspace{-2mm}
\begin{equation}\label{ChannelUR}
{\bf{h}}_i^{{\rm{UR}}}\left[ n \right] = \sqrt {\frac{\beta _0}{{\left( {d^{{\rm{UR}}}\left[ n \right]} \right)}^{2}}} {e^{ - j2\pi i \Delta f \frac{{{d^{{\rm{UR}}}}\left[ n \right]}}{c}}}{\bf{h}}_{\mathrm{LoS}}^{{\rm{UR}}}\left[ n \right] \vspace{-2mm}
\end{equation}
with
\vspace{-2mm}
\begin{align}\label{ChannelURLoS}
{\bf{h}}_{\mathrm{LoS}}^{{\rm{UR}}}\left[ n \right] = &\left[ {1,{e^{- j2\pi {f_\mathrm{c}}\frac{d_{\mathrm{r}} \sin \theta^{\mathrm{UR}}[n] \cos \xi^{\mathrm{UR}}[n]}{c}}}, \ldots ,{e^{- j2\pi {f_\mathrm{c}}\left( {{M_{\rm{r}}} - 1} \right)\frac{d_{\mathrm{r}} \sin \theta^{\mathrm{UR}}[n] \cos \xi^{\mathrm{UR}}[n]}{c}}}} \right]^{\mathrm{T}}  \notag\\[-1mm]
&\otimes\left[ {1,{e^{- j2\pi {f_\mathrm{c}}\frac{d_{\mathrm{c}} \sin \theta^{\mathrm{UR}}[n] \sin \xi^{\mathrm{UR}}[n]}{c}}}, \ldots ,{e^{- j2\pi {f_\mathrm{c}}\left( {{M_{\mathrm{c}}} - 1} \right)\frac{d_{\mathrm{c}} \sin \theta^{\mathrm{UR}}[n] \sin \xi^{\mathrm{UR}}[n]}{c}}}} \right]^{\mathrm{T}},
\end{align}
\vspace{-8mm}\par\noindent
where ${\beta_0}$ denotes the channel power gain at the reference distance $d_0 = 1$ m, $c$ denotes the speed of light, and $f_\mathrm{c}$ is the carrier frequency.
Variables $\theta^{\mathrm{UR}}[n]$ and $\xi^{\mathrm{UR}}[n]$ denote the vertical and horizontal angles-of-arrival (AoAs) at the IRS\footnote{Within one time slot, we can assume that $\theta^{\mathrm{UR}}[n]$ and $\xi^{\mathrm{UR}}[n]$ do not change as $\left|x\left[ n+1 \right] - x\left[ n \right]\right| \ll {{{d^{{\rm{UR}}}}\left[ n \right]}}$, $\left|y\left[ n+1 \right] - y\left[ n \right]\right| \ll {{{d^{{\rm{UR}}}}\left[ n \right]}}$, and $\left|z\left[ n+1 \right] - z\left[ n \right]\right| \ll {{{d^{{\rm{UR}}}}\left[ n \right]}}$ generally hold \cite{RuideLi,You3DUAV,SunUAV3D}.}, respectively, with $ \sin \theta^{\mathrm{UR}}[n] = \frac{z[n] - H_{\mathrm R}}{d^{\mathrm{UR}}[n]}$, $ \sin \xi^{\mathrm{UR}}[n] = \frac{x_{\mathrm R} - x[n]}{\sqrt{(x_{\mathrm R}-x[n])^2+(y_{\mathrm R}-y[n])^2}}$, and $ \cos \xi^{\mathrm{UR}}[n] = \frac{y[n] - y_{\mathrm R}}{\sqrt{(x_{\mathrm R}-x[n])^2+(y_{\mathrm R}-y[n])^2}}$.}
We assume a far-field array response vector model at the IRS since ${d^{{\rm{UR}}}\left[ n \right]} \gg \max\left(M_{\rm r}d_{\mathrm{r}},M_{\rm c}d_{\mathrm{c}}\right)$ holds in practice.
Additionally, we note that the IRS deals with a pass-band signal with a carrier frequency $f_\mathrm{c}$ and a bandwidth $B$ while $B \ll f_\mathrm{c}$ holds usually, i.e., a narrow-band signal in pass band.
Therefore, the array response vector in \eqref{ChannelURLoS} only depends on the corresponding AoAs and thus is frequency-flat\cite{AlkhateebWideBandChannel}, i.e., the beamforming gain is independent of the subcarrier index.
In contrast, the phase shift term ${e^{ - j2\pi i \Delta f \frac{{{d^{{\rm{UR}}}}\left[ n \right]}}{c}}}$ in \eqref{ChannelUR} depends on the subcarrier index\footnote{The frequency domain channel for a discrete LoS channel $\delta\left[n - {n_{\mathrm{\tau}}}\right]$ with a delay of $\tau = \frac{n_{\mathrm{\tau}}}{B}$ can be obtained via performing an $N_{\mathrm{F}}$-point discrete Fourier transform (DFT), i.e.,  $\mathrm{DFT}\left\{\delta\left[n - {n_{\mathrm{\tau}}}\right]\right\} = e^{-j2\pi \frac{i{n_{\mathrm{\tau}}}}{{N_{\mathrm{F}}}}} = e^{-j2\pi i \Delta f \tau},\; \forall i \in \{0,\ldots,{N_{\mathrm{F}}}-1\}$, where $\delta \left[ \cdot \right]$ denotes the delta function.} even if all subcarriers' signal  share the same delay $\frac{{{d^{{\rm{UR}}}}\left[ n \right]}}{c}$.
{In other words, a non-uniform phase is introduced to all subcarriers.}
We note that this is fundamentally different from the existing literature considering only narrow-band IRS communications\cite{WuIRS,WuIRSDiscrete,ChongWenIRS,yu2019robust,zhang2019reflections,li2019reconfigurable}, where the channel between each PRU and each user can be characterized by a single complex number.

On the other hand, due to the possible local scattering around the ground users, we adopt the Rician fading channel model for the UAV-to-user and IRS-to-user links\cite{zeng2019accessing}.
{In time slot $n$, the channel vector between the IRS and ground user $k$ on subcarrier $i$ is given by
\vspace{-2mm}
\begin{equation}\label{ChannelRG}
{\bf{h}}_{k,i}^{{\rm{RG}}}\left[ n \right] = \sqrt {\frac{\beta _0}{{\left( {{d^{{\rm{RG}}}_k}} \right)}^{\alpha^{\mathrm{RG}}_k}}} \left({\sqrt{\frac{\kappa^{\mathrm{RG}}_k}{\kappa^{\mathrm{RG}}_k + 1}}{e^{ - j2\pi i \Delta f \frac{{{d^{{\rm{RG}}}_k}}}{c}}} {\bf{h}}_{k,\mathrm{LoS}}^{{\rm{RG}}}} + {\sqrt{\frac{1}{\kappa^{\mathrm{RG}}_k + 1}}\tilde{\bf{h}}_{k,i}^{{\rm{RG}}}\left[ n \right]}\right),\vspace{-2mm}
\end{equation}
with the scattering component denoted as $\tilde{\bf{h}}_{k,i}^{{\rm{RG}}}\left[ n \right] \sim {\cal CN}(\mathbf{0},\mathbf{I}_{M_{\rm r} M_{\rm c}})$ and the LoS component given by
\vspace{-2mm}
\begin{align}\label{ChannelRGLoS}
{\bf{h}}_{k,\mathrm{LoS}}^{{\rm{RG}}} = & \left[ {1,{e^{ -j2\pi {f_\mathrm{c}}\frac{d_{\mathrm{r}} \sin \theta_{k}^{\mathrm{RG}} \cos \xi_{k}^{\mathrm{RG}}}{c}}}, \ldots ,{e^{ -j2\pi {f_\mathrm{c}}\left( {{M_{\rm{r}}} - 1} \right)\frac{d_{\mathrm{r}} \sin \theta_{k}^{\mathrm{RG}} \cos \xi_{k}^{\mathrm{RG}}}{c}}}} \right]^{\mathrm{T}} \notag\\[-1mm]
& \otimes \left[ {1,{e^{ -j2\pi {f_\mathrm{c}}\frac{d_{\mathrm{c}} \sin \theta_{k}^{\mathrm{RG}} \sin \xi_{k}^{\mathrm{RG}}}{c}}}, \ldots ,{e^{ -j2\pi {f_\mathrm{c}}\left( {{M_{\mathrm{c}}} - 1} \right)\frac{d_{\mathrm{c}} \sin \theta_{k}^{\mathrm{RG}} \sin \xi_{k}^{\mathrm{RG}}}{c}}}} \right]^{\mathrm{T}},
\end{align}
\vspace{-8mm}\par\noindent
where $\alpha^{\mathrm{RG}}_k$ denotes the path loss exponent of the IRS-to-user link for user $k$ and $\kappa^{\mathrm{RG}}_k$ is the corresponding Rician factor.
In the LoS component in \eqref{ChannelRGLoS}, $\theta_{k}^{\mathrm{RG}}$ and $\xi_{k}^{\mathrm{RG}}$ denote the vertical and horizontal angles-of-departure (AoDs) from the IRS to ground user $k$, respectively.
Note that we have $\sin \theta_k^{\mathrm{RG}} = \frac{H_{\mathrm R}}{d^{\mathrm{RG}}_k}$, $\sin \xi_{k}^{\mathrm{RG}} = \frac{x_k-x_{\mathrm R} }{\sqrt{(x_{\mathrm R}-x_k)^2+(y_{\mathrm R}-y_k)^2}}$, and $\cos \xi_{k}^{\mathrm{RG}} = \frac{y_k-y_{\mathrm R}}{\sqrt{(x_{\mathrm R}-x_k)^2+(y_{\mathrm R}-y_k)^2}}$.}
In time slot $n$, the channel between the UAV and ground user $k$ on subcarrier $i$ is given by
\vspace{-2mm}
\begin{equation}\label{ChannelUG}
h_{k,i}^{{\rm{UG}}}\left[ n \right] = \sqrt{\frac{\beta _0}{\left( {d_k^{{\rm{UG}}}\left[ n \right]} \right)^{\alpha^{\mathrm{UG}}_k}}} \left({ { \sqrt{\frac{\kappa^{\mathrm{UG}}_k}{\kappa^{\mathrm{UG}}_k + 1}}e^{-j2\pi i \Delta f \frac{{d_k^{{\rm{UG}}}\left[ n \right]}}{c}}} + {\sqrt{\frac{1}{\kappa^{\mathrm{UG}}_k + 1}} \tilde{h}_{k,i}^{{\rm{UG}}}\left[ n \right]}}\right),\vspace{-2mm}
\end{equation}
where $\alpha^{\mathrm{UG}}_k$ represents the path loss exponent of the UAV-to-user link for user $k$, $\kappa^{\mathrm{UG}}_k$ denotes the corresponding Rician factor,  and $\tilde{h}_{k,i}^{{\rm{UG}}}\left[ n \right] \sim {\cal CN}(0,1)$ is the scattering component of user $k$ on subcarrier $i$ in time slot $n$.

%\begin{Remark}
%\textit{	We note that an altitude/angle-dependent Rician fading channel model or a probabilistic LoS channel model was proposed most recently\cite{zeng2019accessing}, where the Rician factor and the LoS probability depend on UAV's altitude and trajectory.
	%
	%In this case, the UAV's trajectory design and the phase control at IRS become more involved.
	%
	%As a first attempt of applying IRS in UAV OFDMA communication systems, we consider a static Rician fading channel model in this work, where the LoS component exists and the Rician factors $\alpha^{\mathrm{RG}}_k$ and $\alpha^{\mathrm{UG}}_k$ remain unchanged in the considered $N$ time slots.
	%
	%Note that this assumption is reasonable for static propagation environments, especially in rural area \cite{zeng2019accessing}.
	%
	%Based on the altitude/angle-dependent Rician fading or a probabilistic LoS channel model, the sum-rate maximization problem for the IRS-assisted UAV communication systems is an interesting topic for our future work.}
%\end{Remark}

In time slot $n$, the IRS reflection phase coefficient matrix can be represented by
\vspace{-3mm}
\begin{equation}\label{IRSPhase}
\mathbf{\Phi}\left[n\right] = \diag\left(\boldsymbol{\phi}\left[n\right]\right) \in \mathbb{C}^{M_{\rm r} M_{\rm c} \times M_{\rm r} M_{\rm c}},\vspace{-3mm}
\end{equation}
where $\boldsymbol{\phi}\left[n\right] = \left[ e^{j\phi_{1,1}\left[n\right]},\ldots, e^{j\phi_{m_{\rm r}, m_{\rm c}}\left[n\right]},\ldots, e^{j\phi_{{M_{\rm r}, M_{\rm c}}}\left[n\right]}\right]^{\mathrm{T}} \in \mathbb{C}^{M_{\rm r} M_{\rm c} \times 1}$.
In time slot $n$, the concatenation channel for the UAV-IRS-user link of user $k$ on subcarrier $i$ is given by
\vspace{-2mm}
\begin{align}\label{Channel_UAV_IRS_user}
{{h}}_{k,i}^{{\rm{URG}}}\left[ n \right] &= a \left({\bf{h}}_{k,i}^{{\rm{RG}}}\right)^{\mathrm{T}}  \mathbf{\Phi}\left[n\right] {\bf{h}}_i^{{\rm{UR}}}\left[ n \right] = \frac{a \beta _0 }{{{d^{{\rm{UR}}}}\left[ n \right]}\left({ {d_k^{{\rm{RG}}}}}\right)^{\frac{\alpha^{\mathrm{RG}}_k}{2}}} \left\{\sqrt{\frac{\kappa^{\mathrm{RG}}_k}{\kappa^{\mathrm{RG}}_k + 1}}{e^{ - j2\pi i \Delta f \frac{{{d^{{\rm{UR}}}}\left[ n \right]} + {d_k^{{\rm{RG}}}}}{c}}}  \right. \notag\\[-1mm]
&\left.\times \left({\bf{h}}_{k,\mathrm{LoS}}^{{\rm{RG}}}\right)^{\mathrm{T}} \mathbf{\Phi}\left[n\right] {\bf{h}}_{\mathrm{LoS}}^{{\rm{UR}}}\left[ n \right] + \sqrt{\frac{1}{\kappa^{\mathrm{RG}}_k + 1}} {e^{ - j2\pi i \Delta f \frac{{{d^{{\rm{UR}}}}\left[ n \right]}}{c}}} \left(\tilde{\bf{h}}_{k,i}^{{\rm{RG}}}\right)^{\mathrm{T}} \mathbf{\Phi}\left[n\right] {\bf{h}}_{\mathrm{LoS}}^{{\rm{UR}}}\left[ n \right] \right\},
\end{align}
\vspace{-10mm}\par\noindent
where
\vspace{-2mm}
\begin{align}
\left({\bf{h}}_{k,\mathrm{LoS}}^{{\rm{RG}}}\right)^{\mathrm{T}}\mathbf{\Phi}\left[n\right] {\bf{h}}_{\mathrm{LoS}}^{{\rm{UR}}}\left[ n \right] \hspace{-1mm} &= \hspace{-1mm}\sum\limits_{{m_{\rm{c}}} = 1}^{{M_{\rm{c}}}} \sum\limits_{{m_{\rm{r}}} = 1}^{{M_{\rm{r}}}} {e^{ - j2\pi {f_\mathrm{c}}\frac{{d_{\mathrm{r}} \left( {{m_{\rm{r}}} -\hspace{-0.5mm} 1} \right){\sin \theta_{k}^{\mathrm{RG}} \cos \xi_{k}^{\mathrm{RG}}} + d_{\mathrm{c}}\left( {{m_{\rm{c}}} - \hspace{-0.5mm}1} \right){\sin \theta_{k}^{\mathrm{RG}} \sin \xi_{k}^{\mathrm{RG}}}} }{c}}}{e^{j{\phi _{{m_{\rm{r}}},{m_{\rm{c}}}}\left[ n \right]}}} \notag\\[-1mm]
&\hspace{20mm}\times {e^{ - j2\pi {f_\mathrm{c}}\frac{{d_{\mathrm{r}} \left( {{m_{\rm{r}}} -\hspace{-0.5mm} 1} \right){\sin \theta^{\mathrm{UR}}[n] \cos \xi^{\mathrm{UR}}[n]} + d_{\mathrm{c}}\left( {{m_{\rm{c}}} -\hspace{-0.5mm} 1} \right){\sin \theta^{\mathrm{UR}}[n] \sin \xi^{\mathrm{UR}}[n]}} }{c}}}  \notag\\[-1mm]
\text{and}\;\tilde{{h}}_{k,i}^{{\rm{URG}}}\left[ n \right] \hspace{-1mm}&=\hspace{-1mm} {e^{ - j2\pi i \Delta f \frac{{{d^{{\rm{UR}}}}\left[ n \right]}}{c}}} \left(\tilde{\bf{h}}_{k,i}^{{\rm{RG}}}\left[ n \right]\right)^{\mathrm{T}}\mathbf{\Phi}\left[n\right] {\bf{h}}_{\mathrm{LoS}}^{{\rm{UR}}}\left[ n \right].\label{ConcagtenationChannel}
\end{align}
\vspace{-10mm}\par\noindent
%\begin{Remark}
%\textit{	We note that each PRU of the IRS serves as a diffuser \cite{Ntontin2019RIS} since its size is on the order of the wavelength of the impinging radio waves, i.e., \emph{electrically small}\cite{Ntontin2019RIS}.
	%
	%As a result, the path loss along the UAV-IRS-user link in \eqref{Channel_UAV_IRS_user} is inversely proportional to the product of ${{{d^{{\rm{UR}}}}\left[ n \right]}}$ and $\left({ {d_k^{{\rm{RG}}}}}\right)^{\frac{\alpha^{\mathrm{RG}}_k}{2}}$.
	%
	%In contrast, when the geometric size of each PRU is \emph{electrically large}\cite{Ntontin2019RIS}, e.g. ten times larger than the wavelength, the path loss along the UAV-IRS-user link follows the geometric scattering model and hence it is inversely proportional to the total propagation distance ${{{d^{{\rm{UR}}}}\left[ n \right]}} + { {d_k^{{\rm{RG}}}}}$ \cite{Ntontin2019RIS}.
	%
	%In general, the electrically large PRU may outperform the electrically small one as it suffers less path loss given the same number of PRUs.
	%
	%However, we adopt electrically small PRU in our considered system model such that equipping massive PRUs is possible and a substantial passive beamforming gain can be achieved via manipulating the phase shifters at the IRS.
	%
	%More importantly, since ${ {d_k^{{\rm{RG}}}}}$ is a constant and is independent of the UAV's trajectory, the joint design for the case of electrically large PRU can be handled via some simple transformation of the results obtained in this work.}
%\end{Remark}

Now, the composite channel from the UAV to ground user $k$ on subcarrier $i$ in time slot $n$ can be given by
\vspace{-2mm}
\begin{equation}\label{FinalChannel}
{{{g}}}_{k,i}^{{\rm{UG}}}\left[ n \right] = h_{k,i}^{{\rm{UG}}}\left[ n \right] + {{h}}_{k,i}^{{\rm{URG}}}\left[ n \right] = {{{g}}}_{k,i,\mathrm{LoS}}^{{\rm{UG}}}\left[ n \right] + \tilde{{{g}}}_{k,i}^{{\rm{UG}}}\left[ n \right],\vspace{-3mm}
\end{equation}
where
\vspace{-2mm}
\begin{align}
{{{g}}}_{k,i,\mathrm{LoS}}^{{\rm{UG}}}\left[ n \right] & = \sqrt{\frac{\beta _0}{\left( {d_k^{{\rm{UG}}}\left[ n \right]} \right)^{\alpha^{\mathrm{UG}}_k}}} \sqrt{\frac{\kappa^{\mathrm{UG}}_k}{\kappa^{\mathrm{UG}}_k + 1}} {e^{-j2\pi i \Delta f \frac{{d_k^{{\rm{UG}}}\left[ n \right]}}{c}}} +\frac{a \beta _0 }{{{d^{{\rm{UR}}}}\left[ n \right]}\left({ {d_k^{{\rm{RG}}}}}\right)^{\frac{\alpha^{\mathrm{RG}}_k}{2}}} \sqrt{\frac{\kappa^{\mathrm{RG}}_k}{\kappa^{\mathrm{RG}}_k + 1}} \label{ChannelCompositeLoS}\\[-1mm]
&\times {e^{ - j2\pi i \Delta f \frac{{{d^{{\rm{UR}}}}\left[ n \right]} + {d_k^{{\rm{RG}}}}}{c}}} \left({\bf{h}}_{k,\mathrm{LoS}}^{{\rm{RG}}}\right)^{\mathrm{T}}\mathbf{\Phi}\left[n\right] {\bf{h}}_{\mathrm{LoS}}^{{\rm{UR}}}\left[ n \right]\;\text{and}\notag\\[-1mm]
%&\times \sum\limits_{{m_{\rm{c}}} = 1}^{{M_{\rm{c}}}} {\sum\limits_{{m_{\rm{r}}} = 1}^{{M_{\rm{r}}}} {{e^{ - j2\pi {f_\mathrm{c}}\frac{{d_{\mathrm{r}} \left( {{m_{\rm{r}}} - 1} \right){\sin \theta_{k}^{\mathrm{RG}} \cos \xi_{k}^{\mathrm{RG}}} + d_{\mathrm{c}}\left( {{m_{\rm{c}}} - 1} \right){\sin \theta_{k}^{\mathrm{RG}} \sin \xi_{k}^{\mathrm{RG}}}} }{c}}}{e^{j{\phi _{{m_{\rm{r}}},{m_{\rm{c}}}}\left[ n \right]}}}{e^{ - j2\pi {f_\mathrm{c}}\frac{{d_{\mathrm{r}} \left( {{m_{\rm{r}}} - 1} \right){\sin \theta^{\mathrm{UR}}[n] \cos \xi^{\mathrm{UR}}[n]} + d_{\mathrm{c}}\left( {{m_{\rm{c}}} - 1} \right){\sin \theta^{\mathrm{UR}}[n] \sin \xi^{\mathrm{UR}}[n]}} }{c}}}} } \;\text{and}\notag\\[-1mm]
\tilde{{{g}}}_{k,i}^{{\rm{UG}}}\left[ n \right] & = \sqrt{\frac{\beta _0}{\left( {d_k^{{\rm{UG}}}\left[ n \right]} \right)^{\alpha^{\mathrm{UG}}_k}}} \sqrt{\frac{1}{\kappa^{\mathrm{UG}}_k + 1}} \tilde{h}_{k,i}^{{\rm{UG}}}\left[ n \right] + \frac{a \beta _0 }{{{d^{{\rm{UR}}}}\left[ n \right]}\left({ {d_k^{{\rm{RG}}}}}\right)^{\frac{\alpha^{\mathrm{RG}}_k}{2}}} \sqrt{\frac{1}{\kappa^{\mathrm{RG}}_k + 1}}\tilde{{h}}_{k,i}^{{\rm{URG}}}\left[ n \right].\label{ChannelCompositeScattering}
\end{align}
\vspace{-6mm}\par\noindent

We can observe that the existence of scattering components of the Rician fading channels in \eqref{ChannelRG} and \eqref{ChannelUG} makes the composite channel non-deterministic.
Hence, to optimally control the phase of the IRS for coherent combining of signals at the ground user and to design the trajectory of the UAV, the instantaneous channels between each PRU, UAV, and each ground users should be estimated and be fed back to the IRS controller and the UAV controller on the fly, where the associated signaling overhead would consume a lot of system resources.
As a comprise approach, we focus on the phase control and trajectory design based on the deterministic LoS component.
In particular, the LoS CSI components change slowly compared to the scattering components and are predictable based on the UAV's trajectory, {which enables an offline design}\cite{CaiUAVEESecure}.
We note that this simplification is valid for a major range of the application scenarios of UAV communications, where a large Rician factor dominates the system performance \cite{zeng2019accessing}.
Note that due to the randomness of fading channels, the proposed LoS-based design may lead to an outage which the data rate is larger than the capacity.
In this work, we first focus on design for the system sum-rate maximization based on LoS channels and illustrate the system outage rate in Rician fading channels in the section of simulations.
%
%Note that a robust design on UAV's trajectory, IRS scheduling, and resource allocation taking into account the outage probability is an interesting and challenging topic, which will be considered in our future work.

%When designing the resource allocation based on the LoS channels, the phase control at IRS is irrelevant to the scattering component $\tilde{\bf{h}}_{k,i}^{{\rm{RG}}}$ in \eqref{ConcagtenationChannel} and thus $\tilde{{h}}_{k,i}^{{\rm{URG}}}\left[ n \right]  \sim {\cal CN}(0,{M_{\rm{c}}}{M_{\rm{r}}})$.
{Combining \eqref{ChannelURLoS}, \eqref{IRSPhase}, and \eqref{ConcagtenationChannel}, we have
	\vspace{-2mm}
	\begin{align}
	\tilde{{h}}_{k,i}^{{\rm{URG}}}\left[ n \right] &= \sum_{m_{\rm{r}}=1}^{M_{\rm{r}}}\sum_{m_{\rm{c}}=1}^{M_{\rm{c}}}{e^{ - j2\pi i \Delta f \frac{{{d^{{\rm{UR}}}}\left[ n \right]}}{c}}} \left\{\tilde{\bf{h}}_{k,i}^{{\rm{RG}}}\left[ n \right]\right\}_{m_{\rm r},m_{\rm c}} e^{j\phi_{m_{\rm r},m_{\rm c}}\left[n\right]} \notag\\[-1mm]
	&\times {e^{- j2\pi {f_\mathrm{c}}\left( {{m_{\rm{r}}} - 1} \right)\frac{d_{\mathrm{r}} \sin \theta^{\mathrm{UR}}[n] \cos \xi^{\mathrm{UR}}[n]}{c}}} {e^{- j2\pi {f_\mathrm{c}}\left( {{m_{\rm{c}}} - 1} \right)\frac{d_{\mathrm{c}} \sin \theta^{\mathrm{UR}}[n] \sin \xi^{\mathrm{UR}}[n]}{c}}},
	\end{align}
	\vspace{-10mm}\par\noindent
	where $\left\{\tilde{\bf{h}}_{k,i}^{{\rm{RG}}}\left[ n \right]\right\}_{m_{\rm r},m_{\rm c}} \sim {\cal CN}(0,1)$ is the $\left(m_{\rm c}-1\right)M_{\rm r} + m_{\rm r}$ element of the vector $\tilde{\bf{h}}_{k,i}^{{\rm{RG}}}\left[ n \right]$.
	Since the phase terms in the equation above are irrelevant to $\tilde{\bf{h}}_{k,i}^{{\rm{RG}}}\left[ n \right]$, we have $\tilde{{h}}_{k,i}^{{\rm{URG}}}\left[ n \right]  \sim {\cal CN}(0,{M_{\rm{c}}}{M_{\rm{r}}})$	\cite{lemons2002introduction}.
	According to \eqref{ChannelCompositeScattering}, the scattering component of the UAV-IRS-user link follows
	\vspace{-2mm}
	\begin{equation}
	\tilde{{{g}}}_{k,i}^{{\rm{UG}}}\left[ n \right] \sim {\cal CN}\left(0,\frac{\beta _0}{\left( {d_k^{{\rm{UG}}}\left[ n \right]} \right)^{\alpha^{\mathrm{UG}}_k}}\frac{1}{\kappa^{\mathrm{UG}}_k + 1} + \frac{a^2 \beta^2 _0 }{\left({{d^{{\rm{UR}}}}\left[ n \right]}\right)^2\left({ {d_k^{{\rm{RG}}}}}\right)^{{\alpha^{\mathrm{RG}}_k}} }\frac{{M_{\rm{c}}}{M_{\rm{r}}}}{\kappa^{\mathrm{RG}}_k + 1} \right).
	\end{equation}}
On the other hand, we refer ${{{g}}}_{k,i,\mathrm{LoS}}^{{\rm{UG}}}\left[ n \right]$ as the LoS component of the composite channel.
In fact, ${{{g}}}_{k,i,\mathrm{LoS}}^{{\rm{UG}}}\left[ n \right]$ is not a single LoS path but consists of two LoS paths in the UAV-to-user link and the UAV-IRS-user link, respectively.
Since ${{{g}}}_{k,i,\mathrm{LoS}}^{{\rm{UG}}}\left[ n \right]$ is deterministic, we still refer it as the LoS component of the composite channel and denote it with a subscript of LoS.
More importantly, the LoS component of the composite channel consists of two dominated paths with different delays and thus is a frequency-selective channel.

{Benefiting from the offline design, the proposed design in this work only requires the information of users' location and the Rician factors of all the involved links in advance, which significantly reduces the required overhead for CSI acquisition.
Besides, for a typical UAV flying speed, i.e., $20$ m/s \cite{zeng2019accessing}, we can assume that the inter-carrier interference (ICI) caused by Doppler spread can be efficiently mitigated by proper receiver design as stated in \cite{MostofiICI}.
We note that IRS has potential to eliminate the Doppler effect by counteracting the channel fluctuation \cite{basar2019reconfigurable}.
Yet, this requires perfect channel tracking and real-time phase control, which impose a challenge in implementation.}
%
%Since the IRS is usually deployed much closer to the ground users compared to the UAV for efficiently reflection, we assume that the delay spread between the direct UAV-to-user link and the composite UAV-IRS-user link is smaller than an orthogonal frequency-division multiplexing (OFDM) symbol duration, i.e., $\frac{{{d^{{\rm{UR}}}}\left[ n \right]} + {d_k^{{\rm{RG}}}} - {d_k^{{\rm{UG}}}\left[ n \right]}}{c} \le \frac{N_{\mathrm{F}}}{B}$, such that the channel response is flat within each subcarrier and it can be characterized by a complex number.
%
%In addition, we assume that the delay spread is smaller than the cyclic prefix inserted in the OFDM symbol, i.e., $\frac{{{d^{{\rm{UR}}}}\left[ n \right]} + {d_k^{{\rm{RG}}}} - {d_k^{{\rm{UG}}}\left[ n \right]}}{c} \le \frac{L_{\mathrm{CP}}}{B}$ with $L_{\mathrm{CP}}$ denoting the length of cyclic prefix, so that current OFDM symbol does not suffer the inter-symbol interference (ISI) from previous OFDM symbol.

\vspace{-4mm}
\subsection{Resource Allocation and IRS Allocation Design}
To serve multiple users concurrently, we adopt OFDMA via scheduling different users exclusively on different subcarriers and optimize the transmit power to the users.
If user $k$ is allocated to subcarrier $i$ in time slot $n$, we denote ${u_{k,i}}\left[ n \right] = 1$.
Otherwise, ${u_{k,i}}\left[ n \right] = 0$.
To guarantee the orthogonality among users on each subcarrier in each time slot, we impose $\sum_{k = 1}^{K} {u_{k,i}}\left[ n \right] \le 1$, $\forall i,n$.
Besides, the power allocated to user $k$ on subcarrier $i$ in time slot $n$ is denoted as ${p_{k,i}}\left[ n \right] \ge 0$ with
$\sum_{i = 1}^{N_{\mathrm{F}}} \sum_{k = 1}^{K} {p_{k,i}}\left[ n \right] \le p_{\mathrm{max}}$, $\forall n$,
where $p_{\mathrm{max}}$ denotes the maximum transmission power in each time slot.
As shown in \eqref{FinalChannel}, the phase shift introduced by each PRU affects the channel of all users in all subcarriers, since each PRU reflects the whole broadband signal to all the ground users \cite{zheng2019intelligent}.
{This is significantly different from existing works on IRS-assisted UAV communications \cite{zhang2019reflections,li2019reconfigurable} that considering a narrow-band and a single-user system.}
To achieve a considerable reflection gain at the IRS, we assume that the phases of the IRS reflection matrix is aligned w.r.t. one selected user in each time slot, which is defined as \textit{IRS-assisted user} in this paper\footnote{Allowing the IRS to align its beamforming phase shift matrix with respect to multiple users in each time slot may further improve the system performance. However, it complicates the UAV's trajectory design, which will be considered in our future work.}.
In particular, when user $k$ is scheduled as an IRS-assisted user in time slot $n$, we have $s_k \left[n\right]= 1$.
Otherwise, $s_k\left[n\right] = 0$.
Therefore, we have $\sum_{k=1}^{K}s_k\left[n\right] \le 1$, $\forall n$.

\vspace{-4mm}
\section{Phase Control at IRS and The Composite Channel Gain}
In this section, we first design the phase control strategy at the IRS based on LoS channels and then derive the composite channel gain of the UAV-IRS-user link.

When allocating the IRS for user $k$ in time slot $n$, i.e., $s_k \left[n\right]= 1$, to maximize the LoS component in its composite channel in \eqref{ChannelCompositeLoS}, the corresponding phase shift at PRU $\left(m_{\rm{r}},m_{\rm{c}}\right)$ is set as
\vspace{-2mm}
\begin{align}\label{IRSControl}
{\phi _{{m_{\rm{r}}},{m_{\rm{c}}}}\left[ n \right]} &= 2\pi \frac{f_\mathrm{c}}{c}\left\{{d_{\mathrm{r}} \left( {{m_{\rm{r}}} - 1} \right){\sin \theta_{k}^{\mathrm{RG}} \cos \xi_{k}^{\mathrm{RG}}} + d_{\mathrm{c}}\left( {{m_{\rm{c}}} - 1} \right){\sin \theta_{k}^{\mathrm{RG}} \sin \xi_{k}^{\mathrm{RG}}}} + \right.\notag\\[-1mm]
&\left.{{d_{\mathrm{r}} \left( {{m_{\rm{r}}} - 1} \right){\sin \theta^{\mathrm{UR}}[n] \cos \xi^{\mathrm{UR}}[n]} + d_{\mathrm{c}}\left( {{m_{\rm{c}}} - 1} \right){\sin \theta^{\mathrm{UR}}[n] \sin \xi^{\mathrm{UR}}[n]}} } \right\}.
\end{align}
\vspace{-10mm}\par\noindent
The adopted phase control strategy above is optimal in the sense that it can maximize the passive beamforming gain at the IRS with respect to (w.r.t.) the IRS-assisted user.
We can observe that the adopted simple phase control strategy in \eqref{IRSControl} only depends on the locations of UAV and ground users, which significantly reduces the required signaling overhead of CSI acquisition and phase control at the IRS \cite{WuIRS}.
As a result, the phase control at IRS can be designed with the UAV's trajectory in an offline manner.
Additionally, it can be seen that the phase control does not depend on the phase terms ${e^{-j2\pi i \Delta f \frac{{d_k^{{\rm{UG}}}\left[ n \right]}}{c}}}$ and ${e^{ - j2\pi i \Delta f \frac{{{d^{{\rm{UR}}}}\left[ n \right]} + {d_k^{{\rm{RG}}}}}{c}}}$ in the LoS component in \eqref{ChannelCompositeLoS}.
In fact, the phase control at IRS has indeed a flat frequency response, which affects all the subcarriers homogeneously.
Therefore, in general, we can only coherently combine the received signals from both the UAV-to-user link and the UAV-IRS-user link in some subcarriers as different subcarriers generally have different channel phases.
Note that although applying phase control at IRS can improve the system performance, it also introduces a frequency-selective fading with a periodic cosine pattern as will be analyzed in the following.

According to \eqref{ChannelCompositeLoS}, the LoS component of the composite channel of IRS-assisted user $k$ on subcarrier $i$ in time slot $n$ can be rewritten as
\vspace{-1mm}
\begin{align}\label{FinalChannelApproxV2}
\hspace{-2mm}{{{g}}}_{k,i, \mathrm{LoS}}^{{\rm{UG}}}\left[ n \right]
& = {e^{-j2\pi i \Delta f \frac{{d_k^{{\rm{UG}}}\left[ n \right]}}{c}}} \left[\sqrt{\frac{\beta _0}{\left( {d_k^{{\rm{UG}}}\left[ n \right]} \right)^{\alpha^{\mathrm{UG}}_k}}} \sqrt{\frac{\kappa^{\mathrm{UG}}_k}{\kappa^{\mathrm{UG}}_k + 1}}  + \sum_{k'=1}^{K}\frac{a \beta _0 s_{k'}\left[n\right]}{{{d^{{\rm{UR}}}}\left[ n \right]}\left({ {d_k^{{\rm{RG}}}}}\right)^{\frac{\alpha^{\mathrm{RG}}_k}{2}}} \sqrt{\frac{\kappa^{\mathrm{RG}}_k}{\kappa^{\mathrm{RG}}_k + 1}} \right. \notag\\[-0.5mm]
&{\left.\times {e^{ - j2\pi i \Delta f \frac{{{d^{{\rm{UR}}}}\left[ n \right]} + {d_{k}^{{\rm{RG}}}} - {d_{k}^{{\rm{UG}}}\left[ n \right]}}{c}}} {{e^{ - j\left( {{M_{\rm{r}}} - 1} \right)\psi _{k',k}^{\rm{r}}}}{e^{ - j\left( {{M_{\rm{c}}} - 1} \right)\psi _{k',k}^{\rm{c}}}}{B_{{M_{\rm{r}}}}}\left( {\psi _{k',k}^{\rm{r}}} \right){B_{{M_{\rm{c}}}}}\left( {\psi _{k',k}^{\rm{c}}} \right)} \right]},
\end{align}
\vspace{-8mm}\par\noindent
where $\psi^{\rm{r}}_{k',k} = \frac{\pi f_\mathrm{c} d_{\rm{r}} \left( {\theta _{k'}} - {\theta _{k}} \right)}{c}$, $\psi^{\rm{c}}_{k',k} = \frac{\pi f_\mathrm{c} d_{\rm{c}} \left( {\varphi _{k'}} - {\varphi _{k}} \right)}{c}$, and the beam pattern function is ${B_M}\left( x \right) = \frac{{\sin \left( {Mx} \right)}}{{\sin \left( x \right)}}$.
We note that the whole term in the second line in \eqref{FinalChannelApproxV2} is the beam pattern response from the IRS to user $k$ when the IRS is allocated to user $k'$ in time slot $n$.
In particular, when user $k$ is scheduled to utilize the IRS in time slot $n$, i.e., $s_k \left[n\right]= 1$, only one term in the summation in \eqref{FinalChannelApproxV2} is selected with $k' = k$ and the full beamforming gain can be achieved.
On the other hand, when $s_k \left[n\right]= 0$, there is also one term been selected in the summation in \eqref{FinalChannelApproxV2} with $s_{k'} \left[n\right]= 1$, $\forall k' \ne k$, and thus the beamforming gain depends on the AoDs' difference between user $k$
and user $k'$ in the azimuth and elevation planes, respectively.
Now, the composite channel power gain from the UAV to user $k$ on subcarrier $i$ in time slot $n$ is given by
\vspace{-2mm}
\begin{align}\label{FinalChannelApproxII}
&\hspace{-7mm}\left|{{{g}}}_{k,i,\mathrm{LoS}}^{{\rm{UG}}}\left[ n \right]\right|^2
= \left[\underbrace{\frac{{\beta _0}}{\left({d_k^{{\rm{UG}}}\left[ n \right]}\right)^{\alpha_k^{\mathrm{UG}}}} \frac{\kappa^{\mathrm{UG}}_k}{\kappa^{\mathrm{UG}}_k + 1}}_{\text{LoS path gain of the UAV-to-user link}} + \underbrace{\sum_{k'=1}^{K} \frac{a^2\beta^2_0 s_{k'}\left[n\right] {B^2_{{M_{\rm{r}}}}}\left( {\psi _{k',k}^{\rm{r}}} \right){B^2_{{M_{\rm{c}}}}}\left( {\psi _{k',k}^{\rm{c}}} \right) }{\left({{d^{{\rm{UR}}}}\left[ n \right]}\right)^2{\left( {d_k^{{\rm{RG}}}}\right)}^{\alpha^{\mathrm{RG}}_k}}\frac{\kappa^{\mathrm{RG}}_k}{\kappa^{\mathrm{RG}}_k + 1}}_{\text{LoS path gain of the UAV-IRS-user link}} \right. \notag\\[-0.5mm]
&\hspace{15mm}+ \sum_{k'=1}^{K} \frac{2a{\beta^{\frac{3}{2}}_0} s_{k'} \left[n\right] {B_{{M_{\rm{r}}}}}\left( {\psi _{k',k}^{\rm{r}}} \right){B_{{M_{\rm{c}}}}}\left( {\psi _{k',k}^{\rm{c}}} \right) }{{\left({d_k^{{\rm{UG}}}\left[ n \right]}\right)^{\frac{\alpha_k^{\mathrm{UG}}}{2}}} {{d^{{\rm{UR}}}}\left[ n \right]} {\left(d_k^{{\rm{RG}}}\right)^{\frac{\alpha_k^{\mathrm{RG}}}{2}}} } \sqrt{\frac{\kappa^{\mathrm{UG}}_k}{\kappa^{\mathrm{UG}}_k + 1}} \sqrt{\frac{\kappa^{\mathrm{RG}}_k}{\kappa^{\mathrm{RG}}_k + 1}} \notag\\[-0.5mm]
&\hspace{15mm}\underbrace{\left.\times\cos \left({2\pi i \Delta f \frac{{{d^{{\rm{UR}}}}\left[ n \right]} + {d_k^{{\rm{RG}}}} - {d_k^{{\rm{UG}}}\left[ n \right]}}{c}} + \left( {{M_{\rm{r}}} - 1} \right)\psi _{k',k}^{\rm{r}} + \left( {{M_{\rm{c}}} - 1} \right)\psi _{k',k}^{\rm{c}} \right)\right]}_{\text{Fluctuation component}}.
\end{align}
\vspace{-5mm}\par\noindent

{In \eqref{FinalChannelApproxII}, the first term represents the LoS path gain of the UAV-to-user link, the second term denotes the LoS path gain of the UAV-IRS-user link, and the third term is caused by the superposition of the two LoS paths in these two links.
Note that the proposed composite channel model is a generalization of related works on IRS \cite{YangIRSOFDMA,li2019reconfigurable} and UAV communications \cite{CaiUAVEESecure,RuideLi}.
In particular, existing work on IRS-aided OFDMA systems, e.g. \cite{YangIRSOFDMA}, ignored the IRS scheduling feature and the  mobility of UAV.
When the IRS is removed from the considered system, i.e., $a = 0$, the composite channel model in \eqref{FinalChannelApproxII} degenerates to the conventional deterministic channel model for UAV-enabled wireless communication systems \cite{CaiUAVEESecure,RuideLi}.
On the other hand, when the direct link from the UAV to the ground user is blocked, only the second term of \eqref{FinalChannelApproxII} retains which is similar to the LoS component of the composite channel model in \cite{li2019reconfigurable}.
Besides, we note that the constructed channel model in \eqref{FinalChannelApproxII} does not depend on the communication directions and thus can be adopted for the reverse link directly.}

From \eqref{FinalChannelApproxII}, we can observe that even if all the subcarriers' signals experience the same delay, different phase shifts are introduced on different subcarriers resulting in a frequency-selective channel.
In particular, the frequency-selective fading in \eqref{FinalChannelApproxII} for both IRS-assisted and non-IRS-assisted users follows a periodic cosine pattern w.r.t. the subcarrier index, which will be demonstrated in Fig. \ref{FrequencySelectivePattern} in Section \ref{Simulation}.
Furthermore, the period of the cosine fading pattern depends on the delay spread between the UAV-to-user and IRS-UAV-user links.
One can imagine that the closer the IRS to the ground users, the longer the period of the frequency-selective fading.
In particular, when the IRS is sufficiently close to the ground user, i.e., ${d_k^{{\rm{RG}}}} \to 0$, we have ${{d^{{\rm{UR}}}}\left[ n \right]} \approx {d_k^{{\rm{UG}}}\left[ n \right]}$.
In this case, the cosine function in \eqref{FinalChannelApproxII} approaches a constant and $\left|{{{g}}}_{k,i,\mathrm{LoS}}^{{\rm{UG}}}\left[ n \right]\right|^2$ becomes frequency-flat fading.
In fact, when the employed IRS is sufficiently close to the ground users, it is expected that the UAV-to-user and IRS-UAV-user links almost merge with each other forming a pure LoS link with a frequency-flat characteristic.
Additionally, the range of fluctuation of the composite channel gains across the subcarriers is determined by both the AoDs' difference between the IRS-assisted user $k$ and non-IRS-assisted user $k'$ as well as the number of PRUs at the IRS.
On the other hand, on each subcarrier, the composite channel gains for both the IRS-assisted and non-IRS-assisted users fluctuate with a cosine pattern w.r.t. the propagation distances' difference between the UAV-to-user and IRS-UAV-user links ${{d^{{\rm{UR}}}}\left[ n \right]} + {d_k^{{\rm{RG}}}} - {d_k^{{\rm{UG}}}\left[ n \right]}$, as shown in \eqref{FinalChannelApproxII}, which is affected by the UAV's trajectory.
As a result, the composite channel gain on one subcarrier experiences also spatial-selective fading, which fluctuates along the UAV trajectory, as will be shown in Fig. \ref{FrequencySelectivePattern} in Section \ref{Simulation}.

{In general, the models in \eqref{FinalChannelApproxV2} and \eqref{FinalChannelApproxII} are accurate but intractable for joint trajectory and resource allocation design.}
In the following, we first define the peak, the trough, and the direct current (DC) level for the composite channel power gains among all subcarriers, given by
\begin{align}
\left|{{{g}}}_{k,\mathrm{LoS}}^{{\rm{UG}}}\left[ n \right]\right|^2_{\mathrm{Peak}}
&= \left[\sqrt{\frac{\beta _0}{\left( {d_k^{{\rm{UG}}}\left[ n \right]} \right)^{\alpha^{\mathrm{UG}}_k}}} \sqrt{\frac{\kappa^{\mathrm{UG}}_k}{\kappa^{\mathrm{UG}}_k + 1}} + \sum_{k'=1}^{K}\frac{a \beta _0 s_{k'}\left[n\right]}{{{d^{{\rm{UR}}}}\left[ n \right]}\left({ {d_k^{{\rm{RG}}}}}\right)^{\frac{\alpha^{\mathrm{RG}}_k}{2}}} \sqrt{\frac{\kappa^{\mathrm{RG}}_k}{\kappa^{\mathrm{RG}}_k + 1}} \right.\notag\\[-1mm]
&\left. \times {{B_{{M_{\rm{r}}}}}\left( {\psi _{k',k}^{\rm{r}}} \right){B_{{M_{\rm{c}}}}}\left( {\psi _{k',k}^{\rm{c}}} \right)} \right]^2,\label{FinalChannelUpper}\\[-1mm]
\left|{{{g}}}_{k,\mathrm{LoS}}^{{\rm{UG}}}\left[ n \right]\right|^2_{\mathrm{Trough}}&= \left[\sqrt{\frac{\beta _0}{\left( {d_k^{{\rm{UG}}}\left[ n \right]} \right)^{\alpha^{\mathrm{UG}}_k}}} \sqrt{\frac{\kappa^{\mathrm{UG}}_k}{\kappa^{\mathrm{UG}}_k + 1}} - \sum_{k'=1}^{K}\frac{a \beta _0 s_{k'}\left[n\right]}{{{d^{{\rm{UR}}}}\left[ n \right]}\left({ {d_k^{{\rm{RG}}}}}\right)^{\frac{\alpha^{\mathrm{RG}}_k}{2}}} \sqrt{\frac{\kappa^{\mathrm{RG}}_k}{\kappa^{\mathrm{RG}}_k + 1}} \right.\notag\\[-1mm]
&\left. \times {{B_{{M_{\rm{r}}}}}\left( {\psi _{k',k}^{\rm{r}}} \right){B_{{M_{\rm{c}}}}}\left( {\psi _{k',k}^{\rm{c}}} \right)} \right]^2,\;\text{and}\label{FinalChannelLower}\\[-1mm]
\left|{{{g}}}_{k,\mathrm{LoS}}^{{\rm{UG}}}\left[ n \right]\right|^2_{\mathrm{DC}}&= \left[\frac{{\beta _0}}{\left({d_k^{{\rm{UG}}}\hspace{-0.5mm}\left[ n \right]}\right)^{\alpha_k^{\mathrm{UG}}}} \frac{\kappa^{\mathrm{UG}}_k}{\kappa^{\mathrm{UG}}_k \hspace{-0.5mm}+\hspace{-0.5mm} 1} \hspace{-0.5mm}+\hspace{-0.5mm} \sum_{k'=1}^{K} \frac{a^2\beta^2_0 s_{k'}\hspace{-0.5mm}\left[n\right] {B^2_{{M_{\rm{r}}}}}\hspace{-0.5mm}\left( {\psi _{k',k}^{\rm{r}}} \right)\hspace{-0.5mm}{B^2_{{M_{\rm{c}}}}}\hspace{-0.5mm}\left( {\psi _{k',k}^{\rm{c}}} \right)\hspace{-0.5mm} }{\left({{d^{{\rm{UR}}}}\hspace{-0.5mm}\left[ n \right]}\right)^2{\left( {d_k^{{\rm{RG}}}}\right)}^{\alpha^{\mathrm{RG}}_k}}\frac{\kappa^{\mathrm{RG}}_k}{\kappa^{\mathrm{RG}}_k \hspace{-0.5mm}+\hspace{-0.5mm} 1}\right],\label{FinalChannelDC}
\end{align}
\vspace{-6mm}\par\noindent
respectively, which are useful in our proposed parametric approximation in the next section.

\vspace{-4mm}
\section{Problem Formulation}
In this section, we first formulate the sum-rate maximization problem and then develop its upper bound and lower bound based on the proposed parametric approximation method.

\vspace{-4mm}
\subsection{Sum-rate Maximization Problem Formulation}
The achievable data rate of user $k$ on subcarrier $i$ in time slot $n$ can be given by
\vspace{-2mm}
\begin{equation}\label{AchievableRate}
R_{k,i,\mathrm{LoS}}\left[ n \right] = u_{k,i}\left[ n \right] \log_2\left( 1 + {p_{k,i}\left[ n \right]\left|{{{g}}}_{k,i,\mathrm{LoS}}^{{\rm{UG}}}\left[ n \right]\right|^2}/{\sigma^2}\right),\vspace{-2mm}
\end{equation}
where $\sigma^2 = N_0 \Delta f$ denotes the noise power in each subcarrier, $\Delta f$ is the subcarrier spacing, and $N_0$ denotes the noise power spectral density at ground users.
In time slot $n$, the individual data rate of user $k$ and the system sum-rate are given by
\vspace{-2mm}
\begin{align}
R_{k,\mathrm{LoS}}\left[ n \right] &= \sum\nolimits_{i=1}^{N_{\mathrm{F}}}u_{k,i}\left[ n \right]\log_2\left( 1 + {p_{k,i}\left[ n \right]\left|{{{g}}}_{k,i,\mathrm{LoS}}^{{\rm{UG}}}\left[ n \right]\right|^2}/{\sigma^2}\right) \;\text{and}\label{AchievableRatek}\\[-0.5mm]
R_{\mathrm{sum},\mathrm{LoS}}\left[ n \right] &= \sum\nolimits_{i=1}^{N_{\mathrm{F}}} \sum\nolimits_{k=1}^{K} u_{k,i}\left[ n \right]\log_2\left( 1 + {p_{k,i}\left[ n \right]\left|{{{g}}}_{k,i,\mathrm{LoS}}^{{\rm{UG}}}\left[ n \right]\right|^2}/{\sigma^2}\right),\label{AchievableRateSum}
\end{align}
\par\vspace{-2mm}\noindent
respectively.
{Note that benefiting from the adopted OFDMA scheme, the inter-user interference is absent in the achievable rates in \eqref{AchievableRate}, \eqref{AchievableRatek}, and \eqref{AchievableRateSum}.}
Now, the sum-rate maximization problem can be formulated as the following optimization problem:
\vspace{-2mm}
\begin{align}\label{ProblemFormulation}
\mathcal{P}\left(\mathbf{{U}},\mathbf{{P}},{\bf{q}}\left[ n \right],\mathbf{{S}}\right)&:\hspace{5mm}\underset{\mathbf{{U}},\mathbf{{P}},{\bf{q}}\left[ n \right],\mathbf{{S}}}{\maxo}\,\, \frac{1}{N} \sum\nolimits_{n=1}^{N} R_{\mathrm{sum},\mathrm{LoS}}\left[ n \right] \left(\mathbf{{U}},\mathbf{{P}},{\bf{q}}\left[ n \right],\mathbf{{S}}\right) \\[-1mm]
 \notag\mbox{s.t.}\;\;
%%%%%
\mbox{{C1}}:\; &{u_{k,i}}\left[ n \right] \in \{0,1\},\;\forall k,i,n, \hspace{10mm} \mbox{{C2}}:\; \sum\nolimits_{k = 1}^{K} {u_{k,i}}\left[ n \right] \le 1,\;\forall i,n, \notag\\[-1mm]
\mbox{{C3}}:\; & {p_{k,i}}\left[ n \right] \ge 0,\;\forall k,i,n, \hspace{18mm} \mbox{{C4}}:\; \sum\nolimits_{i = 1}^{N_{\mathrm{F}}} \sum\nolimits_{k = 1}^{K} {p_{k,i}}\left[ n \right] \le p_{\mathrm{max}}, \forall n, \notag\\[-1mm]
\mbox{{C5}}:\; & s_k \left[ n \right]\in \{0,1\}, \forall k, n, \hspace{17mm}\mbox{{C6}}:\; \sum\nolimits_{k=1}^{K}s_k\left[ n \right] \le 1, \forall n, \notag\\[-1mm]
\hspace{-10mm} \mbox{{C7}}:\; & \frac{1}{N}\sum\nolimits_{n=1}^{N}R_{k,\mathrm{LoS}}\left[ n \right]\left(\mathbf{{U}},\mathbf{{P}},{\bf{q}}\left[ n \right],\mathbf{{S}}\right) \ge R_{\mathrm{min},k}, \forall k, \notag\\[-1mm]
\mbox{{C8}}:\; & \left\|{\bf{q}}\left[ n \right] - {\bf{q}}\left[ n -1 \right] \right\| \le \delta_{\mathrm{t}} V_{\mathrm{max}}, \forall n, \notag\\[-1mm]
\mbox{{C9}}:\;& {\bf{q}}\left[ 0 \right] = {\bf{q}}_{\mathrm{Initial}}, \hspace{27mm}
\mbox{{C10}}:\; {\bf{q}}\left[ N \right] = {\bf{q}}_{\mathrm{Final}}, \notag\\[-1mm]
\mbox{{C11}}:\;& {H^{\mathrm{min}}_{\mathrm{U}} \le z\left[n\right] \le H^{\mathrm{max}}_{\mathrm{U}}, \forall n}.\notag
%& \mbox{{C10}}:\; M_k \left[ n \right]\ge M_{\rm{min}}, \forall k,n, \notag\\
%& \mbox{{C11}}:\; \sum_{k=1}^{K}M_k \left[ n \right]\le M_{\rm{c}} M_{\rm{r}}, \forall n, \notag\\
%& \mbox{{C13}}:\; u_{k,i}\left[ n \right] = 1-I\left(s_k \cos \left({2\pi i \Delta f \frac{{{d^{{\rm{UR}}}}\left[ n \right]} + {d_k^{{\rm{RG}}}} - {d_k^{{\rm{UG}}}\left[ n \right]}}{c}}\right) \ge \rho\right), \forall i,n. \notag
\end{align}
\par\vspace{-2mm}\noindent
In the formulated problem in \eqref{ProblemFormulation}, { constraints C1, C3, and C5 define the user scheduling, power allocation, and IRS scheduling variables, respectively.
	C2 guarantees that at most one user can be scheduled on each subcarrier in each time slot.
	C4 limits the total transmit power of the UAV in each time slot.
	C6 denotes that the IRS can adjust its beamforming matrix w.r.t. at most one user in each time slot.}
Constant $R_{\mathrm{min},k}$ in C7 denotes the  minimum required average data rate for user $k$ during the whole flight period, which is introduced to guarantee the QoS requirement of user $k$.
Constraint C8 is imposed to make sure that the UAV's displacement in adjacent time slots is less than its maximum speed constraint $V_{\mathrm{max}}$.
Constraints C9 and C10 indicate the required UAV's initial location ${\bf{q}}_{\mathrm{Initial}}$ and final location ${\bf{q}}_{\mathrm{Final}}$, respectively.
{In C11, $H^{\mathrm{min}}_{\mathrm{U}}$ and $H^{\mathrm{max}}_{\mathrm{U}}$ denote the minimum and maximum altitudes for the UAV, respectively.}
The formulated problem is a non-convex mixed-integer optimization problem, which is generally difficult to solve.
In particular, the non-convexity arises from the binary variables ${u_{k,i}}\left[ n \right]$ and $s_k \left[ n \right]$ as well as the non-convex achievable rate function in the objective and constraint C7.
More importantly, as analyzed before, both the spatial and frequency-selective fading arise from the cosine function in the composite channel power gain in \eqref{FinalChannelApproxII}, which has not been studied in the literature.
Although introducing an IRS to UAV communication systems provides the flexibility in trajectory design via the new degrees of freedom, it also makes the trajectory design as a challenging problem due to the multipath propagation.
In the following, we aim to find an upper bound and a lower bound of the formulated problem in \eqref{ProblemFormulation} to facilitate our design.

\begin{Remark}
	{There are a few possible directions for extending this work.
		Firstly, it is worth to investigate the joint resource allocation and trajectory design for multi-IRS-assisted UAV communications.
		%
		%Machine learning \cite{LetaiefAI,LiuMLUAV} might be a good candidate as a black box to learn the statistics of frequency selective channel fading in a multi-IRS-assisted UAV communication system, i.e., obtaining the probability density functions (PDFs) of channel gain with respect to the UAV's location and subcarrier index.
		%
		%Then, the average system sum-rate might be expressed as a function of UAV's trajectory, which enables tractable UAV's trajectory design.
		%
		Besides, introducing multiple UAVs to the considered system with concurrent transmission has the potential to further improve the system performance \cite{RahmatiUAVInterference,HosseinalipourUAV}.
		On the other hand, a practical design to address the vulnerability to the potential jamming/eavesdropping attacks of IRS-assisted UAV communication systems is also interesting and desired.}
\end{Remark}

\vspace{-4mm}
\subsection{Parametric Bounds for the Formulated Problem}

\begin{figure}[t]
	\centering\vspace{-7mm}
	\includegraphics[width=4.0in]{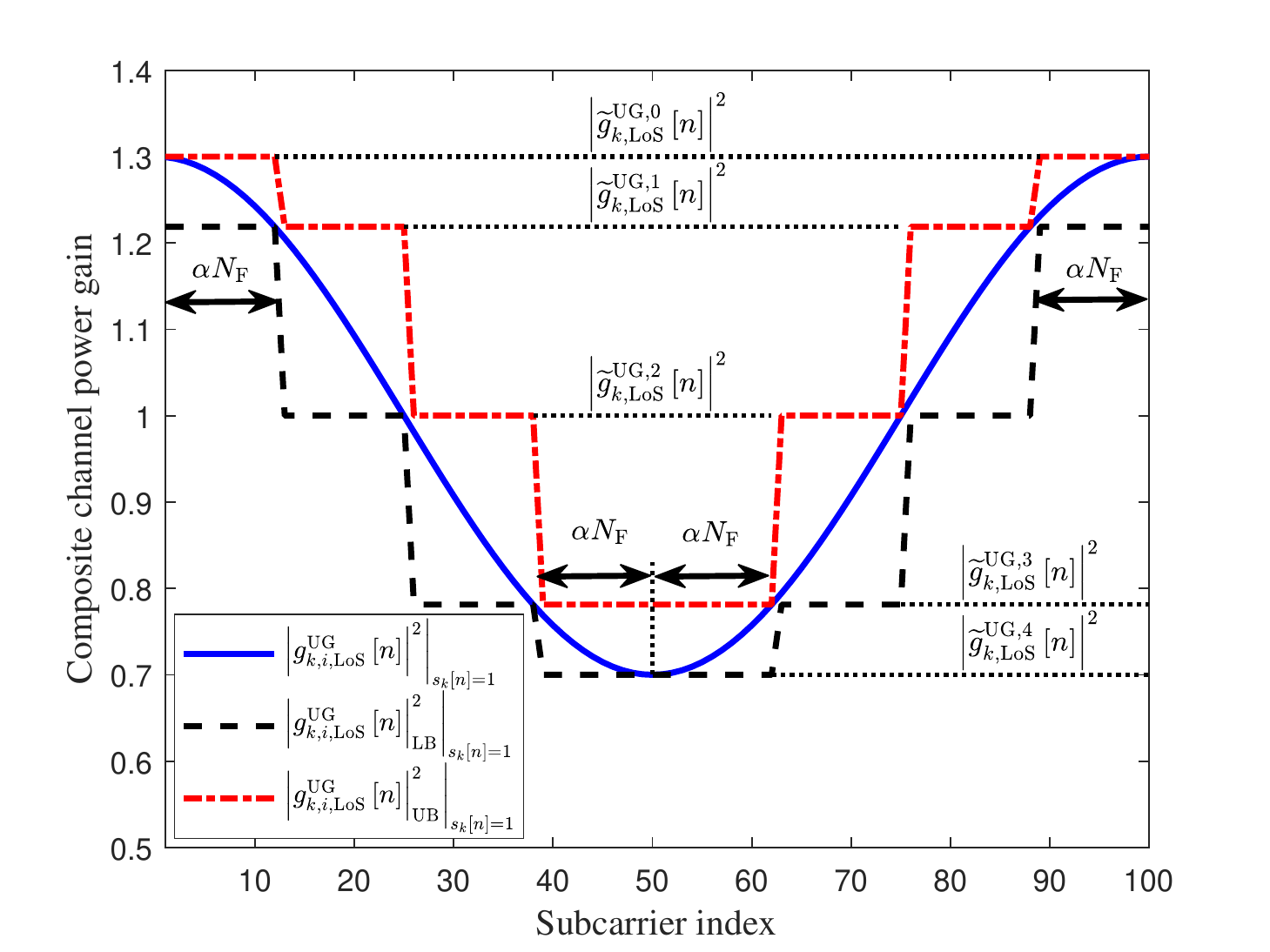}\vspace{-7mm}
	\caption{An illustration of the proposed parametric bounds for the cosine function in the frequency-selective fading of the IRS-assisted user. We assume that there are $100$ subcarriers within one period of the cosine function. Additionally, we assume that the DC level, the peak level, and the trough level of the cosine function are given by ${\left|{{{g}}}_{k,\mathrm{LoS}}^{{\rm{UG}}}\left[ n \right]\right|^2_{\mathrm{DC}}} = 1$, ${\left|{{{g}}}_{k,\mathrm{LoS}}^{{\rm{UG}}}\left[ n \right]\right|^2_{\mathrm{Peak}}} = 1.3$, and
		${\left|{{{g}}}_{k,\mathrm{LoS}}^{{\rm{UG}}}\left[ n \right]\right|^2_{\mathrm{Trough}}} = 0.7$, respectively. A parameter $0<\alpha<\frac{1}{4}$ is introduced to obtain the parametric bounds.}\vspace{-10mm}
	\label{CosineApproximationForIRSAssistedUser}%
\end{figure}

In \eqref{ProblemFormulation}, we can observe that the objective function monotonically increases with the channel power gain on each subcarrier.
Inspired by this observation, we aim to develop two problems via adopting parametric lower bound and upper bound for the composite channel power gains of both IRS-assisted and non-IRS-assisted users which facilitate the development of an upper bound and a lower bound for problem \eqref{ProblemFormulation}, respectively.
In particular, for an IRS-assisted user, we introduce an approximation parameter $\alpha$ to quantize the cosine pattern of frequency-selective fading into four-mode fading channels\footnote{Note that introducing more approximation parameters results in more fading modes and a higher accuracy in approximating the composite channel gains. However, optimizing multiple approximation parameters requires a much higher computational complexity. Therefore, we just consider a single approximation parameter and four-fading modes for an IRS-assisted user in this paper for an illustration purpose.}, as illustrated in Fig. \ref{CosineApproximationForIRSAssistedUser}.
As for a non-IRS-assisted user, as will be shown in Fig. \ref{FrequencySelectivePattern} in Section \ref{Simulation}, its channel fluctuation range is much smaller compared to that of the IRS-assisted user.
This is because the UAV-IRS-user link is generally very weak for a non-IRS-assisted user, especially with a massive number of PRUs at IRS.
Therefore, for simplicity, one fading mode (frequency flat fading) is sufficient for approximating the compositing channel gain of the non-IRS-assisted user.

For a non-IRS-assisted user, we can obtain its composite channel gain's upper bound and lower bound by the corresponding peak and trough levels, respectively, i.e.,
\vspace{-2mm}
\begin{align}
{\left. \left|{{{g}}}_{k',i,\mathrm{LoS}}^{{\rm{UG}}}\left[ n \right]\right|^2 \right|_{{s_k \left[ n \right] = 1}}} &\le {\left. \left|{{{g}}}_{k',i,\mathrm{LoS}}^{{\rm{UG}}}\left[ n \right]\right|^2_{\mathrm{UB}} \right|_{{s_k \left[ n \right] = 1}}} = { \left|{{{g}}}_{k',\mathrm{LoS}}^{{\rm{UG}}}\left[ n \right]\right|^2_{\mathrm{Peak}}}\;\text{and}\\[-1mm]
{\left. \left|{{{g}}}_{k',i,\mathrm{LoS}}^{{\rm{UG}}}\left[ n \right]\right|^2 \right|_{{s_k \left[ n \right] = 1}}} &\ge {\left. \left|{{{g}}}_{k',i,\mathrm{LoS}}^{{\rm{UG}}}\left[ n \right]\right|^2_{\mathrm{LB}} \right|_{{s_k \left[ n \right] = 1}}} = { \left|{{{g}}}_{k',\mathrm{LoS}}^{{\rm{UG}}}\left[ n \right]\right|^2_{\mathrm{Trough}}}, \forall k'\neq k,
\end{align}
\par\vspace{-2mm}\noindent
where ${ \left|{{{g}}}_{k',\mathrm{LoS}}^{{\rm{UG}}}\left[ n \right]\right|^2_{\mathrm{Peak}}}$ and ${ \left|{{{g}}}_{k',\mathrm{LoS}}^{{\rm{UG}}}\left[ n \right]\right|^2_{\mathrm{Trough}}}$ are given by \eqref{FinalChannelUpper} and \eqref{FinalChannelLower}, respectively.
We can observe that the bounds of the composite channel gains of non-IRS-assisted users are frequency-flat.
More importantly, by replacing the composite channel with their corresponding bounds, the design of UAV's trajectory becomes easier as the cosine function is no longer involved.

On the other hand, we develop the bounds for the composite channel gain of the IRS-assisted user.
As illustrated in Fig. \ref{CosineApproximationForIRSAssistedUser}, a cosine function can be bounded by below with a piece-wise step function via introducing an approximation parameter $0 < \alpha < \frac{1}{4}$.
Hence, the composite channel power gain for an IRS-assisted user that follows a cosine pattern as revealed in \eqref{FinalChannelApproxII} can be bounded by below via the following four-mode fading pattern:
\vspace{-2mm}
\begin{equation}\label{LBIRSassistedUser}
	\left. \left|{{{g}}}_{k,i,\mathrm{LoS}}^{{\rm{UG}}}\left[ n \right]\right|^2\right|_{{s_k \left[ n \right] = 1}} \ge
	\left. \left|{{{g}}}_{k,i,\mathrm{LoS}}^{{\rm{UG}}}\left[ n \right]\right|^2_{\mathrm{LB}}\right|_{{s_k \left[ n \right] = 1}} = {\left|{{{g}}}_{k,\mathrm{LoS}}^{{\mathrm{UG},j}}\left[ n \right]\right|^2}, \;{\rm{if}}\; i \in \mathcal{F}_{j}\left[ n \right],\vspace{-2mm}
\end{equation}
where $\mathcal{F}_{j}\left[ n \right]$ denotes the subcarrier index set belonging to the fading mode $j$ in time slot $n$, $\forall j \in \{1,2,3,4\}$.
Variable ${\left|{{{g}}}_{k,\mathrm{LoS}}^{{\mathrm{UG},j}}\left[ n \right]\right|^2}$ denotes the channel power gain of user $k$ in fading mode $j$ in time slot $n$, $\forall j \in \{1,2,3,4\}$, and they are given by
\vspace{-2mm}
\begin{align}
	{\left|{{{g}}}_{k,\mathrm{LoS}}^{{\mathrm{UG},1}}\left[ n \right]\right|^2} & = {\left|{{{g}}}_{k,\mathrm{LoS}}^{{\rm{UG}}}\left[ n \right]\right|^2_{\mathrm{DC}}} + \left({\left|{{{g}}}_{k,\mathrm{LoS}}^{{\rm{UG}}}\left[ n \right]\right|^2_{\mathrm{Peak}}} - {\left|{{{g}}}_{k,\mathrm{LoS}}^{{\rm{UG}}}\left[ n \right]\right|^2_{\mathrm{DC}}}\right)\cos\left( 2 \pi\alpha\right),\\[-1mm]
	{\left|{{{g}}}_{k,\mathrm{LoS}}^{{\mathrm{UG},2}}\left[ n \right]\right|^2} & = {\left|{{{g}}}_{k,\mathrm{LoS}}^{{\rm{UG}}}\left[ n \right]\right|^2_{\mathrm{DC}}}, \\[-1mm]
	{\left|{{{g}}}_{k,\mathrm{LoS}}^{{\mathrm{UG},3}}\left[ n \right]\right|^2} & = {\left|{{{g}}}_{k,\mathrm{LoS}}^{{\rm{UG}}}\left[ n \right]\right|^2_{\mathrm{DC}}} + \left({\left|{{{g}}}_{k,\mathrm{LoS}}^{{\rm{UG}}}\left[ n \right]\right|^2_{\mathrm{Trough}}} - {\left|{{{g}}}_{k,\mathrm{LoS}}^{{\rm{UG}}}\left[ n \right]\right|^2_{\mathrm{DC}}}\right)\cos\left( 2 \pi\alpha\right),\;\text{and}\\[-1mm]
	{\left|{{{g}}}_{k,\mathrm{LoS}}^{{\mathrm{UG},4}}\left[ n \right]\right|^2} & = {\left|{{{g}}}_{k,\mathrm{LoS}}^{{\rm{UG}}}\left[ n \right]\right|^2_{\mathrm{Trough}}},
\end{align}
\par\vspace{-2mm}\noindent
respectively.
Similarly, we can develop an upper bound for the channel power gain of the IRS-assisted user as follows:
\vspace{-4mm}
\begin{equation}\label{UBIRSassistedUser}
\left. \left|{{{g}}}_{k,i,\mathrm{LoS}}^{{\rm{UG}}}\left[ n \right]\right|^2\right|_{{s_k \left[ n \right] = 1}} \le
\left. \left|{{{g}}}_{k,i,\mathrm{LoS}}^{{\rm{UG}}}\left[ n \right]\right|^2_{\mathrm{UB}}\right|_{{s_k \left[ n \right] = 1}} = {\left|{{{g}}}_{k,\mathrm{LoS}}^{{\mathrm{UG},j-1}}\left[ n \right]\right|^2,\;}{{\rm{if}}\;i \in \mathcal{F}_{j}\left[ n \right],} \vspace{-2mm}
\end{equation}
where ${\left|{{{g}}}_{k,\mathrm{LoS}}^{{\mathrm{UG},0}}\left[ n \right]\right|^2} = {\left|{{{g}}}_{k,\mathrm{LoS}}^{{\rm{UG}}}\left[ n \right]\right|^2_{\mathrm{Peak}}}$.

It can be observed that the channel power gain is frequency-flat within each mode based on our proposed parametric approximation.
In addition, compared to \eqref{FinalChannelApproxII}, distance variables ${{d^{{\rm{UR}}}}\left[ n \right]}$ and $d_k^{{\rm{UG}}}\left[ n \right]$ are taken out from the cosine function, which is more tractable for the design of UAV's trajectory.
However, the subcarrier index set for each mode $\mathcal{F}_{j}\left[ n \right]$ still depends on the original cosine pattern in the composite channel gains and thus keep changing along the UAV's trajectory.
In this case, the system sum-rate in \eqref{ProblemFormulation} is still a very complicated function of the UAV's trajectory.
In the following, to further simplify the four-mode fading channel and to facilitate our design, we asymptotically analyze the size of $\mathcal{F}_{j}\left[ n \right]$.
According to \eqref{FinalChannelApproxII}, the cosine pattern in the composite channel gain experiences many cycles when the system bandwidth is sufficiently large, $B = N_F\Delta f \to \infty$, as will be verified in Fig. \ref{FrequencySelectivePattern} in Section \ref{Simulation}.
Hence, the subcarrier index set sizes of the four-mode fading are asymptotically deterministic and they are given by
\vspace{-2mm}
\begin{equation}
\hspace{-2mm}\mathop {\lim }\limits_{B \to \infty} \left|\mathcal{F}_{1}\left[ n \right]\right| = \mathop {\lim }\limits_{B \to \infty}\left|\mathcal{F}_{4}\left[ n \right]\right| = 2 \alpha {N_{\mathrm{F}}}\;\text{and} \;
\mathop {\lim }\limits_{B \to \infty}\left|\mathcal{F}_{2}\left[ n \right]\right| = \mathop {\lim }\limits_{B \to \infty}\left|\mathcal{F}_{3}\left[ n \right]\right| = \left(\frac{1}{2} - 2 \alpha\right) {N_{\mathrm{F}}},\vspace{-2mm}
\end{equation}
respectively, where we assume that $2 \alpha {N_{\mathrm{F}}}$ and $\left(\frac{1}{2} - 2 \alpha\right) {N_{\mathrm{F}}}$ are integers without loss of generality.
Furthermore, since only one user is assisted by the IRS and thus only the IRS-assisted user has a frequency-selective fading channel based on the developed bounds, the subcarrier index set in each fading mode does not matter for the resource allocation design.
Therefore, we can consider a fixed subcarrier index set for each mode along the UAV's trajectory as $\mathcal{F}_{1} = \{1,\ldots,2 \alpha {N_{\mathrm{F}}}\}$, $\mathcal{F}_{2} = \{2 \alpha {N_{\mathrm{F}}} + 1,\ldots,\frac{1}{2} {N_{\mathrm{F}}}\}$, $\mathcal{F}_{3} = \{\frac{1}{2} {N_{\mathrm{F}}}+1,\ldots,{N_{\mathrm{F}}} - 2 \alpha {N_{\mathrm{F}}}\}$, and $\mathcal{F}_{4} = \{{N_{\mathrm{F}}} - 2 \alpha {N_{\mathrm{F}}}+1,\ldots, {N_{\mathrm{F}}}\}$.

Now, the individual data rate and the system sum-rate can be bounded by
\vspace{-3mm}
\begin{equation}
R^{\mathrm{LB}}_{k,\mathrm{LoS}}\left[ n \right] \le R_{k,\mathrm{LoS}}\left[ n \right] \le R^{\mathrm{UB}}_{k,\mathrm{LoS}}\left[ n \right] \;\text{and}\;
R^{\mathrm{LB}}_{\mathrm{sum},\mathrm{LoS}}\left[ n \right] \le R_{\mathrm{sum},\mathrm{LoS}}\left[ n \right] \le R^{\mathrm{UB}}_{\mathrm{sum},\mathrm{LoS}}\left[ n \right],\label{Rate_BoundIndividualSumrate} \vspace{-3mm}
\end{equation}
respectively, with $R^{\mathrm{LB}}_{\mathrm{sum},\mathrm{LoS}}\left[ n \right] = \sum_{k=1}^{K}R^{\mathrm{LB}}_{k,\mathrm{LoS}}\left[ n \right]$, $R^{\mathrm{UB}}_{\mathrm{sum},\mathrm{LoS}}\left[ n \right] = \sum_{k=1}^{K}R^{\mathrm{UB}}_{k,\mathrm{LoS}}\left[ n \right]$,
\vspace{-2mm}
\begin{equation}\label{IndividualRateLB}
	R^{\mathrm{LB}}_{k,\mathrm{LoS}}\left[ n \right] = \sum\nolimits_{j=1}^{4}\sum\nolimits_{i=1}^{{N_{\mathrm{F}}}}R^{\mathrm{LB},j}_{k,i,\mathrm{LoS}}\left[ n \right] I_{i,j},\;\text{and} \;R^{\mathrm{UB}}_{k,\mathrm{LoS}}\left[ n \right] = \sum\nolimits_{j=1}^{4}\sum\nolimits_{i=1}^{{N_{\mathrm{F}}}}R^{\mathrm{UB},j}_{k,i,\mathrm{LoS}}\left[ n \right] I_{i,j},\vspace{-2mm}
\end{equation}
where $I_{i,j}$ is one if $i \in \mathcal{F}_{j}$ and is zero otherwise.
Variables $R^{\mathrm{LB},j}_{k,i,\mathrm{LoS}}\left[ n \right]$ and $R^{\mathrm{UB},j}_{k,i,\mathrm{LoS}}\left[ n \right]$ represent the achievable data rate of user $k$ in fading mode $j$ at time slot $n$ based on the developed lower bound and upper bound in \eqref{LBIRSassistedUser} and \eqref{UBIRSassistedUser}, respectively, and they are given by
\vspace{-2mm}
\begin{align}
\hspace{-3mm}R^{\mathrm{LB},j}_{k,i}\hspace{-0.5mm}\left[ n \right] \hspace{-0.5mm}&= \hspace{-0.5mm}u_{k,i}\hspace{-0.5mm}\left[ n \right] \log_2\hspace{-0.5mm}\left(\hspace{-0.5mm}1 \hspace{-1mm}+\hspace{-1mm} {p_{k,i}\hspace{-0.5mm}\left[ n \right]\hspace{-0.5mm}\left(\hspace{-1mm}s_k\left[ n \right]\left|{{{g}}}_{k,\mathrm{LoS}}^{{\mathrm{UG},j}}\left[ n \right]\right|^2 \hspace{-1mm}+\hspace{-1mm} (1\hspace{-1mm}-\hspace{-1mm}s_k\left[ n \right]){ \left|{{{g}}}_{k,\mathrm{LoS}}^{{\rm{UG}}}\left[ n \right]\right|^2_{\mathrm{Trough}}}\hspace{-1mm}\right)}/{{\sigma^2}}\right) \;\text{and}\label{LowerBoundRateFunction}\\[-1mm]
\hspace{-3mm}R^{\mathrm{UB},j}_{k,i}\hspace{-0.5mm}\left[ n \right] \hspace{-0.5mm}&= \hspace{-0.5mm} u_{k,i}\hspace{-0.5mm}\left[ n \right] \log_2\hspace{-0.5mm}\left(\hspace{-0.5mm} 1 \hspace{-1mm}+\hspace{-1mm} {p_{k,i}\left[ n \right]\left(\hspace{-1mm}s_k\left[ n \right]\left|{{{g}}}_{k,\mathrm{LoS}}^{{\mathrm{UG}},j-1}\left[ n \right]\right|^2 \hspace{-1mm}+\hspace{-1mm} (1\hspace{-1mm}-\hspace{-1mm}s_k\left[ n \right]){ \left|{{{g}}}_{k,\mathrm{LoS}}^{{\rm{UG}}}\left[ n \right]\right|^2_{\mathrm{Peak}}}\hspace{-1mm}\right)}/{{\sigma^2}}\right),
\end{align}
\vspace{-9mm}\par\noindent
respectively.
Substituting the bounds in \eqref{Rate_BoundIndividualSumrate} into  \eqref{ProblemFormulation}, the resulting optimization problems $\mathcal{P}_{\mathrm{LB}}$ and $\mathcal{P}_{\mathrm{UB}}$ as follows provide a lower bound and an upper bound for the formulated problem in \eqref{ProblemFormulation}, respectively:
\vspace{-7mm}
\begin{align}\label{ProblemFormulationLB}
\mathcal{P}_{\mathrm{LB}}\left(\mathbf{{U}},\mathbf{{P}},{\bf{q}}\left[ n \right],\mathbf{{S}}\right)&:\hspace{5mm}\underset{\mathbf{{U}},\mathbf{{P}},{\bf{q}}\left[ n \right],\mathbf{{S}}}{\maxo}\,\, \frac{1}{N} \sum\nolimits_{n=1}^{N} R^{\mathrm{LB}}_{\mathrm{sum},\mathrm{LoS}}\left[ n \right]\left(\mathbf{{U}},\mathbf{{P}},{\bf{q}}\left[ n \right],\mathbf{{S}}\right) \\[-1mm]
\notag\mbox{s.t.}\;\;
%%%%%
& \mbox{{C1-C6, C8-C10}}, \;\;\underline{\mbox{{C7}}}:\; \frac{1}{N} \sum\nolimits_{n=1}^{N} R^{\mathrm{LB}}_{k,\mathrm{LoS}}\left[ n \right]\left(\mathbf{{U}},\mathbf{{P}},{\bf{q}}\left[ n \right],\mathbf{{S}}\right) \ge R_{\mathrm{min},k}, \forall k, \notag
\end{align}
\vspace{-10mm}\par\noindent
\begin{align}\label{ProblemFormulationUB}
\mathcal{P}_{\mathrm{UB}}\left(\mathbf{{U}},\mathbf{{P}},{\bf{q}}\left[ n \right],\mathbf{{S}}\right)&:\hspace{5mm}\underset{\mathbf{{U}},\mathbf{{P}},{\bf{q}}\left[ n \right],\mathbf{{S}}}{\maxo}\,\, \frac{1}{N} \sum\nolimits_{n=1}^{N} R^{\mathrm{UB}}_{\mathrm{sum},\mathrm{LoS}}\left[ n \right]\left(\mathbf{{U}},\mathbf{{P}},{\bf{q}}\left[ n \right],\mathbf{{S}}\right) \\[-1mm]
\notag\mbox{s.t.}\;\;
%%%%%
& \mbox{{C1-C6, C8-C10}}, \;\; \overline{\mbox{{C7}}}:\; \frac{1}{N} \sum\nolimits_{n=1}^{N} R^{\mathrm{UB}}_{k,\mathrm{LoS}}\left[ n \right]\left(\mathbf{{U}},\mathbf{{P}},{\bf{q}}\left[ n \right],\mathbf{{S}}\right) \ge R_{\mathrm{min},k}, \forall k. \notag
\end{align}
\vspace{-10mm}\par\noindent

We note that both problems $\mathcal{P}_{\mathrm{UB}}$ and $\mathcal{P}_{\mathrm{LB}}$ are non-convex optimization problems and there is generally no systematic and computationally efficient approach to solve them.
Note that a suboptimal solution of the upper bound problem $\mathcal{P}_{\mathrm{UB}}$ cannot guarantee to provide an upper bound of the original formulated problem $\mathcal{P}$.
In contrast, a suboptimal solution of the lower bound problem $\mathcal{P}_{\mathrm{LB}}$ provides a pessimistic but achievable solution.
Therefore, in the following, we focus on the lower bound problem $\mathcal{P}_{\mathrm{LB}}$ and propose an efficient alternating optimization approach to achieve a suboptimal solution for the joint trajectory, IRS scheduling, and resource allocation design\footnote{Note that the upper bound problem is proposed to verify the approximation accuracy via evaluating the gap between the upper bound and lower bound problems.
	{Although only a suboptimal solution of the lower bound problem can be achieved by the proposed design, its gap to the optimal solution is smaller than that between the upper bound problem and upper problem, which will be evaluated in Section VI-C.}}.

The lower bound rate functions in \eqref{IndividualRateLB} and \eqref{LowerBoundRateFunction} for the lower bound problem $\mathcal{P}_{\mathrm{LB}}$ are still quite difficult to handle.
To facilitate the trajectory and resource allocation design, we introduce an auxiliary binary variable $t_{k,k',i} \left[n\right] = u_{k,i}\left[n\right] s_{k'}\left[n\right]$ to decouple the binary variables $u_{k,i}\left[n\right]$ and $s_{k'}\left[n\right]$.
If subcarrier $i$ is allocated to user $k$ and IRS is allocated to user $k'$ in time slot $n$, we have $t_{k,k',i} \left[n\right] = u_{k,i}\left[n\right] s_{k'}\left[n\right] = 1$, otherwise, it is zero.
The lower bound rate function in \eqref{IndividualRateLB} can be rewritten as
\vspace{-2mm}
\begin{equation}\label{LowerbOUNDrATEfUNCTION}
R^{\mathrm{LB}}_{k,\mathrm{LoS}}\left[ n \right]\hspace{-1mm} =\hspace{-1mm} \sum_{k'=1}^{K}\sum_{i=1}^{{N_{\mathrm{F}}}}\hspace{-1mm}R^{\mathrm{LB}}_{k,k',i,\mathrm{LoS}}\left[ n \right], \vspace{-2mm}
\end{equation}
where $R^{\mathrm{LB}}_{k,k',i,\mathrm{LoS}}\left[ n \right] \hspace{-1mm}= \hspace{-1mm}{{t_{k,k',i}}\left[ n \right]{{\log }_2}\hspace{-1mm}\left(\hspace{-0.5mm} {1 \hspace{-1mm}+\hspace{-1mm} {{{p_{k,i}}\left[ n \right]\left|{{{g}}}_{k,k',i,\mathrm{LoS}}^{{\mathrm{UG}}}\left[ n \right]\right|^2_{\mathrm{LB}}}}/{{{\sigma^2}}}}\hspace{-0.5mm} \right)}$ and
\vspace{-2mm}
\begin{equation}\label{ChannelGian}
\left|{{{g}}}_{k,k',i,\mathrm{LoS}}^{{\mathrm{UG}}}\left[ n \right]\right|^2_{\mathrm{LB}} ={\frac{A_{k}}{{{\left({d_k^{{\rm{UG}}}\left[ n \right]}\right)^{{\alpha_k^{\mathrm{UG}}}}}}} + \frac{B_{k,k'}}{{{{\left( {{d^{{\rm{UR}}}}\left[ n \right]} \right)}^2}}} + \frac{C_{k,k',i}}{{\left({d_k^{{\rm{UG}}}\left[ n \right]}\right)^{\frac{\alpha_k^{\mathrm{UG}}}{2}}{d^{{\rm{UR}}}}\left[ n \right]}}}, \vspace{-2mm}
\end{equation}
with $A_{k} =  {\beta_0}\frac{\kappa^{\mathrm{UG}}_k}{\kappa^{\mathrm{UG}}_k + 1}$, $B_{k,k'} = \frac{a^2\beta^2_0 {B^2_{{M_{\rm{r}}}}}\left( {\psi _{k',k}^{\rm{r}}} \right){B^2_{{M_{\rm{c}}}}}\left( {\psi _{k',k}^{\rm{c}}} \right) }{{\left( {d_k^{{\rm{RG}}}}\right)}^{\alpha^{\mathrm{RG}}_k}}\frac{\kappa^{\mathrm{RG}}_k}{\kappa^{\mathrm{RG}}_k + 1}$, $C_{k,k',i} = \sum_{j=1}^{4} {D_{k,k',j}} I_{i,j}$, and ${D_{k,k',j}}$ is given by equation \eqref{DD_Derivate} at the top of this page.

\begin{figure*}[!t]
	% ensure that we have normalsize text
	\normalsize
	\vspace{-8mm}
	% Store the current equation number.
	%\setcounter{mytempeqncnt}{\value{equation}}
	% Set the equation number to one less than the one
	% desired for the first equation here.
	% The value here will have to changed if equations
	% are added or removed prior to the place these
	% equations are referenced in the main text.
	\begin{align} \label{DD_Derivate}
	{D_{k,k',j}} = \left\{ {\begin{array}{*{20}{c}}
		{ - \frac{{2a\beta _0^{\frac{3}{2}}{B_{{M_{\rm{r}}}}}\left( {\psi _{k',k}^{\rm{r}}} \right){B_{{M_{\rm{c}}}}}\left( {\psi _{k',k}^{\rm{c}}} \right)}}{{{\left({d_k^{{\rm{RG}}}}\right)^{\frac{\alpha_k^{\mathrm{UG}}}{2}}}}}} \sqrt{\frac{\kappa^{\mathrm{UG}}_k}{\kappa^{\mathrm{UG}}_k + 1}} \sqrt{\frac{\kappa^{\mathrm{RG}}_k}{\kappa^{\mathrm{RG}}_k + 1}}&{{\rm{if}}\;k' \ne k},\\[-0.5mm]
		{\frac{{2a\beta _0^{\frac{3}{2}}{M_{\rm{r}}}{M_{\rm{c}}}}}{{{\left({d_k^{{\rm{RG}}}}\right)^{\frac{\alpha_k^{\mathrm{UG}}}{2}}}}} \sqrt{\frac{\kappa^{\mathrm{UG}}_k}{\kappa^{\mathrm{UG}}_k + 1}} \sqrt{\frac{\kappa^{\mathrm{RG}}_k}{\kappa^{\mathrm{RG}}_k + 1}}} \cos \left( {2\pi \alpha } \right)&{{\rm{if}}\;k' = k,\;j = 1,}\\[-0.5mm]
		0&{{\rm{if}}\;k' = k,\;j = 2,}\\[-0.5mm]
		{ - \frac{{2a\beta_0^{\frac{3}{2}}{M_{\rm{r}}}{M_{\rm{c}}}}}{{{\left({d_k^{{\rm{RG}}}}\right)^{\frac{\alpha_k^{\mathrm{UG}}}{2}}}}}}\sqrt{\frac{\kappa^{\mathrm{UG}}_k}{\kappa^{\mathrm{UG}}_k + 1}} \sqrt{\frac{\kappa^{\mathrm{RG}}_k}{\kappa^{\mathrm{RG}}_k + 1}}\cos \left( {2\pi \alpha } \right)&{{\rm{if}}\;k' = k,\;j = 3,}\\[-0.5mm]
		{ - \frac{{2a\beta _0^{\frac{3}{2}}{M_{\rm{r}}}{M_{\rm{c}}}}}{{{\left({d_k^{{\rm{RG}}}}\right)^{\frac{\alpha_k^{\mathrm{UG}}}{2}}}}}}\sqrt{\frac{\kappa^{\mathrm{UG}}_k}{\kappa^{\mathrm{UG}}_k + 1}} \sqrt{\frac{\kappa^{\mathrm{RG}}_k}{\kappa^{\mathrm{RG}}_k + 1}}&{{\rm{if}}\;k' = k,\;j = 4.}
		\end{array}} \right.
	\end{align}
	% Restore the current equation number.
	\vspace{-10mm}
	% IEEE uses as a separator
	\hrulefill
	% The spacer can be tweaked to stop underfull vboxes.
\end{figure*}

\vspace{-4mm}
\section{Solution of the Lower Bound Problem}
In this section, we aim to obtain a suboptimal solution of the lower bound problem $\mathcal{P}_{\mathrm{LB}}$ by dividing it into two subproblems, where we alternatingly solve the two subproblems until converge.
In particular, in the $\mathrm{iter}$-th iteration, subproblem 1 focuses on resource allocation and IRS scheduling design given the obtained UAV's trajectory while subproblem 2 aims to design the UAV's trajectory given the obtained resource allocation and IRS scheduling strategy.

\vspace{-4mm}
\subsection{Subproblem 1: Resource Allocation and IRS Scheduling Design}

Given the trajectory of the UAV ${\bf{q}}^{\mathrm{iter}}\left[ n \right]$ in the $\mathrm{iter}$-th iteration, substituting the lower bound rate functions in \eqref{LowerbOUNDrATEfUNCTION} into \eqref{ProblemFormulationLB} yields subproblem 1 as follows:
\vspace{-2mm}
\begin{align}\label{ProblemFormulationApproxSubproblem1}
\hspace{2mm}&\underset{\mathbf{{T}},\mathbf{{U}},\mathbf{{P}},\mathbf{{S}}}{\maxo}\,\, \frac{1}{N}\sum\nolimits_{n = 1}^N {\sum\nolimits_{k = 1}^K {R_{k,\mathrm{LoS}}^{{\rm{LB}}}} \left[ n \right]\left( {{\bf{T}},{\bf{U}},{\bf{P}},{\bf{S}}}\left|{\bf{q}}^{\mathrm{iter}}\left[ n \right]\right. \right)} \\[-1mm]
\hspace{-6mm}\notag\mbox{s.t.}\;\;
%%%%%
& \mbox{{C1-C6}},\;\underline{\mbox{{C7}}}:\;  \frac{1}{N} \sum\nolimits_{n=1}^{N} R^{\mathrm{LB}}_{k,\mathrm{LoS}}\left[ n \right]\left({\bf{T}},\mathbf{{U}},\mathbf{{P}},\mathbf{{S}}\left|{\bf{q}}^{\mathrm{iter}}\left[ n \right]\right.\right) \ge R_{\mathrm{min},k}, \forall k, \notag\\[-1mm]
& \mbox{{C12}}:\; 0 \le t_{k,k',i}\left[ n \right] \le 1, \forall k, k', n, i, \hspace{7.2mm}\mbox{{C13}}:\; t_{k,k',i}\left[ n \right] \le {{s_{k'} \left[ n \right]}}, \forall k, k', n, i,\notag\\[-1mm]
& \mbox{{C14}}:\; t_{k,k',i}\left[ n \right] \le u_{k,i}\left[ n \right], \forall k, k', n, i, \hspace{5mm}\mbox{{C15}}:\; t_{k,k',i}\left[ n \right] \ge {{s_{k'} \left[ n \right]}} + u_{k,i}\left[ n \right] -1,\forall k, k', n, i, \notag
\end{align}
\vspace{-8mm}\par\noindent
where ${R_{k,\mathrm{LoS}}^{{\rm{LB}}}} \left[ n \right]\left( {{\bf{T}},{\bf{U}},{\bf{P}},{\bf{S}}}\left|{\bf{q}}^{\mathrm{iter}}\left[ n \right]\right. \right)$ denotes the achievable data rate of user $k$ in time slot $n$ given the trajectory of UAV as ${\bf{q}}^{\mathrm{iter}}\left[ n \right]$.
Constraints C12-C15 are introduced to illustrate the relationship between $t_{k,k',i} \left[n\right]$, $u_{k,i} \left[n\right]$, and $s_{k'}\left[n\right]$.
In particular, $t_{k,k',i} \left[n\right] = 1$ if and only if both $u_{k,i} \left[n\right] = 1$ and $s_{k'}\left[n\right] = 1$.
%
%It is worth to note that we cannot ignore the IRS scheduling variable $s_{k'}\left[n\right]$ even the objective function does not depends on $s_{k'}\left[n\right]$.
%
%This is because that IRS scheduling is frequency-flat and thus we cannot allocate IRS to different users on different subcarriers, i.e., if $t_{k,k',i} \left[n\right] = 1$, we have $t_{k,k'',i'} \left[n\right] = 0$, $\forall k''\neq k'$ and $\forall i' \neq i$.

Given the trajectory of UAV, the subproblem 1 in \eqref{ProblemFormulationApproxSubproblem1} is still a mixed-integer non-convex optimization problem.
{To solve subproblem 1, we adopt the Lagrangian dual method since it can unveil some important insights about power allocation, power scaling law, and IRS scheduling as detailed in the following.}
The binary variables $s_{k'} \left[ n \right]$ and ${u_{k,i}}\left[ n \right]$ span a disjoint feasible solution set which is a hurdle for solving the problem via computationally efficient tools from convex optimization theory.
Therefore, we relax the subcarrier allocation variable ${u_{k,i}}\left[ n \right]$ and the IRS scheduling variable $s_{k'} \left[ n \right]$ to be a real between zero and one instead of a Boolean.
In fact, $u_{k,i} \left[ n \right]$ and $s_{{k'}} \left[ n \right]$ can be interpreted as time-sharing factors for subcarrier allocation and IRS scheduling, respectively\cite{DerrickEESWIPT}.
In the following, we will prove that the optimal solution for $u_{k,i} \left[ n \right]$ and $s_{{k'}} \left[ n \right]$ are still binary, despite the use of binary constraint relaxation.
In other words, the time sharing relaxation is tight and does not lose any optimality.
Besides, the coupling between optimization variables $t_{k,k',i} \left[ n \right]$ and power allocation variables $p_{k,i}$ in the objective function and constraint in $\underline{\mbox{{C7}}}$ is generally intractable.
Fortunately, a dual decomposition method \cite{DerrickEESWIPT} can be employed to handle this issue and to obtain some insights about resource allocation design in an IRS-assisted UAV OFDMA communication system.
In particular, we introduce the auxiliary time-shared power allocation variables $\tilde{p}_{k,k',i} \left[ n \right] =  t_{k,k',i} \left[ n \right]p_{k,i} \left[ n \right]$.
The problem in \eqref{ProblemFormulationApproxSubproblem1} can be rewritten as
\vspace{-2mm}
\begin{align}\label{ProblemFormulationApproxSubproblem1Dual}
&\underset{\mathbf{{T}},\;\mathbf{{U}},\;\tilde{\mathbf{{P}}},\;\mathbf{{S}}}{\maxo}\,\, \frac{1}{N}\sum\nolimits_{n = 1}^N {\sum\nolimits_{k = 1}^K {R_{k,\mathrm{LoS}}^{{\rm{LB}}}} \left[ n \right]\left( {{\bf{T}},{\bf{U}},\tilde{\bf{P}},{\bf{S}}}\left|{\bf{q}}^{\mathrm{iter}}\left[ n \right]\right. \right)} \\[-1mm]
\hspace{-10mm}\notag\mbox{s.t.}\;\;
%%%%%
& \mbox{{C2-C4}},\mbox{{C6}},\mbox{{C12-C15}},\;\;\;\;\;\;\;\;\;\;\;\mbox{{C1}}:\; 0 \le {u_{k,i}}\left[ n \right] \le 1,\;\forall k,i,n, \notag\\[-1mm]
&\mbox{{C5}}:\; 0 \le s_{k'} \left[ n \right]\le 1, \forall {k'}, n, \; \underline{\mbox{{C7}}}:\; \frac{1}{N} \sum\nolimits_{n=1}^{N} R^{\mathrm{LB}}_{k,\mathrm{LoS}}\left[ n \right]\left({\bf{T}},\mathbf{{U}},\tilde{\mathbf{{P}}},\mathbf{{S}}\left|{\bf{q}}^{\mathrm{iter}}\left[ n \right]\right.\right) \ge R_{\mathrm{min},k}, \forall k, \notag
\end{align}
\vspace{-6mm}\par\noindent
where ${R_{k,\mathrm{LoS}}^{{\rm{LB}}}} \left[ n \right]\left( {{\bf{T}},{\bf{U}},\tilde{\bf{P}},{\bf{S}}}\left|{\bf{q}}^{\mathrm{iter}}\left[ n \right]\right. \right) = {R_{k,\mathrm{LoS}}^{{\rm{LB}}}} \left[ n \right]\left( {{\bf{T}},{\bf{U}},{\bf{P}},{\bf{S}}}\left|{\bf{q}}^{\mathrm{iter}}\left[ n \right]\right. \right)\left|_{p_{k,i} \left[ n \right] = \frac{\tilde{p}_{k,k',i} \left[ n \right]}{t_{k,k',i} \left[ n \right]} }\right.$.

\begin{figure*}[!t]
	% ensure that we have normalsize text
	\normalsize
	\vspace{-8mm}
	% Store the current equation number.
	%\setcounter{mytempeqncnt}{\value{equation}}
	% Set the equation number to one less than the one
	% desired for the first equation here.
	% The value here will have to changed if equations
	% are added or removed prior to the place these
	% equations are referenced in the main text.
	%\setcounter{equation}{41}
	\begin{align}\label{LLL_Figure}
	&\mathcal{L}\left(\mathbf{{T}},\mathbf{{U}},\mathbf{\tilde{P}},\mathbf{{S}},\bm{\zeta},\bm{\varrho},\bm{\gamma},\bm{\nu},\bm{\varsigma},\bm{\varpi},\bm{\xi}\right) \notag\\[-1.5mm]
	&= \frac{1}{N}\sum\limits_{n = 1}^N  \sum_{i=1}^{{N_{\mathrm{F}}}}\sum\limits_{k = 1}^K \sum\limits_{k' = 1}^K \left(\nu_k \hspace{-0.5mm}+\hspace{-0.5mm} 1\right){{t_{k,k',i}}\left[ n \right] {{\log }_2}\left( {1 \hspace{-0.5mm}+\hspace{-0.5mm} \frac{{\tilde{p}_{k,k',i} \left[ n \right]\left|{{{g}}}_{k,k',i,\mathrm{LoS}}^{{\mathrm{UG}}}\left[ n \right]\right|^2_{\mathrm{LB}}}}{{{{t_{k,k',i}}\left[ n \right]\sigma^2}}}} \right)} - \sum_{k=1}^{K} \nu_k R_{\mathrm{min},k} \notag\\[-1mm]
	& - \sum_{n=1}^{N} \sum_{i=1}^{N_{\mathrm{F}}} \zeta_{i,n} \left(\hspace{-0.5mm}\sum_{k=1}^{K}  u_{k,i}\left[ n \right] \hspace{-1mm}-\hspace{-1mm} 1\hspace{-0.5mm}\right)  \hspace{-1mm}-\hspace{-1mm} \sum_{n=1}^{N}\varrho_{n} \left(\hspace{-0.5mm} \sum_{i=1}^{N_{\mathrm{F}}} \sum_{k=1}^{K} \sum_{k'=1}^{K} {\tilde{p}_{k,k',i}}\left[ n \right] \hspace{-1mm}-\hspace{-1mm} p_{\mathrm{max}}  \hspace{-0.5mm}\right) \hspace{-1mm}-\hspace{-1mm} \sum_{n=1}^{N}\gamma_{n} \left(\hspace{-0.5mm}\sum_{k'=1}^{K}  s_{k'}\left[ n \right] \hspace{-1mm}- \hspace{-1mm}1\hspace{-0.5mm}\right) \notag\\[-1mm]
	& - \sum_{n=1}^{N} \sum_{i=1}^{N_{\mathrm{F}}} \sum_{k=1}^{K} \sum\limits_{k' = 1}^K \varsigma_{k,k',i,n} 	\left(t_{k,k',i}\left[ n \right] - s_{k'}\left[ n \right] \right) - \sum_{n=1}^{N} \sum_{i=1}^{N_{\mathrm{F}}} \sum_{k=1}^{K} \sum\limits_{k' = 1}^K \varpi_{k,k',i,n} \left(t_{k,k',i}\left[ n \right] - u_{k,i}\left[ n \right]\right) \notag\\[-1mm]
	&- \sum_{n=1}^{N} \sum_{i=1}^{N_{\mathrm{F}}} \sum_{k=1}^{K} \sum\limits_{k' = 1}^K \xi_{k,k',i,n} \left( u_{k,i}\left[ n \right] + s_{k'}\left[ n \right] - t_{k,k',i}\left[ n \right] - 1\right),
	\end{align}
	% Restore the current equation number.
	%\setcounter{equation}{42}
	\vspace{-10mm}
	% IEEE uses as a separator
	\hrulefill
	% The spacer can be tweaked to stop underfull vboxes.
\end{figure*}

The transformed problem in \eqref{ProblemFormulationApproxSubproblem1Dual} is convex w.r.t. to $\mathbf{{T}}$,$\mathbf{{U}}$,$\mathbf{\tilde{P}}$, and $\mathbf{{S}}$, while satisfying the Slater’s constraint qualification \cite{Boyd2004}.
Therefore, we can solve the primal problem by solving its dual problem.
To this end, the Lagrangian function of the primal problem in \eqref{ProblemFormulationApproxSubproblem1Dual} is given by equation \eqref{LLL_Figure} on the top of this page, where $\zeta_{i,n} \ge 0$, $\varrho_{n} \ge 0$, $\gamma_{n} \ge 0$, $\nu_k \ge 0$, $\varsigma_{k,k',i,n} \ge 0$, $\varpi_{k,k',i,n}$, and $\xi_{k,k',i,n} \ge 0$ are the Lagrange multipliers corresponding to constraints C2, C4, C6, C7, C12, C13, and C14, respectively.
Boundary constraints C1, C3, C5, and C12 will be absorbed in the Karush-Kuhn-Tucker (KKT) conditions when deriving the optimal resource allocation policy of subproblem 1 in the following equation.
Therefore, the dual problem for the primal problem in \eqref{ProblemFormulationApproxSubproblem1Dual} is given by
\vspace{-2mm}
\begin{equation} \label{ResourceAllocationDual}
\underset{\bm{\zeta},\bm{\varrho},\bm{\gamma},\bm{\nu},\bm{\varsigma},\bm{\varpi},\bm{\xi}}{\mino} \;\; \underset{\mathbf{{T}},\mathbf{{U}},\mathbf{\tilde{P}},\mathbf{{S}}}{\maxo} \;{\mathcal L}\left(\mathbf{{T}},\mathbf{{U}},\mathbf{\tilde{P}},\mathbf{{S}},\bm{\zeta},\bm{\varrho},\bm{\gamma},\bm{\nu},\bm{\varsigma},\bm{\varpi},\bm{\xi}\right).\vspace{-2mm}
\end{equation}

Since the dual problem is convex, the Lagrange dual decomposition can be employed to solve the dual problem in \eqref{ResourceAllocationDual} iteratively.
In particular, the dual problem in \eqref{ResourceAllocationDual} is decomposed into two-layer optimization problems and is solved iteratively.
Specifically, the inner layer problem maximizes the Lagrangian ${\mathcal L}$ over $\left(\mathbf{{T}},\mathbf{{U}},\mathbf{\tilde{P}},\mathbf{{S}}\right)$ for given Lagrangian multipliers $\left(\bm{\zeta},\bm{\varrho},\bm{\gamma},\bm{\nu},\bm{\varsigma},\bm{\varpi},\bm{\xi}\right)$, while the outer layer optimization problem minimizes ${\mathcal L}$ over $\left(\bm{\zeta},\bm{\varrho},\bm{\gamma},\bm{\nu},\bm{\varsigma},\bm{\varpi},\bm{\xi}\right)$ for given $\left(\mathbf{{T}},\mathbf{{U}},\mathbf{\tilde{P}},\mathbf{{S}}\right)$.
For a fixed set of Lagrange multipliers $\left(\bm{\zeta},\bm{\varrho},\bm{\gamma},\bm{\nu},\bm{\varsigma},\bm{\varpi},\bm{\xi}\right)$, the inner maximization problem is a convex optimization problem w.r.t. $\left(\mathbf{{T}},\mathbf{{U}},\mathbf{\tilde{P}},\mathbf{{S}}\right)$.
Applying the convex optimization techniques and the KKT conditions, the optimal power allocation for user $k$ on subcarrier $i$ in time slot $n$ can be obtained by
\vspace{-2mm}
\begin{equation}\label{PwerUpdate}
	{\tilde{p}_{k,k',i}}^*\left[ n \right] = t_{k,k',i}\left[ n \right] p^*_{k,k',i}\left[ n \right] = t_{k,k',i}\left[ n \right]  \left[ \frac{\left(\nu_k+1\right)}{ \varrho_n \ln (2) N} - \frac{\sigma^2}{\left|{{{g}}}_{k,k',i,\mathrm{LoS}}^{{\rm{UG}}}\left[ n \right]\right|^2_{\mathrm{LB}}}\right]^{+}.\vspace{-1mm}
\end{equation}
%
%As a result, we have
%\begin{align}
%	p_{k,i}^*\left[ n \right] = \left\{ {\begin{array}{*{20}{c}}
%		{{{\left[ {\frac{{1 + {\nu _k}}}{{{\varrho _n}\ln 2N}} - \frac{1}{{{B_k}\left[ n \right]}}} \right]}^ + },}&{{\rm{if }}\;{u_{k,i}}\left[ n \right] = {t_{k,i}}\left[ n \right] = 1}\\
%		{{{\left[ {\frac{{1 + {\nu _k}}}{{{\varrho _n}\ln 2N}} - \frac{1}{{{A_k}\left[ n \right]}}} \right]}^ + },}&{{\rm{if }}\;{u_{k,i}}\left[ n \right] = 1\;{\rm{and}}\;{t_{k,i}}\left[ n \right] = 0}
%		\end{array}} \right.
%\end{align}
It implies that the optimal power allocation follows a multi-level water-filling principle.
The first term in the bracket in \eqref{PwerUpdate} can be interpreted as a water level for allocating power to user $k$ in time slot $n$, which can be determined by both Lagrangian multipliers $\nu _k$ and ${\varrho _n}$ associated with the minimum rate requirement constraint $\underline{\mbox{C7}}$ and the sum-power constraint C4, respectively.
In particular, the larger Lagrangian multiplier $\nu _k$ is, the higher power would allocate to user $k$ to satisfy its minimum rate requirement.
In contrast, the larger Lagrangian multiplier ${\varrho _n}$ is, the less power would allocate in time slot $n$ to satisfy the sum-power constraint.
When $k = k'$, ${\tilde{p}_{k,k,i}}^*\left[ n \right]$ denotes the power allocated to user $k$ on subcarrier $i$ in time slot $n$ if user $k$ is scheduled to be an IRS-assisted user.
In this case, ${\left|{{{g}}}_{k,k,i,\mathrm{LoS}}^{{\rm{UG}}}\left[ n \right]\right|^2_{\mathrm{LB}}}$ is proportional to $M_{\rm r}^2 M_{\rm c}^2$ when $M_{\rm r} \to \infty$ and $M_{\rm c} \to \infty$ as predicted in \eqref{ChannelGian}.
It implies that with a fixed data rate requirement, employing an IRS can scale down the transmit power of the UAV to $\frac{1}{M_{\rm r}^2 M_{\rm c}^2}$ compared to that of the system without IRS, which is consistent with the power scaling law obtained in \cite{WuIRS}.

To obtain the optimal user scheduling, we take the derivative of the Lagrangian function w.r.t. ${u_{k,i}}\left[ n \right]$, ${t_{k,k',i}}\left[ n \right]$, and $s_k\left[ n \right]$, respectively, which yields
\vspace{-2mm}
\begin{align}
\frac{{\partial {\cal L}}}{{\partial {u_{k,i}}\left[ n \right]}} &= M_{k,i,n}^{\left( u \right)} = - {\zeta _{i,n}} + \sum\nolimits_{k'=1}^{K} \left({\varpi _{k,k',i,n}} - {\xi _{k,k',i,n}}\right), \label{SubcarrierAllocation}\\[-1mm]
\frac{{\partial {\cal L}}}{{\partial {t_{k,k',i}}\left[ n \right]}} &= M_{k,k',i,n}^{\left( t \right)} = \frac{1}{N}\left( {{\nu _k} + 1} \right)\left[ {{\log }_2}\left( {1 + \frac{p^*_{k,k',i}\left[ n \right] \left|{{{g}}}_{k,k',i,\mathrm{LoS}}^{{\rm{UG}}}\left[ n \right]\right|^2_{\mathrm{LB}}}{\sigma^2} } \right) \right.\notag\\[-1mm]
&\hspace{-10mm}\left. - \frac{1}{{\ln 2}}\frac{{p^*_{k,k',i}\left[ n \right] \left|{{{g}}}_{k,k',i,\mathrm{LoS}}^{{\rm{UG}}}\left[ n \right]\right|^2_{\mathrm{LB}}}}{{p^*_{k,k',i}\left[ n \right]\left|{{{g}}}_{k,k',i,\mathrm{LoS}}^{{\rm{UG}}}\left[ n \right]\right|^2_{\mathrm{LB}} + \sigma^2}} \right] - \varsigma_{k,k',i,n} - {\varpi _{k,k',i,n}} + {\xi _{k,k',i,n}}, \;\text{and}\label{IRSSubcarrierAllocation}\\[-1mm]
\frac{{\partial {\cal L}}}{{\partial {s_{k'}}\left[ n \right]}} &= M_{k',n}^{\left( s \right)}= -\gamma_{n} +  \sum\nolimits_{k=1}^{K}\sum\nolimits_{i=1}^{{N_{\mathrm{F}}}}\left(\varsigma_{k,k',i,n} - \xi_{k,k',i,n}\right).\label{IRSAllocation}
\end{align}
\vspace{-8mm}\par\noindent

We can observe that the derivatives of the Lagrangian function w.r.t. ${u_{k,i}}\left[ n \right]$, ${t_{k,k',i}}\left[ n \right]$, and $s_{k'}\left[ n \right]$ are all constants, which implies that the Lagrangian function grows linearly w.r.t. ${u_{k,i}}\left[ n \right]$, ${t_{k,k',i}}\left[ n \right]$, and $s_{k'}\left[ n \right]$.
In particular, the derivatives $M_{k,i,n}^{\left( u \right)}$, $M_{k,k',i,n}^{\left( t \right)}$, and $M_{k',n}^{\left( s \right)}$ can be interpreted as a marginal benefit for the system performance when assigning ${u_{k,i}}\left[ n \right] = 1$, ${t_{k,k',i}}\left[ n \right] = 1$, and ${s_{k'}}\left[ n \right] = 1$, respectively.
{In \eqref{ProblemFormulationApproxSubproblem1Dual}, constraints C1, C2, C5, C6, C12-C15 are all affine constraints, which constitute a polyhedron feasible solution set.}
Therefore, the optimal solution for $\mathbf{{T}}$, $\mathbf{{U}}$, $\mathbf{{S}}$ should lie at a vertex of the feasible solution set, i.e., they must be binary.
In addition, C2 and C6 imply that $\sum_{k=1}^{K}\sum_{k'=1}^{K} {t_{k,k',i}}\left[ n \right] = \sum_{k=1}^{K} {u_{k,i}}\left[ n \right] \sum_{k'=1}^{K} s_{k'}\left[ n \right] \le 1$.
It means that at most one user can be assigned on subcarrier $i$ in time slot $n$, no matter whether this user is an IRS-assisted user or a non-IRS-assisted user.
Now, to maximize the Lagrangian given the dual variables, we have
\vspace{-2mm}
\begin{align} \label{BinaryUpdate}
u_{k,i}^*\left[ n \right] &= \left\{ {\begin{array}{*{20}{c}}
	1&{k = \mathop {\arg \max }\limits_{k'} M_{k,i,n}^{\left( u \right)}}\\[-1mm]
	0&{{\rm{otherwise}}}
	\end{array}} \right.,\forall i,n, \;\;
s_k^*\left[ n \right] = \left\{ {\begin{array}{*{20}{c}}
	1&{k = \mathop {\arg \max }\limits_{k'} M_{k',n}^{\left( s \right)}}\\[-1mm]
	0&{{\rm{otherwise}}}
	\end{array}} \right.,\forall n, \;\text{and}\notag\\[-1mm]
t_{k,k',i}^*\left[ n \right] &= \left\{ {\begin{array}{*{20}{c}}
	1&{i \in \mathop {\arg {{\max }}}\limits_{k,k'} M_{k,k',i,n}^{\left( t \right)}}\\[-1mm]
	0&{{\rm{otherwise}}}
	\end{array}} \right.,\forall i, n.
\end{align}
\vspace{-6mm}\par\noindent
Recall that $t_{k,k',i}\left[ n \right] = 1$ means user $k$ is allocated to subcarrier $i$ in time slot $n$ when user $k'$ is scheduled to be an IRS-assisted user.
Therefore, the selection in \eqref{BinaryUpdate} for $t_{k,k',i}\left[ n \right]$ can determine both the subcarrier allocation and IRS scheduling at the same time.
Furthermore, when $M_{\rm r} \to \infty$ and $M_{\rm c} \to \infty$, the composite channel gain of the IRS-assisted user is significantly larger than that of the non-IRS-assisted user, i.e., $\left|{{{g}}}_{k,k,i,\mathrm{LoS}}^{{\rm{UG}}}\left[ n \right]\right|^2_{\mathrm{LB}} \gg \left|{{{g}}}_{k,k',i,\mathrm{LoS}}^{{\rm{UG}}}\left[ n \right]\right|^2_{\mathrm{LB}}$, $\forall k' \neq k$, $\forall i, n$.
As a result, the maximization operation in \eqref{BinaryUpdate} can be simplified as
\vspace{-2mm}
\begin{align} \label{BinaryUpdateII}
\lim_{M_{\rm r} \to \infty, M_{\rm c} \to \infty} t_{k,k,i}^*\left[ n \right] &= \left\{ {\begin{array}{*{20}{c}}
		1&{i \in \mathop {\arg {{\max }}}\limits_{k} M_{k,k,i,n}^{\left( t \right)}}\\[-1mm]
		0&{{\rm{otherwise}}}
\end{array}} \right..
\end{align}
\vspace{-6mm}\par\noindent
{Besides, when $k = k'$, the first term is significantly larger than the remaining terms in \eqref{IRSSubcarrierAllocation} and it dominates the derivatives $M_{k,k,i,n}^{\left( t \right)}$.}
It implies that a user with a higher composite channel power gain or a more stringent data rate requirement has a higher chance to be scheduled as an IRS-assisted user.
Furthermore, gradient methods can be used for updating the Lagrange multipliers in the outer layer:
\vspace{-2mm}
\begin{align}
	\nu _k^{{l_{\mathrm{I}}} + 1} &= \left[ \nu _k^{{l_{\mathrm{I}}}} - \tau _1 ^{{l_{\mathrm{I}}}} \times \left( \frac{1}{N} \sum\nolimits_{n=1}^{N} R^{\mathrm{LB}}_{k,\mathrm{LoS}}\left[ n \right]\left({\bf{T}},\;\mathbf{{U}},\;\tilde{\mathbf{{P}}},\;\mathbf{{S}}\left|{\bf{q}}^{\mathrm{iter}}\left[ n \right]\right.\right) - R_{\mathrm{min},k} \right) \right]^ + , \label{DualUpdatingI}\\[-1mm]
	%\zeta _{i,n}^{{l_{\mathrm{I}}} + 1} &= {\left[ {\zeta _{i,n}^{{l_{\mathrm{I}}}} - \tau_2 ^{{l_{\mathrm{I}}}} \times \left( {1 - \sum\nolimits_{k = 1}^K {{u_{k,i}}} \left[ n \right]} \right)} \right]^ + },\\[-1mm]
	\varrho_{n}^{{l_{\mathrm{I}}} + 1} & = {\left[ {\zeta _{i,n}^{{l_{\mathrm{I}}}} - \tau_2 ^{{l_{\mathrm{I}}}} \times 	 \left(p_{\mathrm{max}} -  \sum\nolimits_{i = 1}^{N_{\mathrm{F}}} \sum\nolimits_{k = 1}^{K} \sum\nolimits_{k' = 1}^{K} \tilde{p}_{k,k',i} \left[ n \right] \right)} \right]^ + },\\[-1mm]
	%\gamma_{n}^{{l_{\mathrm{I}}} + 1} & =  {\left[ {\gamma_{n}^{{l_{\mathrm{I}}}} - \tau_4 ^{{l_{\mathrm{I}}}} \times \left(1 - \sum\nolimits_{k'=1}^{K}  s_{k'}\left[ n \right]\right)} \right]^ + },\\[-1mm]
	\varsigma_{k,k',i,n}^{{l_{\mathrm{I}}} + 1} & = {\left[ {\varsigma_{k,k',i,n}^{{l_{\mathrm{I}}}} - \tau_3 ^{{l_{\mathrm{I}}}} \times \left(s_{k'}\left[ n \right] - t_{k,k',i}\left[ n \right] \right)} \right]^ + }, \\[-1mm]
	\varpi_{k,k',i,n}^{{l_{\mathrm{I}}} + 1} & = {\left[ {\varpi_{k,k',i,n}^{{l_{\mathrm{I}}}} - \tau_4 ^{{l_{\mathrm{I}}}} \times \left(u_{k,i}\left[ n \right] - t_{k,k',i}\left[ n \right] \right)} \right]^ + } , \;\text{and}\\[-1mm]
	\xi_{k,k',i,n}^{{l_{\mathrm{I}}} + 1} & = {\left[ {\xi_{k,k',i,n}^{{l_{\mathrm{I}}}} - \tau_5 ^{{l_{\mathrm{I}}}} \times \left(1+ t_{k,k',i}\left[ n \right] - u_{k,i}\left[ n \right] - s_{k'}\left[ n \right] \right)} \right]^ + } ,\label{DualUpdatingVII}
\end{align}
\vspace{-8mm}\par\noindent
where $\tau_m ^{{l_{\mathrm{I}}}} \ge 0$, $m\in \{1,\ldots,5\}$ denote positive step size for the dual variables in the $l_{\mathrm{I}}$-th iteration for solving subproblem 1.
Note that Lagrange multipliers $\zeta _{i,n}$ and $\gamma_{n}$ remain unchanged as constraints C2 and C6 always hold with equality when updating ${{u_{k,i}}} \left[ n \right]$ and $s_{k'}\left[ n \right]$ according to \eqref{BinaryUpdate}, respectively.
The primal variables and the dual variables are updated iteratively according to \eqref{PwerUpdate}, \eqref{BinaryUpdate}, and \eqref{DualUpdatingI}-\eqref{DualUpdatingVII},  respectively.
As the primal problem in \eqref{ProblemFormulationApproxSubproblem1Dual} is a convex optimization problem, it is guaranteed that the primal optimal solution can be achieved by solving the problem in outer and inner layer iteratively, when the updating step size $\tau_m ^{{l_{\mathrm{I}}}}$ satisfy the infinite travel conditions \cite{DerrickEESWIPT}.

\vspace{-4mm}
\subsection{Subproblem 2: UAV Trajectory Design}
Given resource allocation and IRS scheduling design $\left({{\bf{T}}^{\mathrm{iter}},{\bf{U}}^{\mathrm{iter}},\tilde{\bf{P}}^{\mathrm{iter}},{\bf{S}}}^{\mathrm{iter}}\right)$  in the $\mathrm{iter}$-th iteration, the trajectory design can be formulated as
\vspace{-2mm}
\begin{align}\label{ProblemFormulationApproxSubproblem2}
&\underset{{\bf{q}}\left[ n \right]}{\maxo}\,\, \frac{1}{N}\sum\nolimits_{n = 1}^N {\sum\nolimits_{k = 1}^K {R_{k,\mathrm{LoS}}^{{\rm{LB}}}} \left[ n \right]\left({\bf{q}}\left[ n \right] \left|{{\bf{T}}^{\mathrm{iter}},{\bf{U}}^{\mathrm{iter}},\tilde{\bf{P}}^{\mathrm{iter}},{\bf{S}}}^{\mathrm{iter}}\right. \right)} \\[-1mm]
\hspace{-10mm}\notag\mbox{s.t.}\;\;
%%%%%
%\mbox{{C1}}:\; {u_{k,i}}\left[ n \right] \in \{0,1\},\;\forall k,i,n, \notag\\
%\hspace{-10mm}& \mbox{{C2}}:\; \sum_{k = 1}^{K} {u_{k,i}}\left[ n \right] \le 1,\;\forall i,n, \notag\\
%\hspace{-10mm}& \mbox{{C3}}:\; {p_{k,i}}\left[ n \right] \ge 0,\;\forall k,i,n, \notag\\
%\hspace{-10mm}& \mbox{{C4}}:\; \sum_{i = 1}^{N_{\mathrm{F}}} \sum_{k = 1}^{K} {p_{k,i}}\left[ n \right] \le p_{\mathrm{max}}, \forall n, \notag\\
%\hspace{-10mm}& \mbox{{C5}}:\; s_k \left[ n \right]\in \{0,1\}, \forall k, n, \notag\\
%\hspace{-10mm}& \mbox{{C6}}:\; \sum_{k=1}^{K}s_k\left[ n \right] \le 1, \forall n, \notag\\
& \mbox{\underline{C7}}:\; \frac{1}{N}\sum\nolimits_{n = 1}^N {{R_{k,\mathrm{LoS}}^{{\rm{LB}}}} \left[ n \right]\left({\bf{q}}\left[ n \right] \left|{{\bf{T}}^{\mathrm{iter}},{\bf{U}}^{\mathrm{iter}},\tilde{\bf{P}}^{\mathrm{iter}},{\bf{S}}}^{\mathrm{iter}}\right. \right)} \ge R_{\mathrm{min},k}, \forall k, \;\; \mbox{{C8-C10}}. \notag
\end{align}
\vspace{-8mm}\par\noindent
%
%A trivial observation is that given any arbitrary resource allocation and IRS scheduling strategy, the UAV flies at its minimum altitude to increase the received signal power, i.e., $z\left[n\right] = H_{\mathrm{U}}$, $\forall n$.
%
To simplify the problem formulation, without loss of generality, we treat $v_k^{{\rm{UG}}}\left[ n \right] = {{{\left({d_k^{{\rm{UG}}}\left[ n \right]}\right)^{{\alpha_k^{\mathrm{UG}}}}}}}$ and $v^{{\rm{UR}}}\left[ n \right] = \left(d^{{\rm{UR}}}\left[ n \right]\right)^2$ as slack variables for trajectory design, which results in the following optimization problem:
\vspace{-2mm}
\begin{align}\label{Subproblem2Transformed}
&\underset{{\bf{q}}\left[ n \right],{\bf{v}}_k^{{\rm{UG}}}\left[ n \right],{\bf{v}}^{{\rm{UR}}}\left[ n \right]}{\maxo}\,\, \frac{1}{N}\sum\limits_{n = 1}^N {\sum\limits_{k = 1}^K {R_{k,\mathrm{LoS}}^{{\rm{LB}}}} \left[ n \right]\left({\bf{v}}_k^{{\rm{UG}}}\left[ n \right],{\bf{v}}^{{\rm{UR}}}\left[ n \right] \left|{{\bf{T}}^{\mathrm{iter}}, {\bf{U}}^{\mathrm{iter}},\tilde{\bf{P}}^{\mathrm{iter}},{\bf{S}}}^{\mathrm{iter}}\right. \right)} \\[-1mm]
\notag\mbox{s.t.}\;\;
%%%%%
& \mbox{\underline{C7}}:\; \frac{1}{N}\sum\limits_{n = 1}^N {{R_{k,\mathrm{LoS}}^{{\rm{LB}}}} \left[ n \right]\left({\bf{v}}_k^{{\rm{UG}}}\left[ n \right],{\bf{v}}^{{\rm{UR}}}\left[ n \right] \left|{{\bf{T}}^{\mathrm{iter}},{\bf{U}}^{\mathrm{iter}},\tilde{\bf{P}}^{\mathrm{iter}},{\bf{S}}}^{\mathrm{iter}}\right. \right)} \ge R_{\mathrm{min},k}, \forall k, \;\;\mbox{{C8-C10}},\notag\\[-1mm]
& \mbox{{C15}}:\; v_k^{{\rm{UG}}}\left[ n \right] \ge  {\left\| {{\bf{q}}\left[ n \right] - {{\bf{w}}_k}} \right\|}^{{\alpha_k^{\mathrm{UG}}}}, \forall n,k, \;\; \mbox{{C16}}:\; v^{{\rm{UR}}}\left[ n \right] \ge  {{{\left\| {{\bf{q}}\left[ n \right] - {{\bf{w}}_{\rm{R}}}} \right\|}^2}}, \forall n, \notag
\end{align}
\vspace{-10mm}\par\noindent
with
\vspace{-4mm}
\begin{align}
	&{{R_{k,\mathrm{LoS}}^{{\rm{LB}}}} \left[ n \right]\left({\bf{v}}_k^{{\rm{UG}}}\left[ n \right],{\bf{v}}^{{\rm{UR}}}\left[ n \right] \left|{{\bf{T}}^{\mathrm{iter}},{\bf{U}}^{\mathrm{iter}},\tilde{\bf{P}}^{\mathrm{iter}},{\bf{S}}}^{\mathrm{iter}}\right. \right)}\\[-1mm]
  = & \sum_{i=1}^{{N_{\mathrm{F}}}}\sum\limits_{k' = 1}^K{t^{\mathrm{iter}}_{k,k',i}}\left[ n \right] {{{\log }_2}\left[ 1 + \frac{\tilde{p}^{\mathrm{iter}}_{k,k',i} \left[ n \right]}{t^{\mathrm{iter}}_{k,k',i} \left[ n \right] \sigma^2}\left({\frac{A_{k}}{{{{ {v_k^{{\rm{UG}}}\left[ n \right]} }}}} + \frac{B_{k,k'}}{{{{ {{v^{{\rm{UR}}}}\left[ n \right]} }}}} + \frac{C_{k,k',i}}{{\sqrt{v_k^{{\rm{UG}}}\left[ n \right]}\sqrt{{v^{{\rm{UR}}}}\left[ n \right]}}}}\right) \right]}.\notag
\end{align}
\vspace{-6mm}\par\noindent
Note that constraints C15 and C16 hold with equality at the optimal solution since the closer the UAV to the ground users and the IRS, the higher the system sum-rate.

The transformed subproblem 2 in \eqref{Subproblem2Transformed} is still non-convex and we employ an iterative algorithm based on successive convex approximation (SCA) technique to achieve a suboptimal solution.
In particular, given a feasible solution $\left({\bf{v}}_{k,l_{\mathrm{II}}}^{{\rm{UG}}}\left[ n \right],{\bf{v}}_{l_{\mathrm{II}}}^{{\rm{UR}}}\left[ n \right]\right)$ in the $l_{\mathrm{II}}$-th iteration, we have
\vspace{-2mm}
\begin{align}\label{Subproblem2TransformedII}
\hspace{-10mm}&\underset{{\bf{q}}\left[ n \right],{\bf{v}}_k^{{\rm{UG}}}\left[ n \right],{\bf{v}}^{{\rm{UR}}}\left[ n \right]}{\maxo}\,\, \frac{1}{N}\sum\nolimits_{n = 1}^N {\sum\nolimits_{k = 1}^K {R_{k,\mathrm{LoS}}^{{\rm{LB}},{l_{\mathrm{II}}}}} \left[ n \right]\left({\bf{v}}_k^{{\rm{UG}}}\left[ n \right],{\bf{v}}^{{\rm{UR}}}\left[ n \right] \left|{{\bf{T}}^{\mathrm{iter}}, {\bf{U}}^{\mathrm{iter}},\tilde{\bf{P}}^{\mathrm{iter}},{\bf{S}}}^{\mathrm{iter}}\right. \right)} \\[-1mm]
\hspace{-10mm}\notag\mbox{s.t.}\;\;
%%%%%
& \mbox{\underline{C7}}:\; \frac{1}{N}\sum\nolimits_{n = 1}^N {{R_{k,\mathrm{LoS}}^{{\rm{LB}},l_{\mathrm{II}}}} \left[ n \right]\left({\bf{v}}_k^{{\rm{UG}}}\left[ n \right],{\bf{v}}^{{\rm{UR}}}\left[ n \right] \left|{{\bf{T}}^{\mathrm{iter}},{\bf{U}}^{\mathrm{iter}},\tilde{\bf{P}}^{\mathrm{iter}},{\bf{S}}}^{\mathrm{iter}}\right. \right)} \ge R_{\mathrm{min},k}, \forall k, \notag\\[-1mm]
&\mbox{{C8-C10}}, \mbox{C15}, \mbox{C16},\notag
\end{align}
\vspace{-10mm}\par\noindent
where ${R_{k,\mathrm{LoS}}^{{\rm{LB}},l_{\mathrm{II}}}} \left[ n \right]$ denotes a lower bound of ${R_{k,\mathrm{LoS}}^{{\rm{LB}}}} \left[ n \right]$ given a feasible solution $\left({\bf{v}}_{k,l_{\mathrm{II}}}^{{\rm{UG}}}\left[ n \right],{\bf{v}}_{l_{\mathrm{II}}}^{{\rm{UR}}}\left[ n \right]\right)$ in the $l_{\mathrm{II}}$-th iteration, i.e., ${R_{k,\mathrm{LoS}}^{{\rm{LB}},l_{\mathrm{II}}}} \left[ n \right] \le {R_{k,\mathrm{LoS}}^{{\rm{LB}}}} \left[ n \right]$.
The lower bound function is obtained by computing the first order Taylor expansion at $\left({\bf{v}}_{k,l_{\mathrm{II}}}^{{\rm{UG}}}\left[ n \right],{\bf{v}}_{l_{\mathrm{II}}}^{{\rm{UR}}}\left[ n \right]\right)$, i.e.,
\vspace{-2mm}
\begin{align}
&{{R_{k,\mathrm{LoS}}^{{\rm{LB}},l_{\mathrm{II}}}} \hspace{-0.5mm}\left[ n \right]\hspace{-0.5mm}\left(\hspace{-0.5mm}{\bf{v}}_k^{{\rm{UG}}}\hspace{-0.5mm}\left[ n \right]\hspace{-0.5mm},{\bf{v}}^{{\rm{UR}}}\hspace{-0.5mm}\left[ n \right]\hspace{-0.5mm} \left|{{\bf{T}}^{\mathrm{iter}},\hspace{-0.2mm}{\bf{U}}^{\mathrm{iter}},\hspace{-0.2mm}\tilde{\bf{P}}^{\mathrm{iter}},\hspace{-0.2mm}{\bf{S}}}^{\mathrm{iter}}\right. \hspace{-0.5mm}\right)} \hspace{-1mm}=\hspace{-1mm}{{R_{k,\mathrm{LoS}}^{{\rm{LB}}}} \hspace{-0.5mm}\left[ n \right]\hspace{-0.5mm}\left(\hspace{-0.5mm}{\bf{v}}_{k,l_{\mathrm{II}}}^{{\rm{UG}}}\hspace{-0.5mm}\left[ n \right]\hspace{-0.5mm},{\bf{v}}_{l_{\mathrm{II}}}^{{\rm{UR}}}\hspace{-0.5mm}\left[ n \right]\hspace{-0.5mm} \left|{{\bf{T}}^{\mathrm{iter}},\hspace{-0.2mm}{\bf{U}}^{\mathrm{iter}},\hspace{-0.2mm}\tilde{\bf{P}}^{\mathrm{iter}},\hspace{-0.2mm}{\bf{S}}}^{\mathrm{iter}}\right. \hspace{-0.5mm}\right)}  \notag\\[-1mm]
&+\sum\limits_{i = 1}^{{N_{\rm{F}}}}\sum_{k'=1}^{K} \frac{{{t^{\mathrm{iter}}_{k,k',i}}\left[ n \right] }}{{\ln 2}}\frac{{ - \frac{\tilde{p}^{\mathrm{iter}}_{k,k',i} \left[ n \right]}{t^{\mathrm{iter}}_{k,k',i} \left[ n \right]\sigma^2}\left(\frac{{A_{k}}}{{{{\left( {v_{k,{l_{\mathrm{II}}}}^{{\rm{UG}}}\left[ n \right]} \right)}^2}}} + \frac{{{C_{k,k',i}}}}{{{{\left( {v_{k,{l_{\mathrm{II}}}}^{{\rm{UG}}}\left[ n \right]} \right)}^{\frac{3}{2}}}\sqrt{v_{{l_{\mathrm{II}}}}^{{\rm{UR}}}\left[ n \right]}}}\right)}}{ 1 + \frac{\tilde{p}^{\mathrm{iter}}_{k,k',i} \left[ n \right]}{t^{\mathrm{iter}}_{k,k',i} \left[ n \right]\sigma^2}\left({\frac{A_{k}}{{{{ {v_{k,l_{\mathrm{II}}}^{{\rm{UG}}}\left[ n \right]} }}}} + \frac{B_{k,k'}}{{{{ {{v_{l_{\mathrm{II}}}^{{\rm{UR}}}}\left[ n \right]} }}}} + \frac{C_{k,k',i}}{{\sqrt{v_{k,l_{\mathrm{II}}}^{{\rm{UG}}}\left[ n \right]}\sqrt{{v_{l_{\mathrm{II}}}^{{\rm{UR}}}}\left[ n \right]}}}}\right) } \left[ {v_{k}^{{\rm{UG}}}\left[ n \right] - v_{k,{l_{\mathrm{II}}}}^{{\rm{UG}}}\left[ n \right]} \right] \notag\\[-1mm]
&+ \sum\limits_{i = 1}^{{N_{\rm{F}}}}\sum_{k'=1}^{K} \frac{{{t^{\mathrm{iter}}_{k,k',i}}\left[ n \right]}}{{\ln 2}}\frac{{ - \frac{\tilde{p}^{\mathrm{iter}}_{k,k',i} \left[ n \right]}{t^{\mathrm{iter}}_{k,k',i} \left[ n \right]\sigma^2}\left(\frac{{B_{k,k'}}}{{{{\left( {v_{l_{\mathrm{II}}}^{{\rm{UR}}}\left[ n \right]} \right)}^2}}} + \frac{{{C_{k,k',i}}}}{{{{\left( {v_{{l_{\mathrm{II}}}}^{{\rm{UR}}}\left[ n \right]} \right)}^{\frac{3}{2}}}\sqrt{v_{k,{l_{\mathrm{II}}}}^{{\rm{UG}}}\left[ n \right]}}}\right)}}{ 1 + \frac{\tilde{p}^{\mathrm{iter}}_{k,k',i} \left[ n \right]}{t^{\mathrm{iter}}_{k,k',i} \left[ n \right]\sigma^2}\left({\frac{A_{k}}{{{{ {v_{k,l_{\mathrm{II}}}^{{\rm{UG}}}\left[ n \right]} }}}} + \frac{B_{k,k'}}{{{{ {{v_{l_{\mathrm{II}}}^{{\rm{UR}}}}\left[ n \right]} }}}} + \frac{C_{k,k',i}}{{\sqrt{v_{k,l_{\mathrm{II}}}^{{\rm{UG}}}\left[ n \right]}\sqrt{{v_{l_{\mathrm{II}}}^{{\rm{UR}}}}\left[ n \right]}}}}\right) } \left[ {{v^{{\rm{UR}}}}\left[ n \right] - v_{{l_{\mathrm{II}}}}^{{\rm{UR}}}\left[ n \right]} \right].
\end{align}
\vspace{-5mm}\par\noindent

\begin{table}
	\vspace{-10mm}
	\begin{algorithm} [H]                    % enter the algorithm environment
		\caption{Proposed joint trajectory, IRS scheduling, and resource allocation algorithm}
		\label{alg2}                             % and a label for \ref{} commands later in the document
		\begin{algorithmic} [1]
			\footnotesize          % enter the algorithmic environment
			\STATE \textbf{Initialization}\\
			Initialize the convergence tolerance $\epsilon$, the iteration index $\mathrm{iter} = 1$, the maximum number of iterations $\mathrm{iter}_\mathrm{max}$, and the trajectory of UAV ${\bf{q}}^{\mathrm{iter}}\left[ n \right]$ according to Fig. \ref{GeometrySetup}.
			\REPEAT
			\STATE
			Solve the problem in \eqref{ProblemFormulationApproxSubproblem1Dual} via the proposed dual decomposition method. Output the IRS scheduling and resource allocation strategy $\left({{\bf{T}}^{\mathrm{iter}}, {\bf{U}}^{\mathrm{iter}},\tilde{\bf{P}}^{\mathrm{iter}},{\bf{S}}}^{\mathrm{iter}}\right)$ and the corresponding average system sum-rate $R^{\mathrm{LB}}_{\mathrm{sum},\mathrm{LoS}}\left(2\times\mathrm{iter}-1\right) = \frac{1}{N} \sum\nolimits_{n=1}^{N} R^{\mathrm{LB}}_{\mathrm{sum},\mathrm{LoS}}\left[ n \right] \left( {{\bf{T}}^{\mathrm{iter}}, {\bf{U}}^{\mathrm{iter}},\tilde{\bf{P}}^{\mathrm{iter}},{\bf{S}}}^{\mathrm{iter}}\left|{\bf{q}}^{\mathrm{iter}}\left[ n \right]\right. \right)$.
			
			\STATE
			Solve the problem in \eqref{Subproblem2TransformedII} iteratively based on SCA with the initialized trajectory as ${\bf{q}}^{\mathrm{iter}}\left[ n \right]$. Output the UAV trajectory ${\bf{q}}^{\mathrm{iter}+1}\left[ n \right]$ and the corresponding average system sum-rate $R^{\mathrm{LB}}_{\mathrm{sum},\mathrm{LoS}}\left(2\times \mathrm{iter}\right) = \frac{1}{N} \sum\nolimits_{n=1}^{N} R^{\mathrm{LB}}_{\mathrm{sum},\mathrm{LoS}}\left[ n \right] \left({\bf{q}}^{\mathrm{iter}+1}\left[ n \right] \left|{{\bf{T}}^{\mathrm{iter}}, {\bf{U}}^{\mathrm{iter}},\tilde{\bf{P}}^{\mathrm{iter}},{\bf{S}}}^{\mathrm{iter}}\right. \right)$.
			
			\STATE $\mathrm{iter} = \mathrm{iter} + 1$
			\UNTIL $\mathrm{iter} = \mathrm{iter}_\mathrm{max}$ or $\frac{\left| {R^{\mathrm{LB}}_{\mathrm{sum},\mathrm{LoS}}\left(2\times \mathrm{iter}\right) - R^{\mathrm{LB}}_{\mathrm{sum},\mathrm{LoS}}\left(2\times (\mathrm{iter} - 1)\right)} \right|}{R^{\mathrm{LB}}_{\mathrm{sum},\mathrm{LoS}}\left(2\times (\mathrm{iter} - 1)\right)} \le \epsilon$
		\end{algorithmic}
	\end{algorithm}
	\vspace{-17mm}
\end{table}

The problem in \eqref{Subproblem2TransformedII} is a convex optimization problem and solving \eqref{Subproblem2TransformedII} provides a lower bound for subproblem 2 in \eqref{Subproblem2Transformed}.
To tighten the obtained lower bound, we iteratively update $\left({\bf{v}}_{k}^{{\rm{UG}}}\left[ n \right],{\bf{v}}^{{\rm{UR}}}\left[ n \right]\right)$ which generates a sequence of feasible solution converging to a stationary point of the problem in \eqref{Subproblem2Transformed}, cf. \cite{wei2018multibeam}.
In particular, given $\left({\bf{v}}_{k,l_{\mathrm{II}}}^{{\rm{UG}}}\left[ n \right],{\bf{v}}_{l_{\mathrm{II}}}^{{\rm{UR}}}\left[ n \right]\right)$ in the $l_{\mathrm{II}}$-th iteration, solving the problem in \eqref{Subproblem2TransformedII} generates a feasible solution for the next iteration $\left({\bf{v}}_{k,l_{\mathrm{II}}+1}^{{\rm{UG}}}\left[ n \right],{\bf{v}}_{l_{\mathrm{II}}+1}^{{\rm{UR}}}\left[ n \right]\right)$.
Such an iterative procedure will stop when the maximum iteration number is reached or the improvement of the system sum-rate is smaller than a predefined convergence tolerance.

Now, the overall algorithm for joint trajectory, IRS scheduling, and resource allocation design can be obtained via solving the subproblems 1 and 2 alternatingly.
Due to the page limitation, a description of the overall algorithm is summarized in \textbf{Algorithm} \ref{alg2}.
The overall algorithm is initialized with a feasible trajectory of UAV as shown in Fig. \ref{GeometrySetup} in Section \ref{Simulation} and terminates when the maximum iteration number is reached or the system sum-rate improvement is less than a predefined threshold.

\vspace{-4mm}
\section{Simulation Results}\label{Simulation}
{In this section, we evaluate the performance of the proposed scheme via simulations.}
\vspace{-4mm}
\subsection{Simulation Setup and Baselines}
In this section, we evaluate the performance of the proposed scheme via simulations.
The simulation setups are summarized in Table \ref{simulation_setting}.
Note that the size of each PRUs along the row and column dimensions are set as $d_{\mathrm{r}} = d_{\mathrm{c}} = \frac{c}{10 f_\mathrm{c}}$, respectively \cite{di2019smart,tang2019wireless}.
The selection of $M_\mathrm{r}$ and $M_\mathrm{c}$ in Table \ref{simulation_setting} result in an IRS area ranging from $1 \sim 25 \; \mathrm{m}^2$.
We note that we consider $K = 3$ for simplicity and it is sufficient to present the benefits of deploying an IRS in a UAV-based OFDMA communication system.
The system layout and the locations of ground users as well as the IRS are illustrated in Fig. \ref{GeometrySetup}.
To demonstrate the performance gain brought by the high flexibility of UAV in trajectory design, we consider baseline scheme 1 with a straight line trajectory, as shown in Fig. \ref{GeometrySetup}, but the UAV is assisted by the IRS.
The system sum-rate can be obtained via solving subproblem 1 with the given trajectory between the initial and end points.
To illustrate the contribution of IRS, we compare our proposed scheme with the UAV OFDMA communication system without the assistance of the IRS, which is referred as baseline scheme 2 in this paper.
The system sum-rate can be obtained via executing the developed algorithms by setting $s_k\left[n\right] = 0$, $\forall k,n$.
{In the following, to unveil the insights of deploying IRS in UAV communication systems, we first show the average system sum-rate in the absence of the scattering components from Fig. \ref{GeometrySetup} to Fig. \ref{SumRateVersusPower}.
Then, in Section \ref{RicianFading}, we extend the proposed design to Rician fading channels and evaluate the system outage rate in Fig. \ref{SumRateVersusRician}.}

\begin{table}[t]
	\vspace{-7mm}
	\scriptsize
	\caption{Simulation parameters \cite{zeng2019accessing,qingqing2019towards}.} \label{simulation_setting}
	\vspace{-8mm}
	\begin{center}
		\begin{tabular}{ c | c |c |c | c | c | c| c}
			\hline			
			Notations         & Simulation value & Notations          & Simulation value      & Notations         & Simulation value  &   Notations          & Simulation value \\ \hline
			$K$               &      $[3,8]$           &   $N_{\mathrm{F}}$  &       $1000$           &     $R_{\mathrm{min},k}$  &   $0.3 \sim 1$ bit/s/Hz     &    $M_{\mathrm{r}}$  &       $100 \sim 500$   \\
			$\delta_{\mathrm{t}}$        & $0.1$ s              &   $B$               &       $100$ MHz        &$V_{\mathrm{max}}$ & $20$ m/s            &    $M_{\mathrm{c}}$  &       $100 \sim 500$   \\
			$H_{\mathrm{U}}$  &      $100$ m       &   $\Delta f$        &  $100$ kHz             & $\beta_0$         &  $-50$ dBW           &   $N$               &      $400\sim 800$      \\
			$H_{\mathrm{R}}$  &      $30$ m        &   $f_{\mathrm{c}}$  &  $3$ GHz               & $a$               &  $0.9 $             &   $p_{\mathrm{max}}$ &  $30 \sim 45$ dBm \\
			${\bf{q}}\left[ 1 \right]$ & $[0,0,H_{\mathrm{U}}]^{\mathrm{T}}$ m        &   $c$               &   $3 \times 10^8$ m/s& $\alpha_{\mathrm{UG}}$              &  $ 2.5 $              &   $\kappa^{\mathrm{UG}}_k$  & $2\sim 14$ dB \\
			${\bf{q}}\left[ N \right]$ & $[500,500,H_{\mathrm{U}}]^{\mathrm{T}}$ m    &   $N_0$        &  $-169$ dBm/Hz    & $\alpha_{\mathrm{RG}}$              &   $2.5 $              &   $\kappa^{\mathrm{RG}}_k$  & $2\sim 14$ dB \\\hline
		\end{tabular}
	\end{center}
	\vspace{-10mm}
\end{table}

%Baseline 3: TDMA System:
%As IRS introduce an additional path in UAV communication systems, OFDMA is potential to improve the system sum-rate as it can exploit the frequency-selective fading for resource allocation.
%
%To reveal the performance gain brought by OFDMA compared to TDMA system, we compared the proposed scheme with the system with TDMA communications.
%
%The system performance is obtained via limiting that $u_{k,i}\left[n\right] = u_{k,j}\left[n\right]$, $\forall i, j$.
\vspace{-4mm}
\subsection{Frequency Selective Fading}

\begin{figure}[t]
	\begin{minipage}{.47\textwidth}
		\centering
		\includegraphics[width=\textwidth]{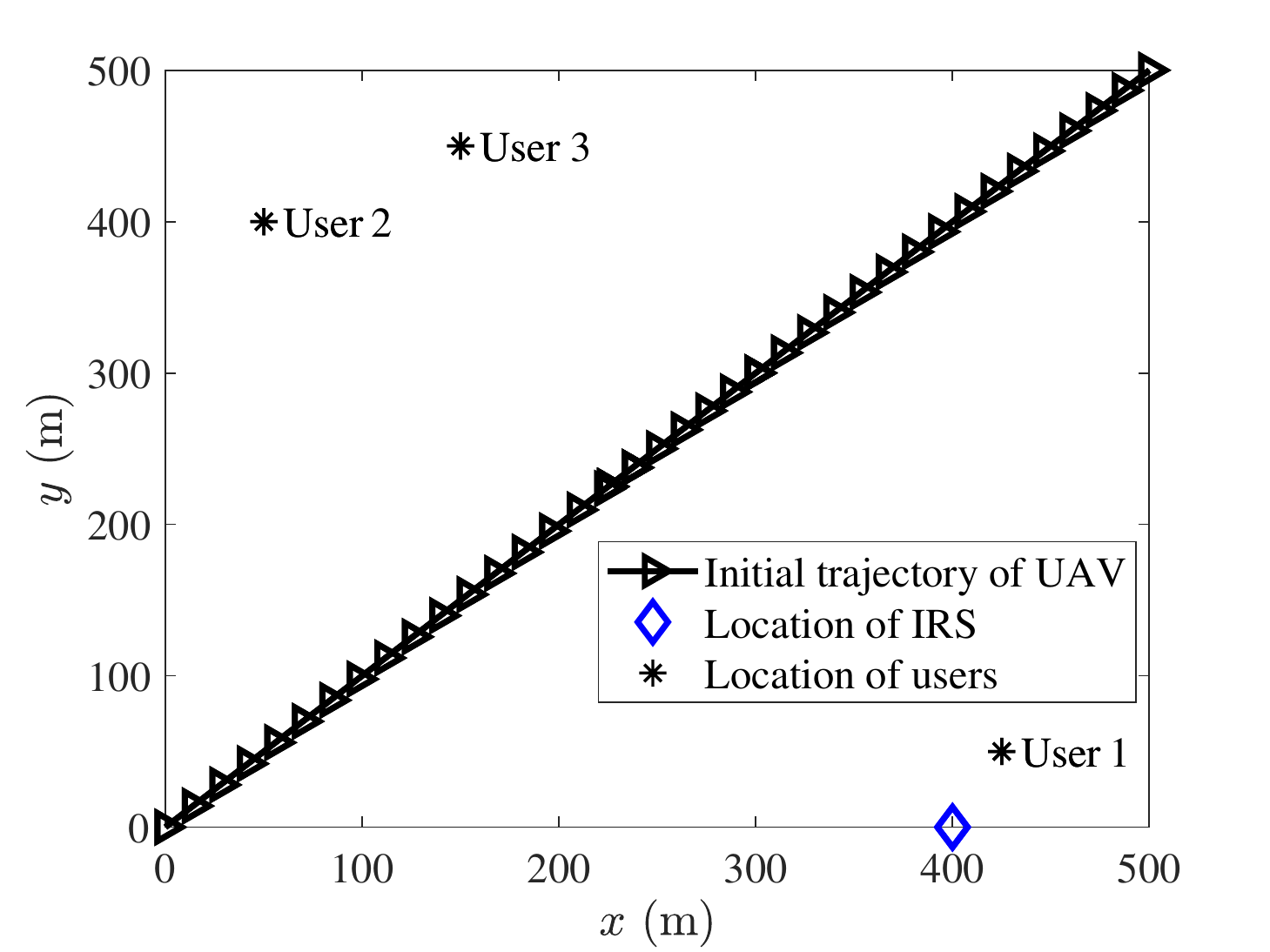}\vspace{-8mm}
		\caption{Geometry setup for the considered IRS-assisted UAV OFDMA communication system.}\vspace{-5mm}
		\label{GeometrySetup}
	\end{minipage}
	\hspace*{1.5mm}
	\begin{minipage}{.47\textwidth}
		\centering
		\includegraphics[width=\textwidth]{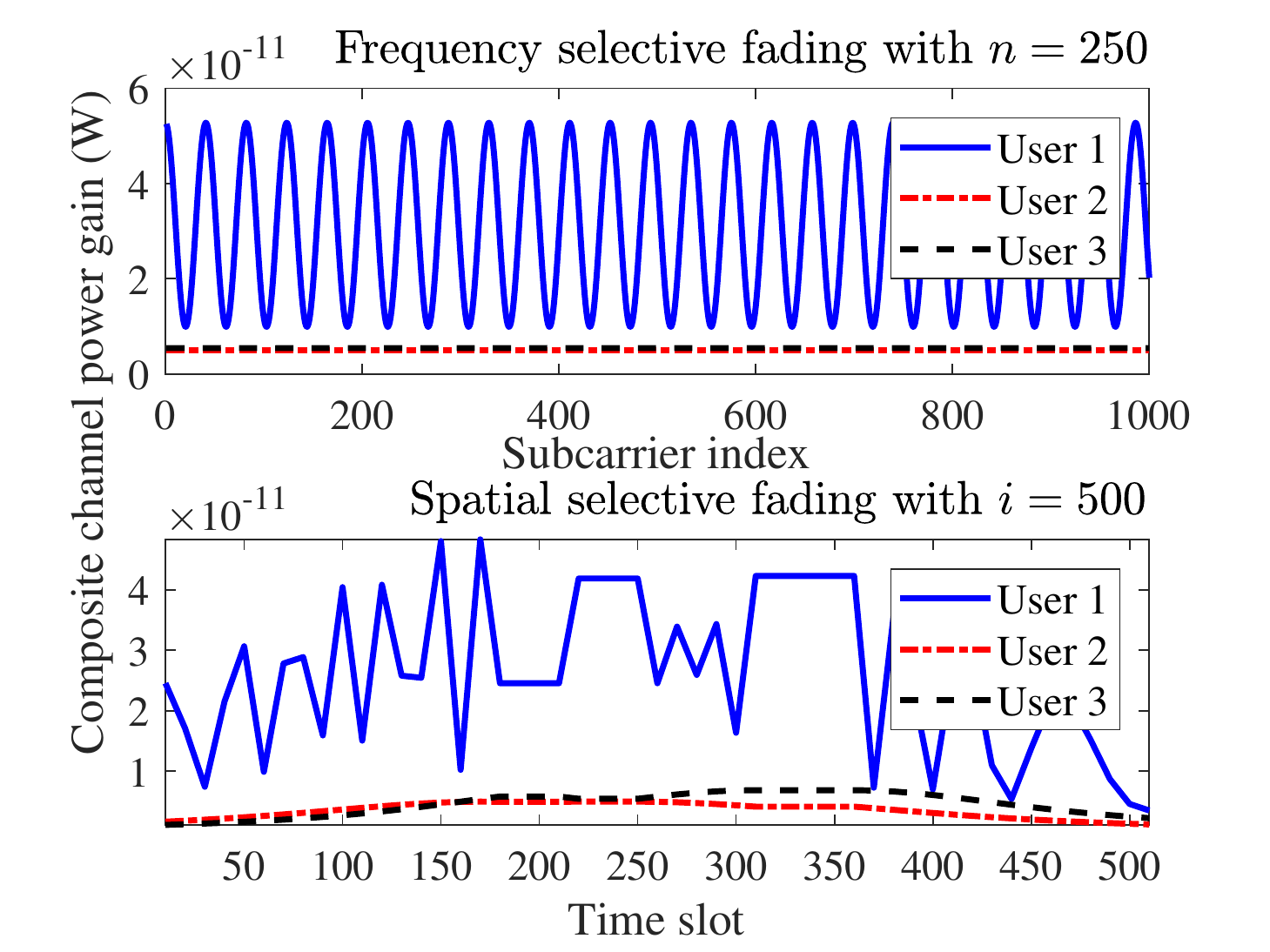}\vspace{-8mm}
		\caption{Channel power gain of IRS-assisted and non-IRS-assisted users versus the subcarrier index.}\vspace{-5mm}
		\label{FrequencySelectivePattern}%
	\end{minipage}
\end{figure}

We first visualize the cosine pattern in the frequency-selective composite channels induced by the introduction of an IRS, i.e., \eqref{FinalChannelApproxII}, which is exploited to serve as a building block for deriving the parametric bounds for the formulated problem.
Fig. \ref{FrequencySelectivePattern} illustrates the snapshots of the composite channel gains for both the IRS-assisted and non-IRS-assisted users with the straight line trajectory of UAV in Fig. \ref{GeometrySetup}.
We assume that user 1 is selected to be assisted by the IRS during the whole flight period, i.e., $s_1\left[n\right] = 1$, $\forall n$.
In the upper figure of Fig. \ref{FrequencySelectivePattern}, we can observe that the channel fading of the IRS-assisted user 1 in time slot $n = 250$ is frequency-selective as predicted in \eqref{FinalChannelApproxII}.
Compared to the non-IRS assisted users 2 and 3, the composite channel power gain of user 1 is significantly increased due to the substantial gain introduced by the passive beamforming performed by the IRS.
Furthermore, we can observe that the composite channel power gains for the non-IRS-assisted users are almost frequency-flat, compared to that of the IRS-assisted user 1.
In fact, due to the limited signal leakage from the IRS to non-IRS-assisted users, the composite channel gain of the non-IRS-assisted user is dominated by its DC component in \eqref{FinalChannelDC}.
On the other hand, in the lower figure of Fig. \ref{FrequencySelectivePattern}, the composite channel power gain on each subcarrier, e.g. subcarrier $i = 500$, also exhibits a spatial fluctuation w.r.t. the time slot due to the induced additional path reflected via the IRS, which complicates the UAV's trajectory design.

\vspace{-4mm}
\subsection{Parametric Bounds and The Optimal Approximation Parameter}

\begin{figure}[t]
	\begin{minipage}{.47\textwidth}
		\centering
		\includegraphics[width=\textwidth]{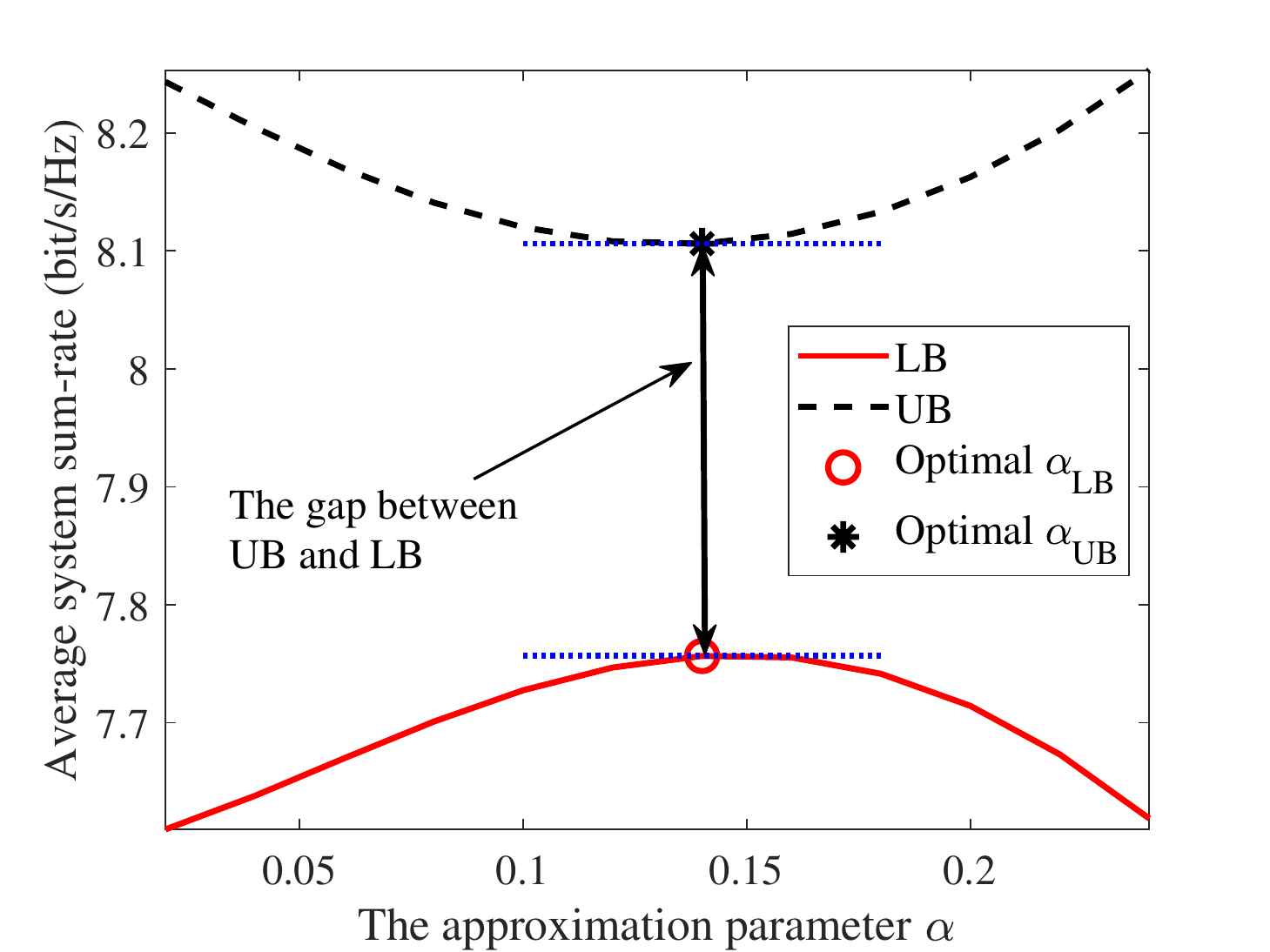}\vspace{-8mm}
		\caption{Average system sum-rate (bit/s/Hz) versus the approximation parameter $\alpha$.}\vspace{-10mm}
		\label{LBUBAlpha}%
	\end{minipage}
	\hspace*{1.5mm}
	\begin{minipage}{.47\textwidth}
		\centering
		\includegraphics[width=\textwidth]{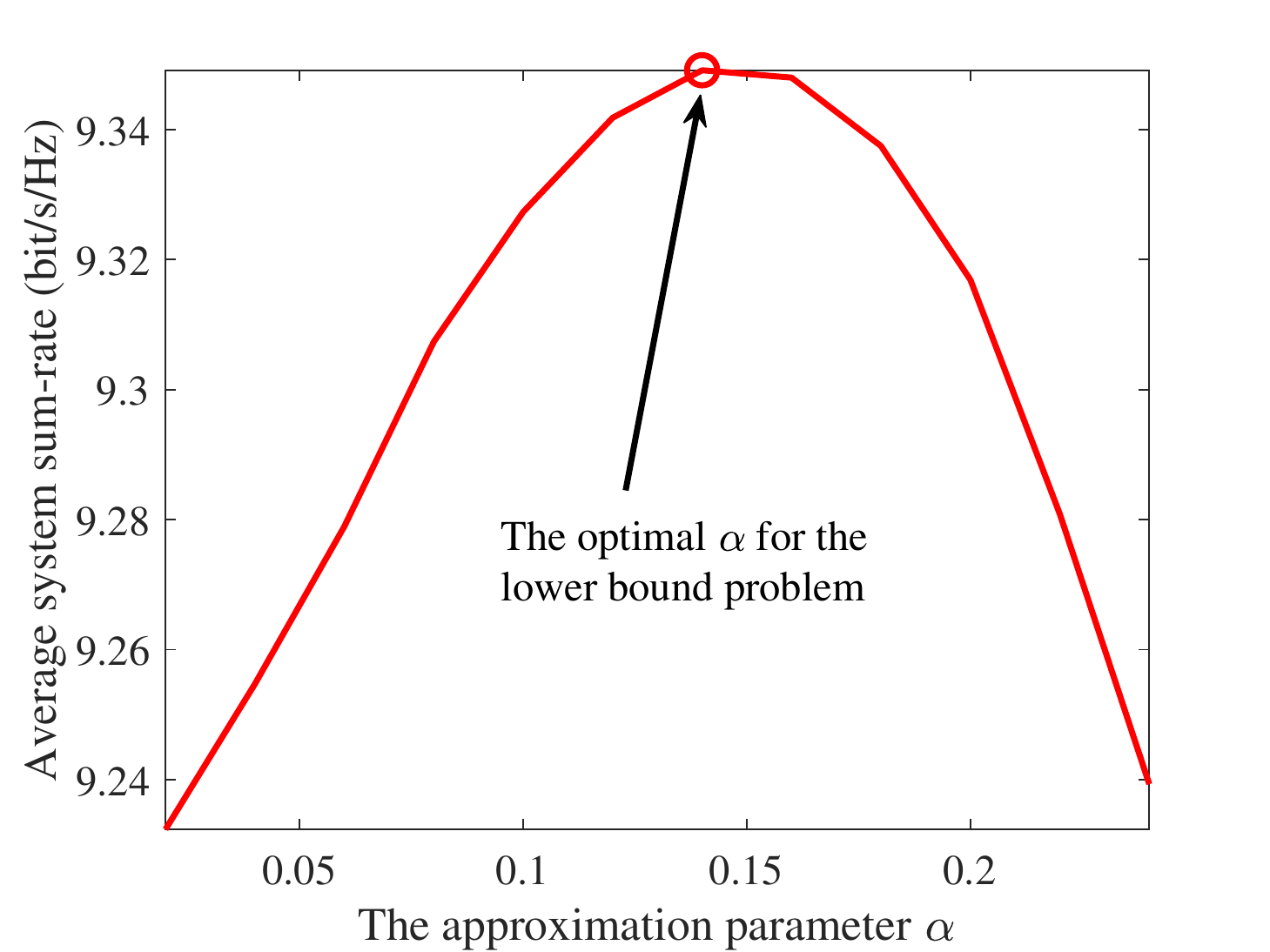}\vspace{-8mm}
		\caption{The optimal approximation parameter for the lower bound problem.}\vspace{-10mm}
		\label{OptimalALPHA}%
	\end{minipage}
\end{figure}

To demonstrate the effectiveness of the proposed parametric bounds in \eqref{ProblemFormulationLB} and \eqref{ProblemFormulationUB}, we consider a simple scenario with only a single intermediate location of UAV, i.e., $N = 3$, and adopt the exhaustive search to find the optimal intermediate location.
Fig. \ref{LBUBAlpha} illustrates the upper bound (UB) and lower bound (LB) performance of the formulated problem versus the approximation parameter $0 < \alpha < 0.25$ with $p_{\mathrm{max}} = 35$ dBm, $R_{\mathrm{min},k} = 0.5$ bit/s/Hz, and $M_{\mathrm{r}} = M_{\mathrm{c}} = 200$.
We can observe that the optimal approximation parameter is the same at $\alpha = 0.14$ for the proposed upper bound and lower bound problems.
Furthermore, the gap between the upper bound and lower bound problems at the optimal $\alpha$ is only 0.35 bit/s/Hz, which is approximately $4.5\%$ of the optimal lower bound performance at $\alpha = 0.14$.
It implies the effectiveness of the proposed parametric bounds and the performance loss caused by solving the lower bound problem is marginal.
Now, we conduct a one-dimensional search to find the optimal approximation parameter for the lower bound problem in our considered practical cases.
Fig. \ref{OptimalALPHA} shows the average system sum-rate achieved by our proposed scheme versus the approximation parameter $\alpha$ with $N = 500$ and $R_{\mathrm{min},k} = 1$ bit/s/Hz. All the other parameters are the same as Fig. \ref{LBUBAlpha}.
It can be observed that the average system sum-rate first increases and then decreases with increasing $\alpha$.
This is because a too small or too large $\alpha$ both yield a loose lower bound for the cosine fading pattern as shown in Fig. \ref{CosineApproximationForIRSAssistedUser}.
Additionally, we can observe that the best approximation parameter is also $\alpha = 0.14$ in this setup.
In the following simulations, we set $\alpha = 0.14$ for simplicity.

\vspace{-4mm}
\subsection{The Impact of IRS on UAV's Trajectory Design}
\begin{figure}[t]
	\centering\vspace{-7mm}
	\includegraphics[width=3.75in]{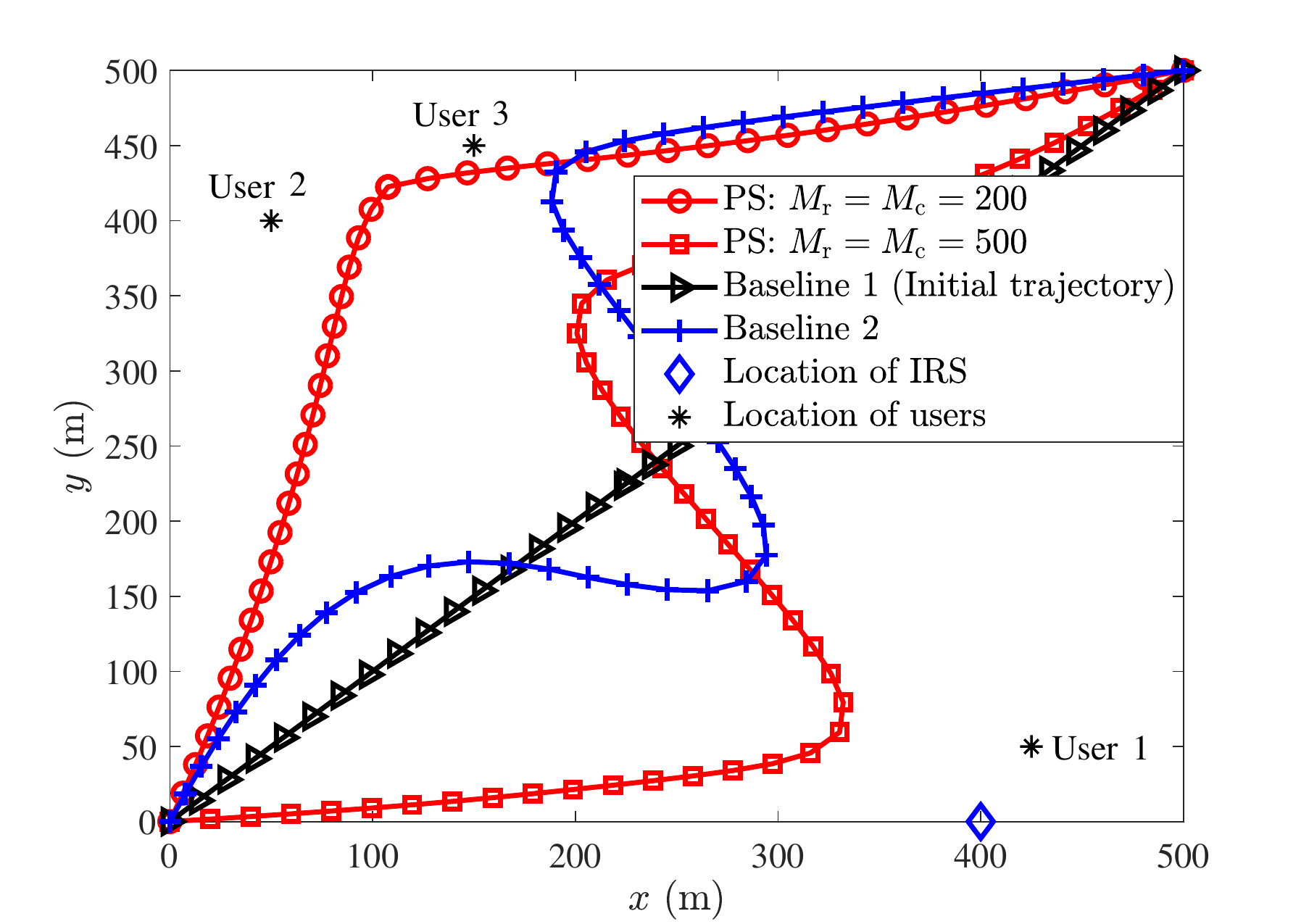}\vspace{-8mm}
	\caption{Trajectory of UAV of the proposed scheme.}\vspace{-10mm}
	\label{TrajectoryUAVFinal}%
\end{figure}

Fig. \ref{TrajectoryUAVFinal} compares the obtained trajectories of UAV for the proposed scheme (PS) and baseline schemes to demonstrate the impact of IRS on UAV's trajectory design with $p_{\mathrm{max}} = 35$ dBm, $R_{\mathrm{min},k} = 1$ bit/s/Hz, and $N = 500$.
For the proposed scheme, two simulation cases with $M_{\mathrm{r}} = M_{\mathrm{c}} = 200$ and $M_{\mathrm{r}} = M_{\mathrm{c}} = 500$, respectively, are conducted.
{We found that the UAV keeps flying at the minimum altitude, i.e., $z\left[n\right] = H^{\mathrm{min}}_{\mathrm{U}}$, $\forall n$.
In fact, in our considered scenario, flying higher results in a larger path loss between the UAV and ground users.}
For baseline 2, the UAV tries to approach each of all the three users in its route to establish strong communication links such that the ground users' minimum data rate requirements can be satisfied.
In contrast, when equipping an IRS with $M_{\mathrm{r}} = M_{\mathrm{c}} = 200$, the UAV in the proposed scheme has a higher flexibility in designing its trajectory.
Instead of flying to user 1, the UAV would directly fly towards a centroid formed by user 2 and user 3 for maximizing the system sum-rate.
This is because the IRS located near user 1 can collect the dissipated radio power from the UAV and reflect it to user 1 through the proposed phase control for enhancing the composite power gain of user 1.
In other words, the minimum data rate constraint of user 1 can still be satisfied even if the UAV is far away from it.
When $M_{\mathrm{r}} = M_{\mathrm{c}} = 500$, the UAV in our proposed scheme would first detour to the IRS and user 1 at the beginning before flying to users 2 and 3.
In fact, equipping more PRUs allows the IRS reflecting the radiated signal more efficiently and thus approaching the IRS and user 1 becomes more beneficial to the system sum-rate performance.
Therefore, compared to baseline 2, the UAV in our proposed scheme flies towards user 1 earlier so as to achieve a higher system sum-rate.

\vspace{-4mm}
\subsection{Average System Sum-rate in Different System Setups}

\begin{figure}[t]
	\vspace{-8mm}
	\begin{minipage}{.47\textwidth}
		\centering
		\includegraphics[width=\textwidth]{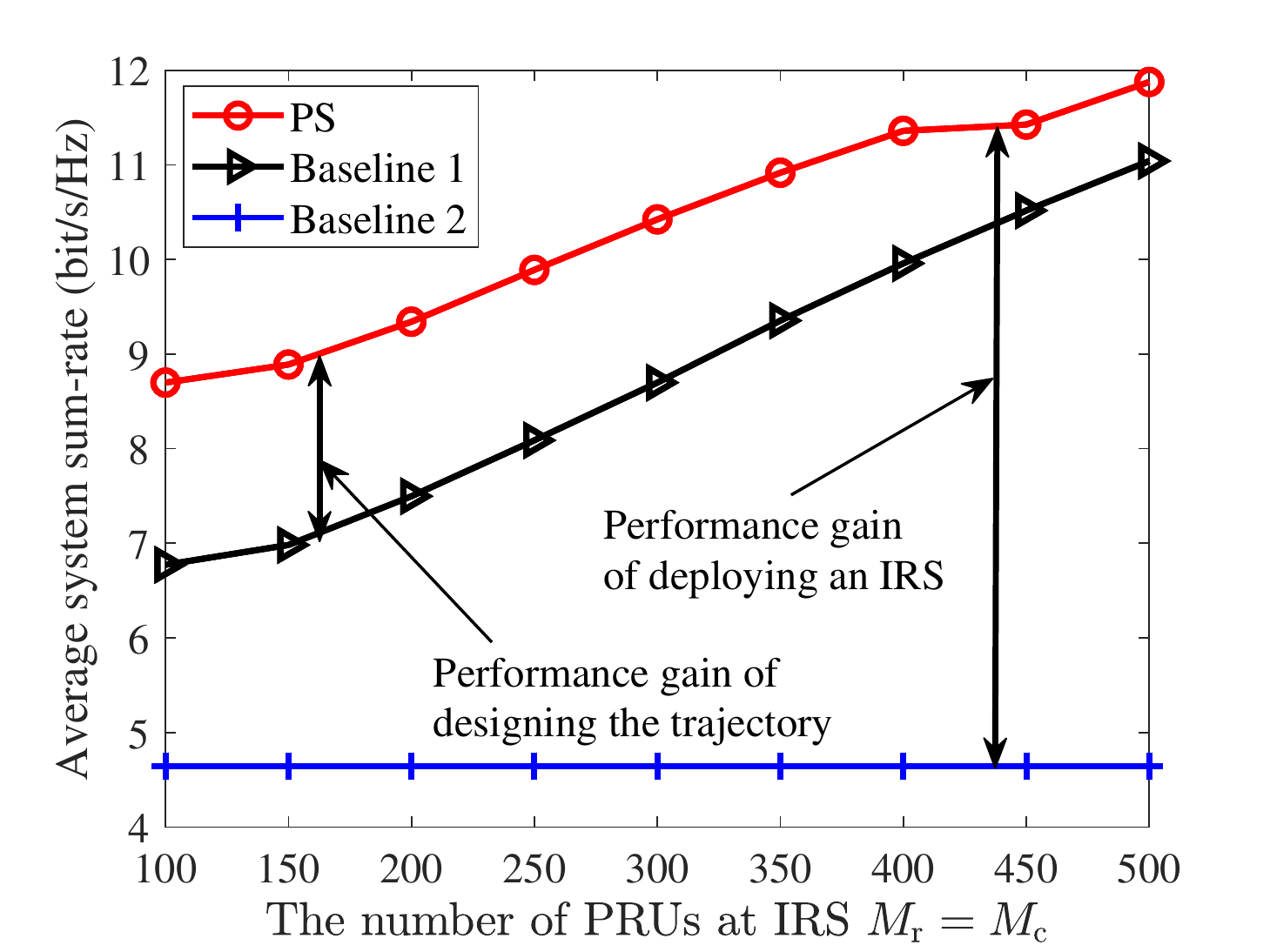}\vspace{-8mm}
		\caption{Average system sum-rate (bit/s/Hz) versus the number of PRUs at IRS.}\vspace{-10mm}
		\label{SumRateVersusIRS}%
	\end{minipage}
	\hspace*{1.5mm}
	\begin{minipage}{.47\textwidth}
		\centering
		\includegraphics[width=\textwidth]{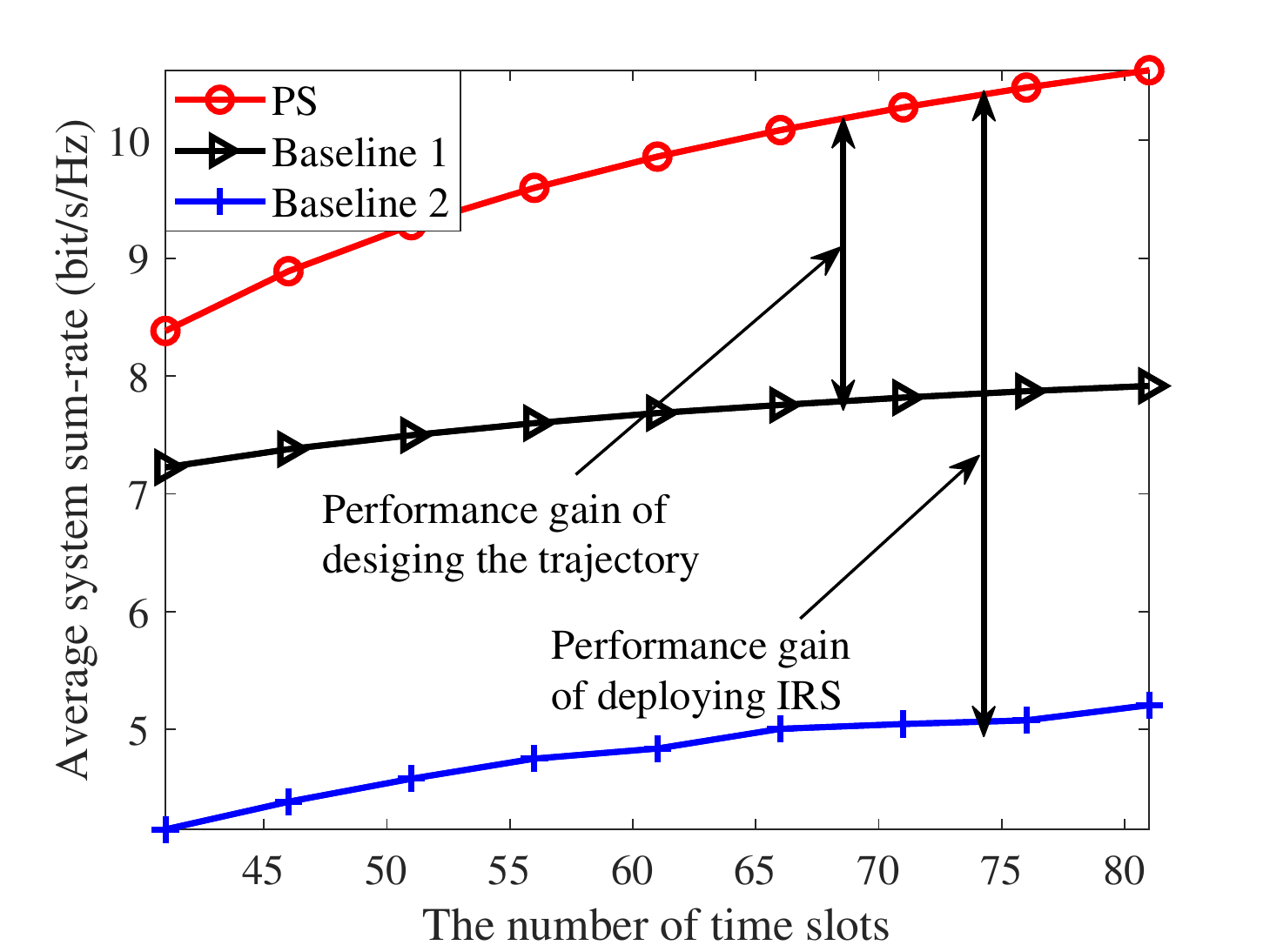}\vspace{-8mm}
		\caption{Average system sum-rate (bit/s/Hz) versus the number of time slots.}\vspace{-10mm}
		\label{SumRateVersusT}%
	\end{minipage}
\end{figure}

Fig. \ref{SumRateVersusIRS} depicts the average system sum-rate versus the number of PRUs at the deployed IRS for the proposed scheme with $p_{\mathrm{max}} = 35$ dBm, $R_{\mathrm{min},k} = 1$ bit/s/Hz, and $N = 500$.
We can observe that the system sum-rate of both the proposed scheme and baseline 1 increase with the increasing number of PRUs  due to the enhanced passive beamforming gain, which can be achieved by our proposed phase control.
Compared to baseline 1, a considerable sum-rate gain can be achieved by the proposed scheme due to the high flexibility of the UAV in trajectory design, as discussed in Fig. \ref{TrajectoryUAVFinal}.
Furthermore, it can be observed that the performance gain of the proposed scheme over baseline 1 slightly decreases with increasing the number of PRUs.
This is because the IRS's passive beamforming gain is magnified by the increasing $M_{\mathrm{r}}$ and $M_{\mathrm{c}}$.
As a result, the IRS can efficiently assist any user in need and the associated performance gain even dominates the counterpart brought by UAV's trajectory optimization.
In addition, a significant sum-rate gain of the proposed scheme over baseline 2 can be observed due to the energy focusing capability of the deployed IRS.
%
%Furthermore, both performance gains increase with increasing the number of PRUs at IRS, which imply the mutual benefits of UAV's flexibility and IRS's beamforming gain brought by the proposed scheme.

Fig. \ref{SumRateVersusT} illustrates the average system sum-rate versus the number of available time slots $N$ for the proposed scheme with $p_{\mathrm{max}} = 35$ dBm, $R_{\mathrm{min},k} = 1$ bit/s/Hz, and $M_{\mathrm{r}} = M_{\mathrm{c}} = 200$.
We can observe that the system sum-rates for all the three schemes increase with increasing $N$.
In fact, the UAV's trajectory design becomes more flexible with more available time slots.
Furthermore, the sum-rate gain of the proposed scheme over baseline 1 is enlarged for a large number of time slots.
It is due to the fact that a longer flying time duration enables the UAV to efficiently exploit the passive beamforming gain of the deployed IRS via a more flexible trajectory optimization.
We note that even for baseline 1 and baseline 2, increasing the total number of time slots allows the UAV to hover above each user for a longer duration to achieve a higher system sum-rate.

Fig. \ref{SumRateVersusPower} shows the average system sum-rate versus the transmit power $p_{\mathrm{max}}$ with $N = 500$, $R_{\mathrm{min},k} = 0.3$ bit/s/Hz, and $M_{\mathrm{r}} = M_{\mathrm{c}} = 200$.
We can observe that all the system sum-rate of the three schemes improve with increasing the total transmit radiated power from the UAV.
More importantly, the sum-rate gain of the proposed scheme compared to baseline 2 enlarges with increasing $p_{\mathrm{max}}$.
In fact, due to the flexibility in UAV's trajectory design and the passive beamforming gain of the deployed IRS, the proposed scheme can exploit the transmit power more efficiently to improve the system performance.
{Additionally, it can be seen that the performance gain of deploying IRS for $K = 8$ is higher than that of $K = 3$ which confirms the effectiveness of the proposed scheme in handling multiple users.}
{To demonstrate the performance gain of adopting OFDMA, we evaluate the system performance for the case of $K = 8$ using the time-division multiple access (TDMA) scheme while the UAV's trajectory is designed by the proposed algorithm.
We can observe that our proposed scheme adopting OFDMA outperforms the TDMA scheme and the corresponding performance gain increases with the total transmit power.
In fact, OFDMA can exploit the inherent multi-user diversity via flexible subcarrier allocation \cite{DerrickEEOFDMA,Sun2016Fullduplex,WeiTCOM2017}, compared with TDMA.
In particular, the multi-user diversity enables a more efficient utilization of the power budget for improving the system performance.}

\vspace{-4mm}
\subsection{Outages in Rician Fading Channels}\label{RicianFading}
%Due to the local scattering around the ground users, the channels in the IRS-to-user and UAV-to-user links may suffer also from random scattering components, as shown in \eqref{ChannelRG} and \eqref{ChannelUG}.
%
{Recall that we design the phase control, trajectory, and resource allocation policies based on the deterministic LoS components of all involved channels.
In other words, the allocated power and rate are adapted to the deterministic LoS components, instead of the instantaneous channels, where an outage may occur as the existence of scattering components in Rician fading channels introduces randomness to the achievable rate\footnote{Note that practical design taking into account of outage event is an interesting research topic which is left for future work.}.}
Firstly, in Rician fading channel, the achievable data rate of user $k$ on subcarrier $i$ in time slot $n$ might be smaller than its counterpart in LoS channels, i.e., $R_{k,i,\mathrm{Rician}}\left[ n \right] <  R_{k,i,\mathrm{LoS}}\left[ n \right]$, which yields a \textit{subcarrier-level outage}.
Secondly, due to the subcarrier-level outage, the minimum data rate requirement of each user in constraint C7 might not be satisfied, which is named as \textit{user-level outage} in this paper.
Therefore, we can introduce a rate control parameter $0 < \eta < 1$ to extend our design to handle the case in Rician fading channels.
In particular, we increase the minimum data rate requirement of each user by $\tilde{R}_{\mathrm{min},k} = \frac{R_{\mathrm{min},k}}{\eta}$ and adopt $\tilde{R}_{\mathrm{min},k}$ to obtain a conservative solution for the joint trajectory and resource allocation design.
After obtaining the achievable rate $R_{k,i,\mathrm{LoS}}\left[ n \right]$ in LoS channels, we only allocate a rate $\eta R_{k,i,\mathrm{LoS}}\left[ n \right]$ for user $k$ on subcarrier $i$ in time slot $n$ to avoid the possible subcarrier-level outage due to channel randomness.
In particular, in the $l_{\mathrm{mc}}$-th Monte Carlo experiment, the individual outage rate of user $k$ can be defined as
\vspace{-2mm}
\begin{equation}
{R^{\mathrm{outage}}_{\mathrm{k},l_{\mathrm{mc}}}} =  \frac{1}{N} \sum\nolimits_{n = 1}^{N} \sum\nolimits_{i = 1}^{N_F}\eta R_{k,i,\mathrm{LoS}}\left[ n \right] I\left(R^{l_{\mathrm{mc}}}_{k,i,\mathrm{Rician}}\left[ n \right] \ge \eta R_{k,i,\mathrm{LoS}}\left[ n \right]\right),\vspace{-2mm}\label{AchievableOutageInvidual}
\end{equation}
where $I\left(\cdot\right)$ denotes an indication function and $R^{l_{\mathrm{mc}}}_{k,i,\mathrm{Rician}}\left[ n \right]$ denotes the achievable data rate of user $k$ on subcarrier in time slot $n$ in $l_{\mathrm{mc}}$-th Monte Carlo experiment.
From \eqref{AchievableOutageInvidual}, we can observe that only the allocated rate $\eta R_{k,i,\mathrm{LoS}}\left[ n \right]$ smaller than the corresponding achievable rate in Rician fading channels is taken account into the individual outage rate.
To further take into account the user-level outage, we define the average system outage rate as:
\vspace{-2mm}
\begin{equation}
\overline{R^{\mathrm{outage}}_{\mathrm{sum}}} = \frac{1}{L_{\mathrm{mc}}}\sum\nolimits_{l_{\mathrm{mc}}=1}^{L_{\mathrm{mc}}} \sum\nolimits_{k=1}^{K} {R^{\mathrm{outage}}_{\mathrm{k},l_{\mathrm{mc}}}} I\left({R^{\mathrm{outage}}_{\mathrm{k},l_{\mathrm{mc}}}} \ge R_{\mathrm{min},k}\right),\vspace{-2mm}\label{AchievableOutageRateSumUser}
\end{equation}
where $L_{\mathrm{mc}}$ denotes the total number of Monte Carlo experiments.
It can be seen that only the individual outage rate larger than the corresponding required minimum data rate contributes to the average system outage rate.
In the following, we evaluate the average system outage rate of our proposed scheme in Rician fading channels with a fixed rate control parameter $\eta = 0.8$.

%Therefore, we perform simulations to demonstrate the average system outage rate which is defined by
%\begin{equation}
%\overline{R^{\mathrm{outage}}_{\mathrm{sum}}} = \frac{1}{L_{\mathrm{mc}}}\sum_{l_{\mathrm{mc}}=1}^{L_{\mathrm{mc}}} \sum_{k=1}^{K} \frac{1}{N} \sum_{n = 1}^{N} \sum_{i = 1}^{N_F}\eta R_{k,i,\mathrm{LoS}}\left[ n \right] I\left(R^{l_{\mathrm{mc}}}_{k,i,\mathrm{Rician}}\left[ n \right] \ge \eta R_{k,i,\mathrm{LoS}}\left[ n \right]\right),\label{AchievableOutageRateSum}
%\end{equation}
%where $0 < \eta < 1$ denotes the rate control parameter based on the achievable rate in LoS channels $R_{k,i,\mathrm{LoS}}\left[ n \right]$ and $I\left(\cdot\right)$ denotes an indication function.
%
%The larger the $\eta$, the higher the outage probability, and vice versa.
%
%In this simulation, we fix $\eta = 0.9$.
%
%Integer $l_{\mathrm{mc}}$ denotes the index of Monte Carlo experiment, $L_{\mathrm{mc}}$ denotes the total number of Monte Carlo experiments, and $R^{l_{\mathrm{mc}}}_{k,i,\mathrm{Rician}}\left[ n \right]$ denotes the achievable data rate of user $k$ on subcarrier in time slot $n$ in $l_{\mathrm{mc}}$-th Monte Carlo experiment.
%
%From \eqref{AchievableOutageRateSum}, we can observe that only the allocated rate $\eta R_{k,i,\mathrm{LoS}}\left[ n \right]$ smaller than the corresponding achievable rate in Rician fading channels contributes to the system outage rate.

\begin{figure}[t]
	\vspace{-8mm}
	\begin{minipage}{.47\textwidth}
		\centering
		\includegraphics[width=\textwidth]{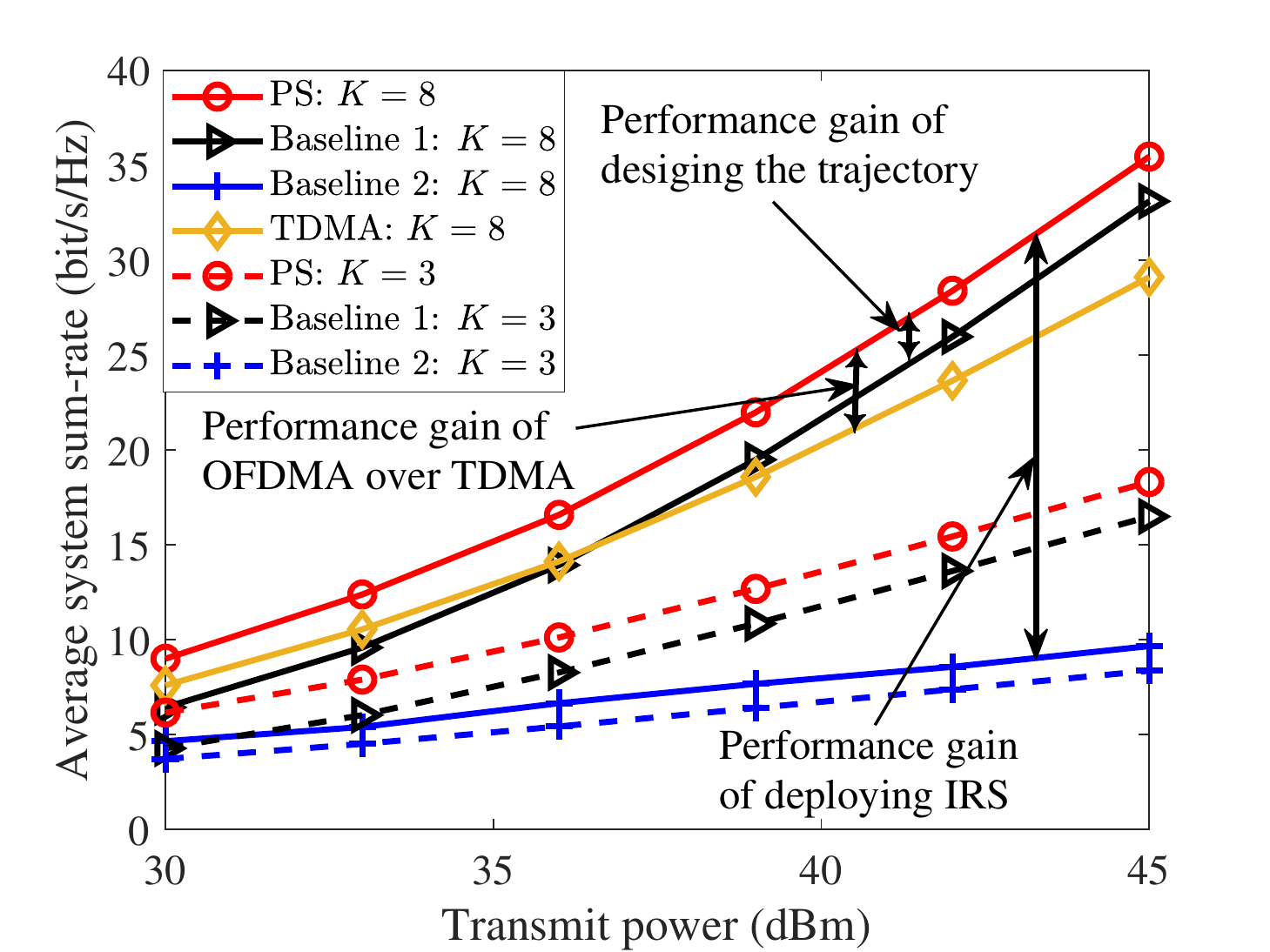}\vspace{-8mm}
		\caption{Average system sum-rate (bit/s/Hz) versus the total transmit power (dBm).}\vspace{-10mm}
		\label{SumRateVersusPower}%
	\end{minipage}
	\hspace*{1.5mm}
	\begin{minipage}{.47\textwidth}
		\centering
		\includegraphics[width=\textwidth]{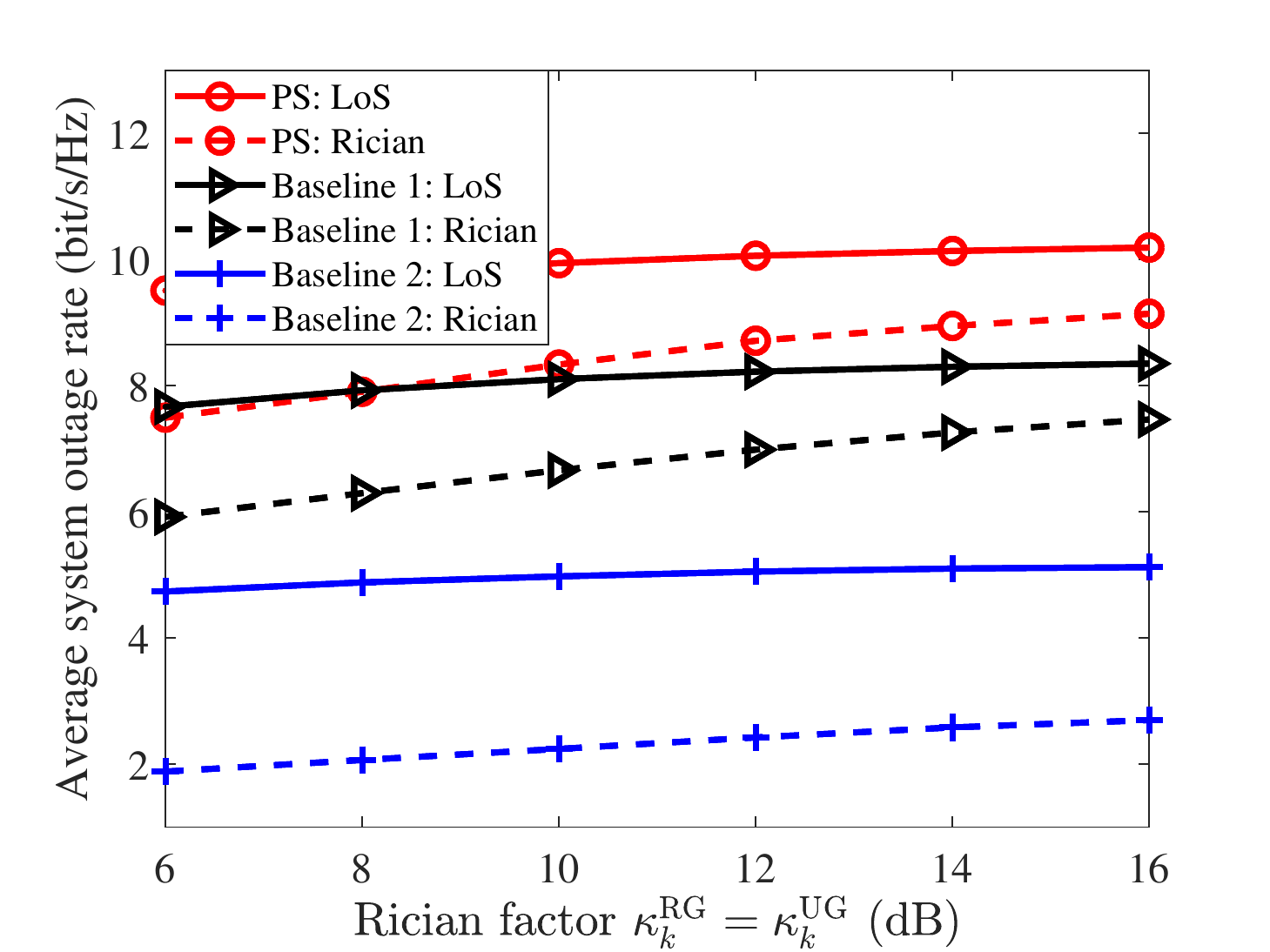}\vspace{-8mm}
		\caption{Average system outage rate (bit/s/Hz) versus the Rician factor (dB).}\vspace{-10mm}
		\label{SumRateVersusRician}%
	\end{minipage}
\end{figure}

Fig. \ref{SumRateVersusRician} illustrates the average system outage rate versus the Rician factor for our proposed scheme with $p_{\mathrm{max}} = 35$ dBm, $M_{\mathrm{r}} = M_{\mathrm{c}} = 200$, $N = 500$, and $R_{\mathrm{min},k} = 0.8$ bit/s/Hz, $\forall k$.
Due to the channel randomness of Rician fading, the average system outage rate is smaller than the system sum-rate in LoS channels for all the three schemes.
Furthermore, we can observe that the larger the Rician factor, the higher the average system outage rate.
This is because with a higher Rician factor, the proposed design based on the LoS can closely approximate the one based on Rician fading channels.
More importantly, the performance loss due to the channel randomness for the proposed scheme and baseline 1 is relatively smaller compared to that of baseline 2.
In fact, the passive beamforming achieved by our proposed phase control at IRS not only focuses the energy on the LoS path, but it can also suppress the signal energy propagating through the scattering paths in Rician fading channels.
In other words, the composite channel from the UAV to ground users in IRS-assisted systems is more deterministic compared to that of baseline 2, which is equivalent to the effect of increasing the Rician factor and thus yields a higher average system outage rate.

\vspace{-4mm}
\subsection{Optimization of IRS's Location}
\begin{figure}
	\centering\vspace{-7mm}
	\includegraphics[width=4in]{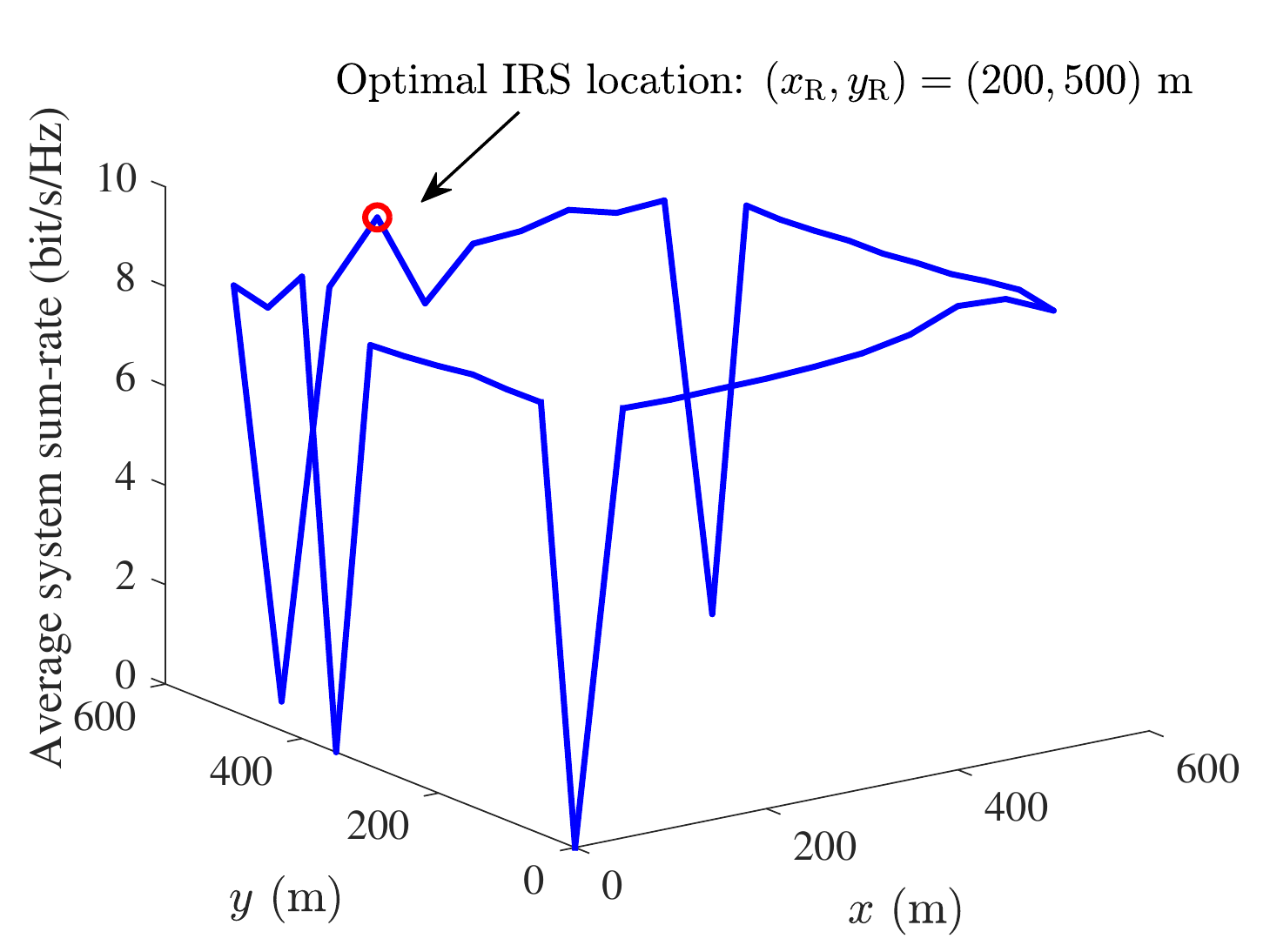}\vspace{-8mm}
	\caption{Average system sum-rate (bit/s/Hz) versus the IRS horizontal location.}\vspace{-10mm}
	\label{SumRateVersusLocation}%
\end{figure}

{Assuming a fixed IRS deployment height of $H_{\rm R} = 30$ m, there are only two horizontal coordinates to be designed and thus the optimal IRS location can be obtained via exhaustive searching.
	In Fig. \ref{SumRateVersusLocation}, we evaluate the system performance w.r.t. the IRS's horizontal location with $p_{\mathrm{max}} = 35$ dBm, $R_{\mathrm{min},k} = 1$ bit/s/Hz, $M_{\mathrm{r}} = M_{\mathrm{c}} = 200$, and $N = 500$.
	Note that the IRS is restricted to be localized on the boundary of the considered service area to be in sight of both the UAV and ground users.
	We can observe that the optimal IRS location is $(x_{\rm R},y_{\rm R}) = (200, 500)$ m.
	Comparing the users' layout in Fig. \ref{GeometrySetup}, we can conclude that deploying IRS at the boundary but close to the area with a high density of users is more beneficial for improving the system sum-rate.}

\vspace{-4mm}
\section{Conclusions}
{This paper proposed a novel IRS-assisted UAV OFDMA communication system and studied its joint trajectory, IRS scheduling, and resource allocation design to maximize the system sum-rate.
	Although the composite channel suffers from both frequency- and spatial-selective fadings due to the existence of the IRS, we proposed a parametric approximation approach to facilitate the tractability of the UAV's trajectory design.
	An alternating optimization approach was adopted to design the resource allocation and IRS scheduling strategy as well as the UAV's trajectory.
	Extensive simulations were conducted to demonstrate the system sum-rate improvement via deploying an IRS in a UAV OFDMA communication system.
	Our results unveil that (1) the substantial beamforming gain offered by the IRS and the high maneuverability of the UAV are both vital for improving the communication performance;
	(2) the size of the IRS significantly affects the trajectory of the UAV in exploiting the degrees of freedom of the system to improve the achievable rate of all the users.}
\vspace{-4mm}

\bibliographystyle{IEEEtran}
\bibliography{NOMA}

\begin{thebibliography}{10}
\providecommand{\url}[1]{#1}
\csname url@samestyle\endcsname
\providecommand{\newblock}{\relax}
\providecommand{\bibinfo}[2]{#2}
\providecommand{\BIBentrySTDinterwordspacing}{\spaceskip=0pt\relax}
\providecommand{\BIBentryALTinterwordstretchfactor}{4}
\providecommand{\BIBentryALTinterwordspacing}{\spaceskip=\fontdimen2\font plus
\BIBentryALTinterwordstretchfactor\fontdimen3\font minus
  \fontdimen4\font\relax}
\providecommand{\BIBforeignlanguage}[2]{{%
\expandafter\ifx\csname l@#1\endcsname\relax
\typeout{** WARNING: IEEEtran.bst: No hyphenation pattern has been}%
\typeout{** loaded for the language `#1'. Using the pattern for}%
\typeout{** the default language instead.}%
\else
\language=\csname l@#1\endcsname
\fi
#2}}
\providecommand{\BIBdecl}{\relax}
\BIBdecl

\bibitem{zeng2019accessing}
Y.~{Zeng}, Q.~{Wu}, and R.~{Zhang}, ``Accessing from the sky: A tutorial on
  {UAV} communications for {5G} and beyond,'' \emph{Proceedings of the IEEE},
  vol. 107, no.~12, pp. 2327--2375, Dec. 2019.

\bibitem{ZengThroughMaxi}
Y.~{Zeng}, R.~{Zhang}, and T.~J. {Lim}, ``Throughput maximization for
  {UAV}-enabled mobile relaying systems,'' \emph{IEEE Trans. Commun.}, vol.~64,
  no.~12, pp. 4983--4996, Dec. 2016.

\bibitem{CaiUAVEESecure}
Y.~{Cai}, Z.~{Wei}, R.~{Li}, D.~W.~K. {Ng}, and J.~{Yuan}, ``Joint trajectory
  and resource allocation design for energy-efficient secure {UAV}
  communication systems,'' \emph{IEEE Trans. Commun.}, vol.~68, no.~7, pp.
  4536--4553, Mar. 2020.

\bibitem{ZengyongEnergyMinimization}
Y.~{Zeng}, J.~{Xu}, and R.~{Zhang}, ``Energy minimization for wireless
  communication with rotary-wing {UAV},'' \emph{IEEE Trans. Wireless Commun.},
  vol.~18, no.~4, pp. 2329--2345, Apr. 2019.

\bibitem{QingqingWuUAV}
Q.~{Wu}, Y.~{Zeng}, and R.~{Zhang}, ``Joint trajectory and communication design
  for multi-{UAV} enabled wireless networks,'' \emph{IEEE Trans. Wireless
  Commun.}, vol.~17, no.~3, pp. 2109--2121, Mar. 2018.

\bibitem{SunUAV3D}
Y.~{Sun}, D.~{Xu}, D.~W.~K. {Ng}, L.~{Dai}, and R.~{Schober}, ``Optimal
  {3D}-trajectory design and resource allocation for solar-powered {UAV}
  communication systems,'' \emph{IEEE Trans. Commun.}, vol.~67, no.~6, pp.
  4281--4298, Jun. 2019.

\bibitem{DongfangUAV}
D.~{Xu}, Y.~{Sun}, D.~W.~K. {Ng}, and R.~{Schober}, ``Multiuser {MISO UAV}
  communications in uncertain environments with no-fly zones: Robust trajectory
  and resource allocation design,'' \emph{IEEE Trans. Commun.}, vol.~68, no.~5,
  pp. 3153--3172, 2020.

\bibitem{di2019smart}
M.~Di~Renzo, M.~Debbah, D.-T. Phan-Huy, A.~Zappone, M.-S. Alouini, C.~Yuen,
  V.~Sciancalepore, G.~C. Alexandropoulos, J.~Hoydis, H.~Gacanin \emph{et~al.},
  ``Smart radio environments empowered by reconfigurable {AI} meta-surfaces: an
  idea whose time has come,'' \emph{EURASIP J. Wireless Commun. and
  Networking}, vol. 2019, no.~1, pp. 1--20, 2019.

\bibitem{zhang2019multiple}
J.~{Zhang}, E.~{Bj{\"o}rnson}, M.~{Matthaiou}, D.~W.~K. {Ng}, H.~{Yang}, and D.~J.
  {Love}, ``Prospective multiple antenna technologies for beyond {5G},''
  \emph{IEEE J. Select. Areas Commun.}, vol.~38, no.~8, pp. 1637--1660, Jun.
  2020.

\bibitem{Ntontin2019RIS}
M.~{Di Renzo}, K.~{Ntontin}, J.~{Song}, F.~H. {Danufane}, X.~{Qian},
  F.~{Lazarakis}, J.~{De Rosny}, D.~{Phan-Huy}, O.~{Simeone}, R.~{Zhang},
  M.~{Debbah}, G.~{Lerosey}, M.~{Fink}, S.~{Tretyakov}, and S.~{Shamai},
  ``Reconfigurable intelligent surfaces vs. relaying: Differences,
  similarities, and performance comparison,'' \emph{IEEE Open J. of the Commun.
  Society}, vol.~1, pp. 798--807, Jul. 2020.

\bibitem{qingqing2019towards}
Q.~{Wu} and R.~{Zhang}, ``Towards smart and reconfigurable environment:
  Intelligent reflecting surface aided wireless network,'' \emph{IEEE Commun.
  Mag.}, vol.~58, no.~1, pp. 106--112, Nov. 2020.

\bibitem{WuIRS}
------, ``Intelligent reflecting surface enhanced wireless network via joint
  active and passive beamforming,'' \emph{IEEE Trans. Wireless Commun.},
  vol.~18, no.~11, pp. 5394--5409, Nov. 2019.

\bibitem{WuIRSDiscrete}
------, ``Beamforming optimization for intelligent reflecting surface with
  discrete phase shifts,'' in \emph{Proc. IEEE Intern. Conf. on Acoust., Speech
  and Signal Process.}, May 2019, pp. 7830--7833.

\bibitem{ChongWenIRS}
C.~{Huang}, A.~{Zappone}, G.~C. {Alexandropoulos}, M.~{Debbah}, and C.~{Yuen},
  ``Reconfigurable intelligent surfaces for energy efficiency in wireless
  communication,'' \emph{IEEE Trans. Wireless Commun.}, vol.~18, no.~8, pp.
  4157--4170, Aug. 2019.

\bibitem{Di2019reflection}
M.~Di~Renzo and J.~Song, ``Reflection probability in wireless networks with
  metasurface-coated environmental objects: An approach based on random spatial
  processes,'' \emph{EURASIP Journal on Wireless Commun. and Networking}, vol.
  2019, no.~1, p.~99, 2019.

\bibitem{kishk2020exploitingV2}
M.~A. {Kishk} and M.~{Alouini}, ``Exploiting randomly-located blockages for
  large-scale deployment of intelligent surfaces,'' \emph{IEEE J. Select. Areas
  Commun.}, early access, 2020.

\bibitem{yu2019robust}
X.~{Yu}, D.~{Xu}, Y.~{Sun}, D.~W.~K. {Ng}, and R.~{Schober}, ``Robust and
  secure wireless communications via intelligent reflecting surfaces,''
  \emph{IEEE J. Select. Areas Commun.}, early access, 2020.

\bibitem{zheng2019intelligent}
B.~{Zheng} and R.~{Zhang}, ``Intelligent reflecting surface-enhanced {OFDM}:
  Channel estimation and reflection optimization,'' \emph{IEEE Wireless Commun.
  Lett.}, vol.~9, no.~4, pp. 518--522, Dec. 2020.

\bibitem{SekanderB5G}
S.~{Sekander}, H.~{Tabassum}, and E.~{Hossain}, ``Multi-tier drone architecture
  for {5G/B5G} cellular networks: Challenges, trends, and prospects,''
  \emph{IEEE Commun. Mag.}, vol.~56, no.~3, pp. 96--103, Mar. 2018.

\bibitem{LiIoT}
B.~{Li}, Z.~{Fei}, and Y.~{Zhang}, ``{UAV} communications for {5G} and beyond:
  Recent advances and future trends,'' \emph{IEEE Internet Things J.}, vol.~6,
  no.~2, pp. 2241--2263, Apr. 2019.

\bibitem{DMTadeoff}
L.~Zheng and D.~N.~C. Tse, ``Diversity and multiplexing: a fundamental tradeoff
  in multiple-antenna channels,'' \emph{IEEE Trans. Inf. Theory}, vol.~49,
  no.~5, pp. 1073--1096, May 2003.

\bibitem{ShaHuLIS}
S.~{Hu}, F.~{Rusek}, and O.~{Edfors}, ``Beyond massive {MIMO}: The potential of
  data transmission with large intelligent surfaces,'' \emph{IEEE Trans. Signal
  Process.}, vol.~66, no.~10, pp. 2746--2758, Mar. 2018.

\bibitem{NadeemIRS}
Q.~{Nadeem}, A.~{Kammoun}, A.~{Chaaban}, M.~{Debbah}, and M.~{Alouini},
  ``Asymptotic max-min {SINR} analysis of reconfigurable intelligent surface
  assisted {MISO} systems,'' \emph{IEEE Trans. Wireless Commun.}, early access,
  2020.

\bibitem{RuideLi}
R.~{Li}, Z.~{Wei}, L.~{Yang}, D.~W.~K. {Ng}, J.~{Yuan}, and J.~{An}, ``Resource
  allocation for secure multi-{UAV} communication systems with
  multi-eavesdropper,'' \emph{IEEE Trans. Commun.}, vol.~68, no.~7, pp.
  4490--4506, Mar. 2020.

\bibitem{You3DUAV}
C.~{You} and R.~{Zhang}, ``{3D} trajectory optimization in rician fading for
  {UAV}-enabled data harvesting,'' \emph{IEEE Trans. Wireless Commun.},
  vol.~18, no.~6, pp. 3192--3207, Jun. 2019.

\bibitem{zhang2019reflections}
Q.~{Zhang}, W.~{Saad}, and M.~{Bennis}, ``Reflections in the sky: Millimeter
  wave communication with {UAV-carried} intelligent reflectors,'' in
  \emph{Proc. IEEE Global Commun. Conf.}, Dec. 2019, pp. 1--6.

\bibitem{li2019reconfigurable}
S.~{Li}, B.~{Duo}, X.~{Yuan}, Y.~{Liang}, and M.~{Di Renzo}, ``Reconfigurable
  intelligent surface assisted {UAV} communication: Joint trajectory design and
  passive beamforming,'' \emph{IEEE Wireless Commun. Lett.}, vol.~9, no.~5, pp.
  716--720, Jan. 2020.

\bibitem{GeIRSUAV}
L.~{Ge}, P.~{Dong}, H.~{Zhang}, J.~{Wang}, and X.~{You}, ``Joint beamforming
  and trajectory optimization for intelligent reflecting surfaces-assisted
  {UAV} communications,'' \emph{IEEE Access}, vol.~8, pp. 78\,702--78\,712,
  2020.

\bibitem{PanMulticellIRS}
C.~{Pan}, H.~{Ren}, K.~{Wang}, W.~{Xu}, M.~{Elkashlan}, A.~{Nallanathan}, and
  L.~{Hanzo}, ``Multicell {MIMO} communications relying on intelligent
  reflecting surfaces,'' \emph{IEEE Trans. Wireless Commun.}, vol.~19, no.~8,
  pp. 5218--5233, May 2020.

\bibitem{DiFinitePS}
B.~{Di}, H.~{Zhang}, L.~{Li}, L.~{Song}, Y.~{Li}, and Z.~{Han}, ``Practical
  hybrid beamforming with finite-resolution phase shifters for reconfigurable
  intelligent surface based multi-user communications,'' \emph{IEEE Trans. Veh.
  Technol.}, vol.~69, no.~4, pp. 4565--4570, Feb. 2020.

\bibitem{Sohrabi2016}
F.~Sohrabi and W.~Yu, ``Hybrid digital and analog beamforming design for
  large-scale antenna arrays,'' \emph{IEEE J. Select. Areas Commun.}, vol.~10,
  no.~3, pp. 501--513, Apr. 2016.

\bibitem{DerrickEESWIPT}
D.~W.~K. Ng, E.~S. Lo, and R.~Schober, ``Wireless information and power
  transfer: Energy efficiency optimization in {OFDMA} systems,'' \emph{IEEE
  Trans. Wireless Commun.}, vol.~12, no.~12, pp. 6352--6370, Dec. 2013.

\bibitem{WuMultiUAV}
Q.~{Wu}, Y.~{Zeng}, and R.~{Zhang}, ``Joint trajectory and communication design
  for multi-{UAV} enabled wireless networks,'' \emph{IEEE Trans. Wireless
  Commun.}, vol.~17, no.~3, pp. 2109--2121, Jan. 2018.

\bibitem{NguyenUAV}
H.~C. {Nguyen}, R.~{Amorim}, J.~{Wigard}, I.~Z. {Kov{\`a}Cs}, T.~B. {S{\o}rensen}, and
  P.~E. {Mogensen}, ``How to ensure reliable connectivity for aerial vehicles
  over cellular networks,'' \emph{IEEE Access}, vol.~6, pp. 12\,304--12\,317,
  2018.

\bibitem{SimunekUAV}
M.~{Simunek}, P.~{Pechac}, and F.~P. {Fontan}, ``Excess loss model for low
  elevation links in urban areas for {UAVs},'' \emph{Radio engineering},
  vol.~20, no.~3, pp. 561--568, Sep. 2011.

\bibitem{AlkhateebWideBandChannel}
A.~{Alkhateeb} and R.~W. {Heath}, ``Frequency selective hybrid precoding for
  limited feedback millimeter wave systems,'' \emph{IEEE Trans. Commun.},
  vol.~64, no.~5, pp. 1801--1818, May 2016.

\bibitem{lemons2002introduction}
D.~S. Lemons and P.~Langevin, \emph{An introduction to stochastic processes in
  physics}.\hskip 1em plus 0.5em minus 0.4em\relax JHU Press, 2002.

\bibitem{MostofiICI}
Y.~{Mostofi} and D.~C. {Cox}, ``{ICI} mitigation for pilot-aided {OFDM} mobile
  systems,'' \emph{IEEE Trans. Wireless Commun.}, vol.~4, no.~2, pp. 765--774,
  Apr. 2005.

\bibitem{basar2019reconfigurable}
E.~Basar and I.~F. Akyildiz, ``Reconfigurable intelligent surfaces for doppler
  effect and multipath fading mitigation,'' \emph{arXiv preprint
  arXiv:1912.04080}, 2019.

\bibitem{YangIRSOFDMA}
Y.~{Yang}, S.~{Zhang}, and R.~{Zhang}, ``{IRS-Enhanced OFDMA}: Joint resource
  allocation and passive beamforming optimization,'' \emph{IEEE Wireless
  Commun. Lett.}, vol.~9, no.~6, pp. 760--764, Jan. 2020.

\bibitem{RahmatiUAVInterference}
A.~{Rahmati}, S.~{Hosseinalipour}, Y.~{Yapici}, X.~{He}, I.~{Guvenc}, H.~{Dai},
  and A.~{Bhuyan}, ``Interference avoidance in {UAV}-assisted networks: Joint
  {3D} trajectory design and power allocation,'' in \emph{Proc. IEEE Global
  Commun. Conf.}, 2019, pp. 1--6.

\bibitem{HosseinalipourUAV}
S.~{Hosseinalipour}, A.~{Rahmati}, and H.~{Dai}, ``Interference avoidance
  position planning in dual-hop and multi-hop {UAV} relay networks,''
  \emph{IEEE Trans. Wireless Commun.}, early access, 2020.

\bibitem{Boyd2004}
S.~Boyd and L.~Vandenberghe, \emph{Convex optimization}.\hskip 1em plus 0.5em
  minus 0.4em\relax Cambridge university press, 2004.

\bibitem{wei2018multibeam}
Z.~Wei, L.~Zhao, J.~Guo, D.~W.~K. Ng, and J.~Yuan, ``Multi-beam {NOMA} for
  hybrid mmwave systems,'' \emph{IEEE Trans. Commun.}, vol.~67, no.~2, pp.
  1705--1719, Feb. 2019.

\bibitem{tang2019wireless}
W.~Tang, M.~Z. Chen, X.~Chen, J.~Y. Dai, Y.~Han, M.~Di~Renzo, Y.~Zeng, S.~Jin,
  Q.~Cheng, and T.~J. Cui, ``Wireless communications with reconfigurable
  intelligent surface: Path loss modeling and experimental measurement,''
  \emph{arXiv preprint arXiv:1911.05326}, 2019.

\bibitem{DerrickEEOFDMA}
D.~W.~K. Ng, E.~S. Lo, and R.~Schober, ``Energy-efficient resource allocation
  in {OFDMA} systems with large numbers of base station antennas,'' \emph{IEEE
  Trans. Wireless Commun.}, vol.~11, no.~9, pp. 3292--3304, Sep. 2012.

\bibitem{Sun2016Fullduplex}
Y.~Sun, D.~W.~K. Ng, Z.~Ding, and R.~Schober, ``Optimal joint power and
  subcarrier allocation for full-duplex multicarrier non-orthogonal multiple
  access systems,'' \emph{IEEE Trans. Commun.}, vol.~65, no.~3, pp. 1077--1091,
  Mar. 2017.

\bibitem{WeiTCOM2017}
Z.~Wei, D.~W.~K. Ng, J.~Yuan, and H.~M. Wang, ``Optimal resource allocation for
  power-efficient {MC-NOMA} with imperfect channel state information,''
  \emph{IEEE Trans. Commun.}, vol.~65, no.~9, pp. 3944--3961, May 2017.

\end{thebibliography}

\end{document}